\documentclass[11pt]{article}
\pdfoutput=1

\usepackage{jheppub}
\usepackage{bigints}
\usepackage{rotating}

\usepackage{tabularx}

\usepackage{tikz}
\usepackage{fancybox}
 
\usepackage{tikz-cd}
\usepackage{ae}
\usepackage{url}

\usepackage{subfig}

\usepackage{tabularx}

\usepackage{graphicx}
\usepackage{dsfont}
\usepackage{setspace}
\usepackage{amsfonts,amsmath,amsthm,amssymb,amsbsy}
\usepackage{longtable}
\usepackage{array}
\usepackage{color}
\usepackage{needspace}
\usepackage[numbers]{natbib}
\usepackage{colortbl}

\usepackage{enumitem}

\usepackage{fancyhdr}

\usepackage{tabularx}

\usepackage{graphicx}
\usepackage{dsfont}
\usepackage{setspace}
\usepackage{amsfonts,amsmath,amsthm,amssymb,amsbsy}
\usepackage{longtable}
\usepackage[latin1]{inputenc}
\usepackage{array}
\usepackage{color}
\usepackage{needspace}
\usepackage[numbers]{natbib}
\usepackage{colortbl}

\usepackage{shuffle}

\usepackage{xcolor}
\definecolor{colourO}{HTML}{4FEEBB}
\definecolor{colourL}{HTML}{5DB197}
\definecolor{colourG}{HTML}{3E7CEF}
\definecolor{colourK}{HTML}{3535E3}

\usepackage{mathtools}

\usepackage{enumitem}

\usepackage{fancyhdr}

\usepackage{booktabs}
\usepackage{tcolorbox}
\usepackage{tabularx}

\usepackage{tensor}
\usepackage{mathtools}
\usepackage{blkarray, bigstrut}
\usepackage{soul}

\title{Kleiss-Kuijf Relations from Momentum Amplituhedron Geometry}

\author[1]{David Damgaard,}\emailAdd{d.damgaard@lmu.de}
\author[1,2]{Livia Ferro,}\emailAdd{livia.ferro@lmu.de}
\author[2]{Tomasz \L ukowski,}\emailAdd{t.lukowski@herts.ac.uk}
\author[2]{and Robert Moerman}\emailAdd{r.moerman@herts.ac.uk}

\affiliation[1]{Arnold--Sommerfeld--Center for Theoretical Physics,\\ Ludwig--Maximilians--Universit\"at, \\ Theresienstra\ss e 37, 80333 M\"unchen, Germany }
\affiliation[2]{Department of Physics, Astronomy and Mathematics, \\ University of Hertfordshire, \\  Hatfield, Hertfordshire, AL10 9AB, United Kingdom}

\abstract{In recent years, it has been understood that color-ordered scattering amplitudes can be encoded as logarithmic differential forms on positive geometries. In particular, amplitudes in maximally supersymmetric Yang-Mills theory in spinor helicity space are governed by the momentum amplituhedron. Due to the group-theoretic structure underlying color decompositions, color-ordered amplitudes enjoy various identities which relate different orderings. In this paper, we show how the Kleiss-Kuijf relations arise from the geometry of the momentum amplituhedron.  
We also show how similar relations can be realised for the kinematic associahedron, which is the positive geometry of bi-adjoint scalar cubic theory.}
\begin{document}
\begin{flushright}
{\small LMU-ASC 07/21}
\end{flushright}
\maketitle

%%%%%%%%%%%%%%%%%%%%%%%%%%%%%%%%%%%%%%%%
%%%%%%%%%%%%%%%%%%%%%%%%%%%%%%%%%%%%%%%%
%Introduction
%%%%%%%%%%%%%%%%%%%%%%%%%%%%%%%%%%%%%%%%
%%%%%%%%%%%%%%%%%%%%%%%%%%%%%%%%%%%%%%%%

\section{Introduction}

Scattering amplitudes are central quantities in fundamental physics. Being the building blocks for cross sections measured in high-energy colliders, amplitudes provide the bridge between theoretical descriptions and experimental observations. Throughout the decades, significant effort has been devoted to searching for increasingly more efficient methods for computing them, and tremendous progress has been achieved which not only provides us with more powerful tools for calculations, but also with an enhanced understanding of the underlying theories. 

One basic, well-established technique which simplifies calculations in gauge theories is  \emph{color decomposition}. This decomposition disentangles the color and kinematic degrees of freedom, and repackages the latter into objects called \emph{color-ordered} or \emph{partial} amplitudes, which are gauge invariant, easier to compute and encode purely kinematic information. 
At tree level, a standard decomposition for amplitudes $\mathcal{A}_n^{\mathrm{tree}}$
 involving $n$ particles which transform in the adjoint representation of the gauge group with generators $T^{a}$ is
\begin{equation}
\label{color1}
\mathcal{A}_n^{\mathrm{tree}} = \sum_{\sigma} \mathrm{Tr}\left(T^{a_{1}} T^{{\sigma(a_2}} \cdots T^{a_n)}\right) A_n^{\mathrm{tree}} [1, \sigma(2 \ldots n)] \,,
\end{equation}
where the sum is over  the $(n-1)!$ non-cyclic permutations of particle labels for which particle $1$ has been fixed to the first position. Here, $A_n^{\mathrm{tree}}$ are the color-ordered partial amplitudes which depend only on the momenta and type of the external particles. 
The partial amplitudes $A_n^{\mathrm{tree}}[i_1 ,i_2,\ldots,i_n]$ are not all independent: they are cyclically symmetric, invariant under reflections, and they obey $U(1)$ decoupling identities. They also satisfy the Kleiss-Kuijf (KK) relations \cite{Kleiss:1988ne,DelDuca:1999rs}, which arise from the fact that the amplitude $\mathcal{A}_n^{\mathrm{tree}}$ can be alternatively expanded in a basis formed from the gauge group structure constants. The KK relations
\begin{equation}
\label{KK-intro}
A_n^{\mathrm{tree}}[1,\{\alpha\},n,\{\beta\}] = (-1)^{n_{\beta}} \sum_{\sigma} A_n^{\mathrm{tree}}[1,\sigma(\{\alpha\}\{\beta^T\}),n] \,,
 \end{equation}
  where the sum is over certain permutations of the sets $\{\alpha\}$ and $\{\beta\}$ of external particles,
 are consistent with all previous identities, and they further reduce the number of linearly independent partial amplitudes to $(n-2)!$.
Importantly, the $U(1)$ decoupling identities are special cases of the KK relations.

In the standard Feynman approach, color-ordered amplitudes are easier to compute, as they carry purely kinematic information and receive contributions only from Feynman diagrams with a particular cyclic ordering. In recent years, it has also become apparent that, at least in some theories, these amplitudes can be calculated using a novel approach based on geometry. To establish the link with geometry, partial amplitudes need to be thought of as differential forms, rather than functions, on kinematic space, i.e.~the space of physical kinematic variables. These differential forms are, in fact,  canonical differential forms of \emph{positive geometries} \cite{Arkani-Hamed:2017tmz}, i.e.~real, oriented, closed geometries with boundaries of all co-dimension  equipped with differential forms which have logarithmic singularities along all boundaries.  The prime example of a positive geometry has been the amplituhedron \cite{Arkani-Hamed:2013jha}, which encodes tree-level and (integrands of) loop-level amplitudes in planar $\mathcal{N}=4$ supersymmetric Yang-Mills (sYM) in momentum twistor space. This geometry is strictly related to the planar sector since the ordering is embedded in the construction of momentum twistors. 
In the spinor helicity space, the tree-level amplitudes in $\mathcal{N}=4$ sYM are described by the momentum amplituhedron \cite{Damgaard:2019ztj}.

The natural question to ask is how the aforementioned relations between partial amplitudes emerge from this geometric description and whether they can be derived using positive geometries. 
As we will show in this paper, the answer is affirmative for tree-level amplitudes in $\mathcal{N}=4$ sYM, as well as for those in bi-adjoint scalar cubic theory,  where positive geometries provide a beautiful geometrical realization of the KK relations!

The momentum amplituhedron provides us with the right framework for studying the KK relations amongst partial amplitudes in $\mathcal{N}=4$ sYM, 
because it is defined in terms of spinor helicity variables where no specific ordering is enforced (in contradistinction with the definition of the amplituhedron in momentum twistor space), and we can freely consider amplitudes for any ordering of external particles, as demanded by \eqref{KK-intro}. We will show that the KK relations can be realised geometrically as collections of momentum amplituhedra which combine to form  bounded regions without vertices, to be defined shortly. Specifically, for each ordering $\sigma$ of $n$ particles, and for a fixed helicity sector $k$, the momentum amplituhedron $\mathcal{M}_{n,k}^{(\sigma)}$ is a subset of a particular $(2n-4)$-dimensional subspace of the spinor helicity kinematic space. It is equipped with a differential form $\Omega_{n,k}^{(\sigma)}$ which has logarithmic singularities on all boundaries of $\mathcal{M}_{n,k}^{(\sigma)}$.  For two orderings $\sigma\neq \tau$, the momentum amplituhedra $\mathcal{M}_{n,k}^{(\sigma)}$ and $\mathcal{M}_{n,k}^{(\tau)}$ do not overlap, but they do share common boundaries. In the simplest scenario, these two regions share a co-dimension one boundary, and the singularity associated with this boundary vanishes in the sum of canonical forms $\Omega_{n,k}^{(\sigma)}+\Omega_{n,k}^{(\tau)}$. More generally, such cancellations can occur deeper into the geometry, after taking a sequence of boundary operations. For the purposes of this paper, we are interested in finding sums of momentum amplituhedra for which these cancellations take place for all possible sequences of $2n-4$ boundary operations. Such sums of positive geometries are 
no longer positive geometries as they do not possess boundaries of all co-dimensions, specifically zero-dimensional boundaries, i.e.~vertices, and the sums of the corresponding  
canonical differential forms vanish. In this paper we will show that all KK relations for $\mathcal{N}=4$ sYM descend from such sums of momentum amplituhedra, providing a geometric origin for these relations. While for maximally-helicity violating (MHV) amplitudes it is possible to describe this geometric construction using the language of polyhedral geometry, for higher helicity sectors we will use a homological approach based on the known boundary structure of the momentum amplituhedron \cite{Ferro:2020lgp}.

A similar construction exists for scattering amplitudes in bi-adjoint scalar cubic theory, for which the positive geometry is the kinematic associahedron \cite{Arkani-Hamed:2017mur}. In order to derive the KK relations in this setting we will provide a new definition of kinematic associahedra which are relevant for orderings different from the standard one. In this new definition, all associahedra for different orderings live on the same affine subspace inside the kinematic space, and therefore we will be able to compare them directly. Since the kinematic associahedron is a polytope, our construction significantly simplifies and mimics the polyhedral construction for MHV amplitudes in $\mathcal{N}=4$ sYM alluded to earlier.   

This paper is organised as follows. In Sec.~\ref{sec:relations}, we review the color structure of scattering amplitudes in gauge theories, and collect the identities which they satisfy. In Sec.~\ref{sec:posgeometries}, after a review of positive geometries, we show how to add them and explain when their sum might fail to be a positive geometry. We recall the definition of the momentum amplituhedron in Sec.~\ref{sec:momampl}. In Sec.~\ref{sec:geometry} we present the main result of this paper: we show how amplitude relations originate geometrically from the momentum amplituhedron. In particular, we present two approaches: a ray-based approach, valid for MHV amplitudes, and a poset-based approach, applicable to any $n$ and $k$. In that section we also present explicit examples illustrating our construction. In Sec.~\ref{sec:associahedron} we expand our results to the KK relations for the bi-adjoint  scalar $\phi^3$ theory and explain how they arise geometrically from the kinematic associahedron. Conclusions and an outlook close the paper.

%%%%%%%%%%%%%%%
% Relations between Amplitudes
%%%%%%%%%%%%%%%

\section{Color Structure for Gauge Theory Amplitudes}
\label{sec:relations}

In this section we give more details on how the color structure of $SU(N)$ gauge amplitudes is organised; we recall the definition of color-ordered amplitudes and the relations between them. For a more extensive review see e.g.~\cite{Dixon:2011xs}.
We start by considering the trace decomposition, where the color factors are written in terms of the generators of the gauge group. In the case of tree-level amplitudes with external states in the adjoint representation, such as is the case for the states in $\mathcal{N}=4$ sYM, we have the following trace-based color decomposition
\begin{equation}
\label{trace-basis}
\mathcal{A}_n^{\mathrm{tree}}(\{p_i,h_i,a_i\}) = g^{n-2} \sum_{\sigma\in \mathcal{O}_n} \mathrm{Tr}\left(T^{a_{1}} T^{a_{\sigma(2)}} \cdots T^{a_{\sigma(n)}}\right) A_n^{\mathrm{tree}} [1^{h_1}, \sigma(2^{h_2}), \ldots, \sigma(n^{h_n})] \,,
\end{equation}
where $\mathcal{O}_n \cong S_n/Z_n \cong S_{n-1}$ is the set of $(n-1)!$ non-cyclic permutations of the $n$ particles, where the position of particle $1$ has been fixed to the first entry using the cyclic invariance of the trace, and $T^{a_i}$ are the generators of $SU(N)$ with adjoint indices $a_i=1,\ldots,N^2-1$. 
The full amplitude $\mathcal{A}_n^{\mathrm{tree}}$ is a function of the momenta $p_i$ and helicities $h_i$ of the external particles, as well as of the color indices $a_i$.
The objects $A_n^{\mathrm{tree}} $ are called color-ordered or partial amplitudes and carry only kinematic information, since the color dependence has been stripped off.  
They receive contributions only from planar diagrams in a particular ordering and therefore have singularities only when the sum of adjacent momenta in this ordering go on-shell.
The $(n-1)!$ color-ordered amplitudes are not all independent and enjoy various  relations:
\begin{itemize}
\item
Cyclicity: $A_n[1,2,\ldots,n] = A_n[2,\ldots,n,1]=\ldots =A_n[n,1,2,\ldots,n-1]$\,,
\item
Reflection symmetry: $A_n[1,2,\ldots,n] =  (-1)^n A_n[n,\ldots,2,1]\,$,
\item
$U(1)$ decoupling identity: $A_n[1,2,\ldots,n] + A_n[2,1,3,\ldots,n] + \ldots + A_n[2,3,\ldots,1,n] = 0$\,.
\end{itemize}

There also exists another color decomposition, in terms of the structure constants $f^{abc}$ of the gauge group rather than the traces of the generators, which reads:
\begin{equation}
\label{f-basis}
\mathcal{A}_n^{\mathrm{tree}}(\{p_i,h_i,a_i\}) =  g^{n-2} \sum_{\sigma\in S_{n-2}} f^{a_1 a_{\sigma_1} b_1}  f^{b_1 a_{\sigma_2} b_2} \ldots  f^{b_{n-3} a_{\sigma_{n-2}} a_n} A_n^{\mathrm{tree}} [1, \sigma(2 \ldots n-1), n]\,,
\end{equation}
where now the sum is over $(n-2)!$ elements, rather than $(n-1)!$. This exposes a larger class of identities for the partial amplitudes called the Kleiss-Kuijf (KK) relations \cite{Kleiss:1988ne,DelDuca:1999rs} 
\begin{equation}
\label{KK}
A_n^{\mathrm{tree}}(1,\{\alpha\},n,\{\beta\}) = (-1)^{n_{\beta}} \sum_{\omega \in \{\alpha\}\shuffle\{\beta^T\}} A_n^{\mathrm{tree}}(1,\{\omega\},n) \,,
 \end{equation}
where $\{\alpha\}$ and $\{\beta\}$ are disjoint sets of external particle lables with $\{\alpha\}\cup \{\beta\} = \{2,3,\ldots,n-1\}$.
$\{\beta^T\}$ denotes the reverse ordering of the labels $\{\beta\}$,
$n_{\beta}$ is the number of elements in $\{\beta\}$ and $\{\alpha\}\shuffle\{\beta^T\}$ denotes the set of all shuffles of $\{\alpha\}$ with $\{\beta^T\}$, i.e.~the set of permutations on  $\{\alpha\}\cup \{\beta\}$ preserving the ordering within $\{\alpha\}$  and $\{\beta^T\}$. These relations can be used to put any two legs next to each other, these being $1$ and $n$ in \eqref{KK}.
Both the reflection symmetry relations and the $U(1)$ decoupling identities are  particular cases of the KK relations.

%%%%%%%%%%%%%%%
% Positive geometries 
%%%%%%%%%%%%%%%

\section{Positive Geometries and How to Add Them}
\label{sec:posgeometries}

In the following sections we will use the momentum amplituhedron and kinematic associahedron to show how the KK relations arise from positive geometries. Both the momentum amplituhedron and the kinematic associahedron  are families of geometries, whose explicit shape can become very intricate. 
In particular, their dimensions grow with the number of particles. This complexity makes it difficult to see the geometric origin of the cancellations between differential forms which must occur in order to produce the KK relations. However, the general strategy we will employ in this paper can be easily explained using examples in two dimensions and this will be the purpose of this section.

Let us start by recalling the definition of a positive geometry \cite{Arkani-Hamed:2017tmz}. We take $X$ to be a complex projective variety of dimension $d\geq 0$ and $X_{\geq 0} \subset X(\mathbb{R})$ to be an oriented $d$-dimensional subset of its real slice. Then the pair $(X,X_{\geq0})$ is a $d$-dimensional positive geometry if it can be equipped with a unique non-zero logarithmic top-form $\Omega(X, X_{\geq 0})$, called the canonical form,  satisfying the following recursive condition:  for $d > 0$ every boundary component $(C, C_{\geq 0})$ of $\Omega(X, X_{\geq 0})$  is again a positive geometry of dimension $d-1$, whose canonical form is constrained by the residue relation 
$\mbox{Res}_C\, \Omega(X, X_{\geq 0}) = \Omega(C, C_{\geq 0})$
 and $\Omega(X, X_{\geq 0})$ has no singularities elsewhere. For $d = 0$,  $X_{\geq 0}$ is a single real point and $\Omega(X, X_{\geq 0})=\pm 1$ depending on the orientation of $X_{\geq 0}$. 
 We will often borrow language from polyhedral geometry and refer to co-dimension-one boundary components as facets, one-dimensional boundaries as edges, zero-dimensional boundaries as vertices, etc.
 To simplify our notation, from now on we will refer to a positive geometry $(X,X_{\geq 0})$ by keeping track only of its real part $X_{\geq 0}$.

Since every positive geometry comes equipped with a differential form, it is possible to combine positive geometries by means of adding their respective differential forms.
However, when two canonical differential forms are added,  the resulting form is not necessarily canonical, with leading singularities $\pm 1$. 
In this section we want to show in various scenarios what the possible outcomes of such sums are, and how to interpret them geometrically.

As was pointed out in \cite{Arkani-Hamed:2017tmz}, if one takes two positive geometries $X_1$ and $X_2$, with differential forms $\Omega_1$ and $\Omega_2$ respectively, such that their intersection is empty, $X_1\cap X_2=\emptyset$,  then their union $X_1 \cup X_2$, with the orientation inherited from $X_1$ and $X_2$, is a positive geometry with logarithmic canonical form $\Omega_{X_1\cup X_2}=\Omega_1+\Omega_2$. Instead, we will be interested in scenarios when positive geometries do intersect. 
We will consider two cases: when two (or more) positive geometries intersect only along their boundaries, and when one geometry is a subset of another. In order to properly account for the orientations of the geometries $X_1,X_2,\ldots X_p$, we will introduce the notion of an {\it oriented sum} of such geometries, which we denote by $X_1 \oplus X_2\oplus \ldots \oplus X_p$.

\begin{figure}[!h]
\begin{center}
\includegraphics[scale=0.5]{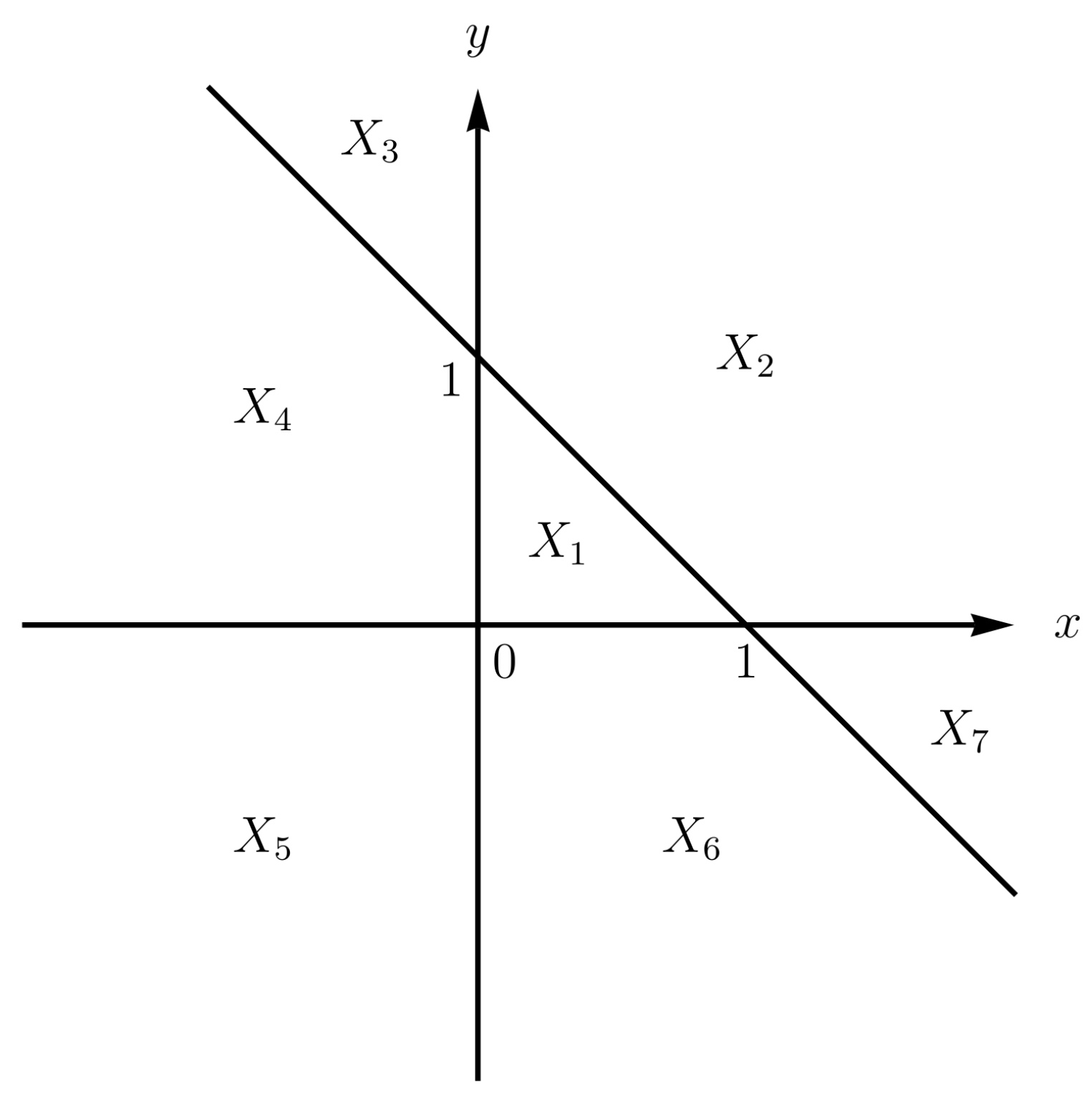}
\caption{Decomposition of the two-dimensional plane into positive geometries used to illustrate examples of oriented sums in the text.}
\label{fig:2dexample}
\end{center}
\end{figure}

To illustrate how the oriented sum is defined, let us consider the decomposition of the two-dimensional plane into regions depicted in Fig.~\ref{fig:2dexample}.
Moreover, let us denote the positive quadrant by $X_8=\{(x,y):x\geq 0,y\geq 0\}$, and $X_9=\{(x,y):x\leq 0,y\geq 0\}$. Each region $X_i$ for $i=1,\ldots,9$ is a positive geometry if we additionally equip it with an orientation. For each $X_i$ we have two possible choices: it can be oriented clockwise or counter-clockwise, which we denote by $X_{i}^{-}$ or $X_{i}^{+}$ respectively. In both cases, the  canonical forms differ only by an overall sign and we will denote them by $\Omega_i^-$ and $\Omega_i^+=-\Omega_i^-$, respectively.

Let us consider different scenarios which arise when we start to combine these differential forms:
\begin{itemize}[label=(I)]
\item
{\bf {Combinations of two geometries giving a positive geometry:}}
\end{itemize}
\begin{itemize}
\item 
$X_1^+\oplus X_2^+=X_8^+$. The resulting positive geometry is just the positive quadrant with the differential form
\begin{equation}
\Omega=\Omega_1^+\oplus \Omega_2^+=d\log x \wedge d\log y \,.
\end{equation}
The common boundary between the regions $X_1^+$ and $X_2^+$ is oriented in opposite ways, and therefore disappears in the sum. This type of behaviour is familiar from when we discuss triangulations of positive geometries, where a bigger positive geometry can be decomposed into a union of smaller geometries, with orientations such that singularities along spurious boundaries cancel in the sum of canonical forms.
\item  
$X_1^+\oplus X_5^-$. The resulting positive geometry is the region shaded in Fig.\ \ref{fig:2dexample_1} with the differential form
\begin{figure}[!h]
\begin{center}
\includegraphics[scale=0.5,clip]{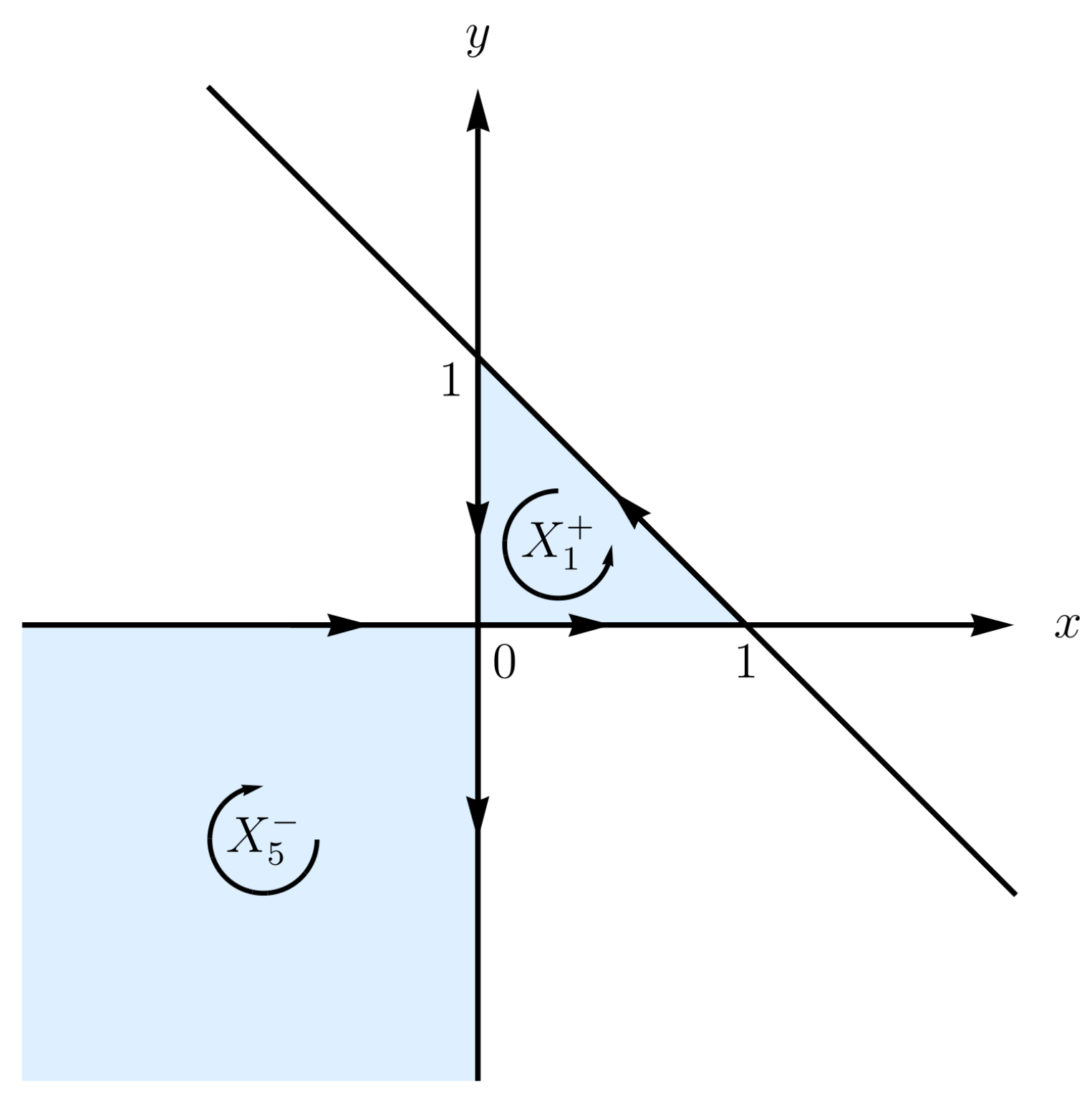}
\caption{Example of an oriented sum of positive geometries which is a positive geometry.}
\label{fig:2dexample_1}
\end{center}
\end{figure}
\begin{equation}
\Omega=\Omega_1^+\oplus \Omega_5^-=d\log x \wedge d\log (1-x-y)+d\log(1-x-y)  \wedge d\log y\,.
\end{equation}
Interestingly, although both geometries $X_1^+$ and $X_5^-$ have the point $(x,y)=(0,0)$ as a vertex, it is not a vertex of their oriented sum.  This can be explained by the residue calculation:
\begin{equation}
\mbox{res}_{x=0}\,\mbox{res}_{y=0}\, \Omega_1^+=-1\,,\qquad \mbox{res}_{x=0}\, \mbox{res}_{y=0}\, \Omega_5^-=1 \,,
\end{equation} 
and	
\begin{equation}
\mbox{res}_{y=0}\,\mbox{res}_{x=0}\, \Omega_1^+=1\,,\qquad \mbox{res}_{y=0}\, \mbox{res}_{x=0}\, \Omega_5^-=-1 \,,
\end{equation}
which implies that, when we arrive at this particular zero-dimensional boundary,  we can approach it from two different directions, and the resulting residues cancel each other, independent of the order in which we take these residues.  Although the origin is not a vertex of the oriented sum,  the combination $X_1^+\oplus X_5^-$ is a positive geometry, with two vertices at $(1,0)$ and $(0,1)$, which have residues $\pm 1$.
\item  
$X_1^+\oplus X_8^-=X^+_2$. In this scenario, one geometry is a subset of another and they also share common boundaries. The boundaries which are shared are oriented oppositely, and therefore they are not present in the oriented sum. 
\end{itemize}
\begin{itemize}[label=(II)]
\item
{\bf{Combinations of two geometries  \emph{not} giving a positive geometry:}}
\end{itemize}
\begin{itemize}
\item 
$X_1^+\oplus X_5^+$. In this case we get a similar picture as in Fig. \ref{fig:2dexample_1}, where the only difference is the orientation of the region $X_5$. The sum of the canonical differential forms $\Omega_1^++\Omega^+_5$ is no longer a canonical differential form. The reason for this is that when we calculate the residue of $\Omega_1^++\Omega^+_5$ at $x=0$, the resulting one-dimensional differential form has a singularity at $y=0$ with residue $\text{res}_{y=0}\text{res}_{x=0}(\Omega_1^++\Omega^+_5)=2$.  Since the residues at the two vertices $(1,0)$ and $(0,1)$ are $\pm1$, it is not possible to rescale the full differential form such that all vertices have residues $\pm 1$. Therefore this combination violates the definition of  positive geometry.
\item  
$X_8^+\oplus X_5^-$. The union of these regions is depicted in Fig. \ref{fig:2dexample_2}. The resulting geometry is the union of the positive and the negative quadrants. The orientations of these regions are aligned in such a way that the lines $x=0$ and $y=0$, i.e.~the one-dimensional boundaries of this geometry, are oriented from $\mp\infty$ to $\pm\infty$. With this particular orientation of the geometries, the sum of the canonical forms vanishes
\begin{equation}
\Omega=\Omega_8^++\Omega_5^-=d\log x\wedge d\log y-d\log x\wedge d\log y=0\,.
\end{equation}
As in a case discussed before, the origin is not a vertex of the oriented sum. Since in this case there are no other zero-dimensional boundaries in either $X_8^+$ or $X^-_5$, the oriented sum $X_8^+ \oplus X^-_5$ has no vertices. Consequently, not only is the oriented sum not a positive geometry, but the sum of the differential forms  $\Omega_8^+ \oplus \Omega^-_5$ must necessarily vanish.
\begin{figure}[!h]
\begin{center}
	\includegraphics[scale=0.5,clip]{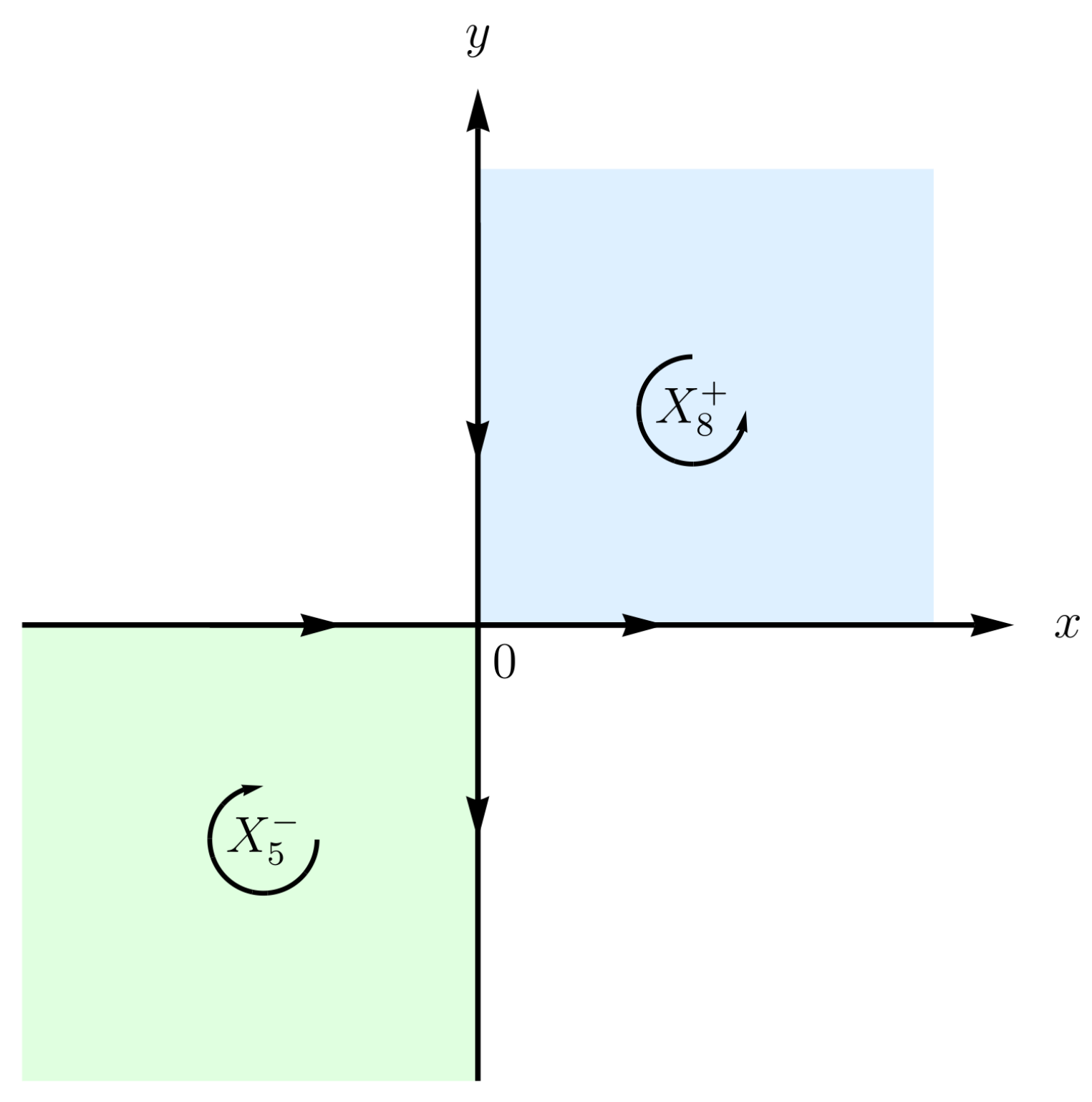}	
\caption{Example of an oriented sum of positive geometries which is not a positive geometry.}
\label{fig:2dexample_2}
\end{center}
\end{figure}
\item 
$X_8^+\oplus X_9^+$. The geometry which we obtain is the upper-half plane as illustrated in Fig. \ref{fig:2dexample_3}. The only one-dimensional boundary of this region is the line $y=0$ oriented from left to right. When treating the two geometries separately, we can evaluate residues along this line and we get 
\begin{equation}
\mbox{res}_{y=0}\, \Omega_8^+=-\frac{dy}{y}\,,\qquad \mbox{res}_{y=0}\, \Omega_9^+=+\frac{dy}{y} \,.
\end{equation}
The boundaries of both geometries $X_8^{+}$ and $X_9^{+}$, when restricted to the line $y=0$, are half-lines with the boundary at $x=0$. The residues at $x=0$ are
\begin{equation}
\mbox{res}_{x=0}\,\mbox{res}_{y=0}\, \Omega_8^+=-1\,,\qquad \mbox{res}_{x=0}\, \mbox{res}_{y=0}\, \Omega_9^+=+1 \,.
\end{equation}
Here we see that as we approach the zero-dimensional boundary, i.e.~the vertex $(x,y)=(0,0)$, from opposite directions along the line $y=0$, the residues of the canonical forms for each geometry produce opposite signs. If we now consider the sum of the two geometries, these residues/zero-dimensional canonical forms cancel and hence the sum $\Omega_8^++\Omega_9^+$ has a vanishing residue at the origin. We can explain this purely in geometric terms by observing that the oriented sum $X^+_8 \oplus X_9^{+}$ again does not have any zero-dimensional boundaries, and the orientations of both regions along the one-dimensional boundary match.
\begin{figure}[!h]
\begin{center}
	\includegraphics[scale=0.5,clip]{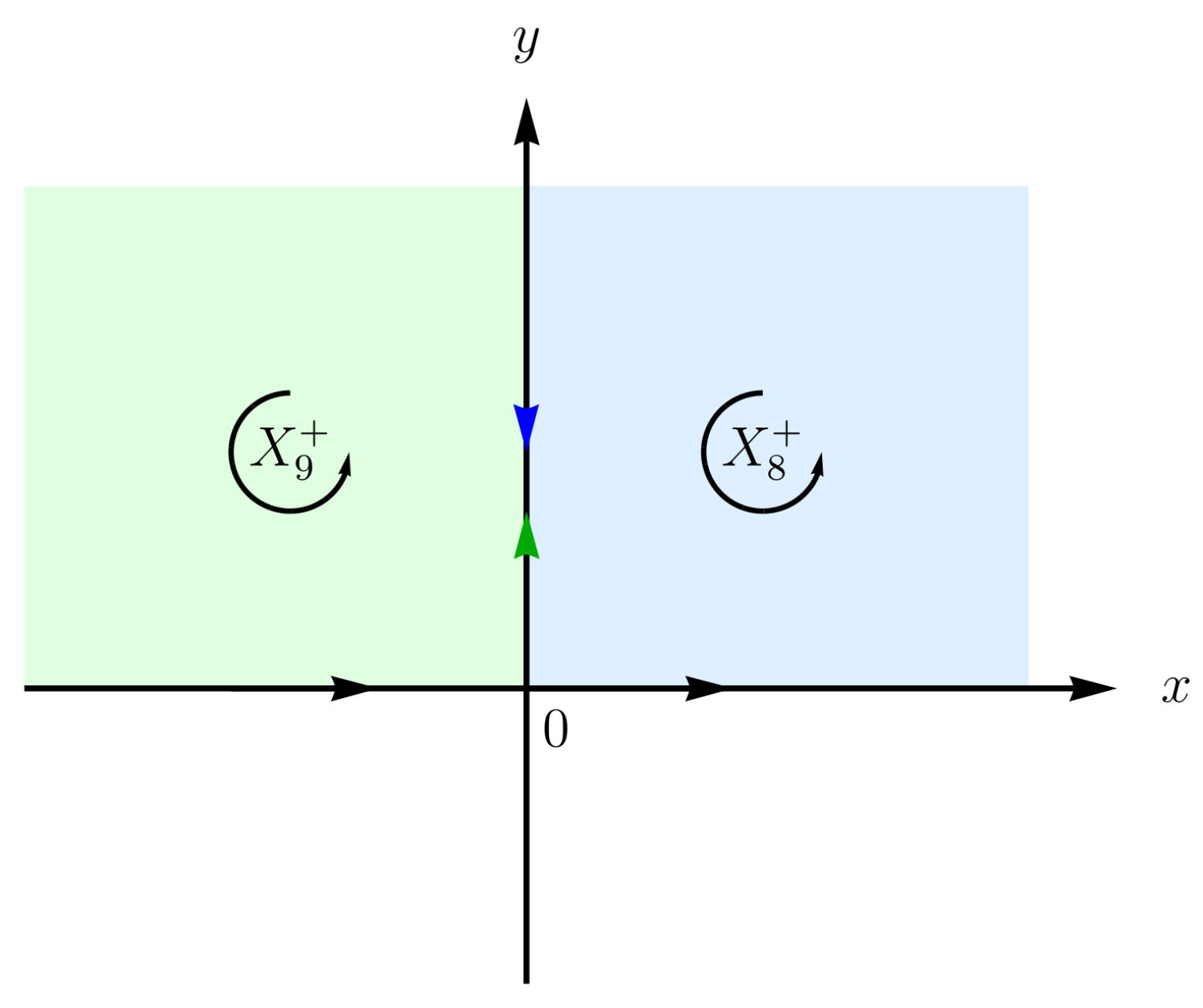}	
\caption{Example of an oriented sum of positive geometries which is not a positive geometry.}
\label{fig:2dexample_3}
\end{center}
\end{figure}
\end{itemize}

Our analysis can easily be extended to higher-dimensional positive geometries and to positive geometries with `curvy' boundaries, as for example is the case for the positive Grassmannian and the momentum amplituhedron. In this paper, we will be primarily interested in cases where the oriented sum of positive geometries is not a positive geometry because the sum of canonical differential forms is zero, as in the last two cases above. This statement can be checked by performing the residue calculation as we did above. However, even for more complicated cases, it is possible to determine for which collections of positive geometries the sum of their canonical forms will vanish by studying the boundary stratifications of the individual geometries we are combining. 
In particular, we claim:
\begin{center}
\emph{If the oriented sum of positive geometries has no vertices in its boundary stratification, \\then the sum of their canonical forms must vanish.}
\end{center}
In the following sections we develop methods to make this statement more precise. This will allow us to find all KK relations for $\mathcal{N}=4$ sYM and for the bi-adjoint $\phi^3$ theory purely from the geometry of momentum amplituhedron and kinematic associahedron respectively.

%%%%%%%%%%%%%%%
% Momentum Amplituhedron
%%%%%%%%%%%%%%%

\section{Momentum Amplituhedron}
\label{sec:momampl}

The momentum amplituhedron is the positive geometry associated with tree-level scattering amplitudes in $\mathcal{N}=4$ sYM in spinor helicity space \cite{Damgaard:2019ztj}. 
In $\mathcal{N}=4$ sYM, an $n$-particle superamplitude $\mathcal{A}_n = \mathcal{A}_n(\Phi_1, \Phi_2, \ldots,\Phi_n)$ -- where $\Phi_i$ are on-shell chiral superfields which collect the on-shell supermultiplet -- can be expanded in terms of helicity sectors, denoted by $k$, as follows:
\begin{equation}
\mathcal{A}_n = \mathcal{A}_{n,2} + \mathcal{ A}_{n,3} + \ldots +  \mathcal{A}_{n,n-2},\quad n\geq 4\,,
\end{equation}
where $\mathcal{A}_{n,2}$ is the maximally-helicity-violating (MHV) amplitude,  $ \mathcal{A}_{n,3}$ is the next-to-MHV (NMHV) amplitude and so on, with $\mathcal{A}_{n,k}$ the amplitude for the $\text{N}^{k-2}\text{MHV}$ sector and having Grassmann degree $4k$. In turn, each of the amplitudes $\mathcal{A}_{n,k}$ can be expanded into different color orderings, as in \eqref{trace-basis}. We denote the partial amplitude 
 with  ordering $\sigma$ by $A_{n,k}[\sigma]$.
In order to make connection to geometry, and therefore to interpret the amplitudes as differential forms, one needs to write them in    
the non-chiral superspace $(\lambda^a,\eta^r \,|\, \widetilde\lambda^{\dot a},\widetilde\eta^{\dot r})$, with indices $a,\dot a,r,\dot r=1,2$, where a Fourier transform for two of the four Grassmann-odd variables is performed. In this way,  via  the replacement
\begin{equation}
\label{lambda_eta}
\eta^a \to d\lambda^a\, ,\qquad\qquad \widetilde\eta^{\dot a}\to d\widetilde\lambda^{\dot a}\,,
\end{equation} 
the tree-level N$^{k-2}$MHV  scattering amplitudes can be written as differential forms of degree $(2(n-k),2k)$ in $(d\lambda,d\widetilde\lambda)$  \cite{He:2018okq}.  Importantly, due to supersymmetric Ward identities, this $2n$-form vanishes and one needs to strip off $(dq)^4$ or $(d\tilde q)^4$ to obtain the non-trivial information relevant for scattering amplitudes \cite{He:2018okq}. In this way, the scattering amplitude $A_{n,k}[\sigma]$ in $\mathcal{N}=4$ sYM can be translated into a differential form of degree $(2n-4)$ which is the canonical differential form of the momentum amplituhedron in ordering $\sigma$, as we will define shortly.

The momentum amplituhedron can be defined directly in terms of kinematic data in spinor helicity space and we start by recalling its definition for the standard ordering \cite{Ferro:2020ygk}. We define an affine subspace of the kinematic space
\begin{equation}\label{subspace.definition}
\mathcal{V}_{n,k} 
\equiv 
 \{(\lambda_i^a,\widetilde\lambda_i^{\dot a}):
     \lambda_i^a = \lambda^{*a}_i+y_{\alpha}^a \,\Delta_i^\alpha ,       \widetilde\lambda_i^{\dot a} = \widetilde\lambda^{*\dot{a}}_i+\widetilde y_{\dot{\alpha}}^{\dot{a}} \,\widetilde\Delta_i^{\dot\alpha},\lambda_i^a \widetilde\lambda_i^{\dot a}=0 
\}\,,
\end{equation}
where $(\lambda^*,\widetilde\lambda^*)$ are two fixed two-planes in $n$ dimensions, $\widetilde\Delta$ is a fixed $k$-plane and $\Delta$ is a fixed $(n-k)$-plane in $n$ dimensions. Moreover, we assume that when we assemble these subspaces as in 
\begin{equation}\label{big.lambda}
\Lambda_i^A = 
\begin{pmatrix} 
\lambda_i^{a *} \\
\Delta^\alpha_i
\end{pmatrix} ,
\qquad \widetilde\Lambda_i^{\dot A} = 
\begin{pmatrix} 
\widetilde\lambda_i^{\dot a *} \\
\widetilde\Delta^{\dot\alpha}_i
\end{pmatrix}  \, ,
\end{equation}
 $\widetilde\Lambda$ is a positive matrix and $\Lambda$ is a twisted positive matrix; see \cite{Lukowski:2020dpn} for a precise definition of the latter. Notice that $\mathcal{V}_{n,k}$ is a co-dimension-four subspace of an affine space of dimension $2n$. Next we define a winding space $\mathcal{W}_{n,k}$ as the subset of kinematic space 
satisfying the conditions \cite{He:2018okq}
\begin{align}
\label{winding.definition}
\mathcal{W}_{n,k}\equiv &\{(\lambda_i^a,\widetilde\lambda_i^{\dot a}):\langle i\ i+1\rangle\geq 0,[i\ i+1]\geq 0, s_{i,i+1,\ldots,i+j} \geq 0\,, \nonumber \\&
\mbox{the sequence } \{\langle 12\rangle,\langle 13\rangle,\ldots,\langle 1n\rangle\} \mbox{ has } k-2 \mbox{ sign flips}\,, \nonumber \\
&\mbox{the sequence } \{[ 12],[ 13],\ldots,[ 1n]\} \mbox{ has } k \mbox{ sign flips}\} \,,
\end{align}
where $s_{i,i+1,\ldots,i+j}$ are planar multiparticle Mandelstam variables: $s_{i,i+1,\ldots,i+j}=(p_i+p_{i+1}+\ldots+p_{i+j})^2$.
 Then the momentum amplituhedron $\mathcal{M}_{n,k}$ for the standard ordering is the intersection
\begin{equation}
\mathcal{M}_{n,k}\equiv \mathcal{V}_{n,k}\cap \mathcal{W}_{n,k}\,.
\end{equation}
The canonical differential form
 $\Omega_{n,k}$ of the momentum amplituhedron $\mathcal{M}_{n,k}$ has degree which is independent of $k$ and equals $2n-4$.
The scattering amplitude in non-chiral superspace can then be obtained as 
\begin{equation}
A_{n,k}[12\ldots n]=\delta^{(4)}(p)d^4p\wedge\Omega_{n,k}\Big|_{d\lambda\to\eta,d\tilde\lambda\to\tilde\eta} \,.
\end{equation}
An important information about the momentum amplituhedron which we will use later is its boundary stratification. This was found in \cite{Ferro:2020lgp} using the Mathematica$\textsuperscript{TM}$ package \texttt{amplituhedronBoundaries} \cite{Lukowski:2020bya} and can be easily generated for all values of $n$ and $k$. Importantly, each boundary of $\mathcal{M}_{n,k}$ is labelled by a cell in the positive Grassmannian $G_+(k,n)$, which in turn is labelled by an affine permutation; see \cite{bourjaily2012positroids} for details.  

In this paper we are interested in scattering amplitudes with various color orderings, we also need to introduce a definition of the momentum amplituhedron for orderings different from the standard one.
 At tree-level, scattering amplitudes in different color orderings can be obtained from the standard one simply by relabelling the momenta, or equivalently the spinor helicity variables, and we can write
\begin{equation}
A_{n,k}[\sigma]=A_{n,k}[12\ldots n]\big|_{\lambda_i\to\lambda_{\sigma(i)},\tilde\lambda_i\to\tilde\lambda_{\sigma(i)}}\,.
\end{equation}
To reflect this, we define the momentum amplituhedron  for the ordering $\sigma$, $\mathcal{M}_{n,k}^{(\sigma)}$, as the following intersection 
 \begin{equation}
\mathcal{M}_{n,k}^{(\sigma)}\equiv \mathcal{V}_{n,k}\cap \mathcal{W}^{(\sigma)}_{n,k}\,,
\end{equation}
where the subspace $\mathcal{V}_{n,k}$ is exactly the same as the one we used for the standard ordering \eqref{subspace.definition}, while the winding space with respect to the color ordering $\sigma$ is 
\begin{equation}
\mathcal{W}_{n,k}^{(\sigma)}=\{(\lambda_i,\tilde \lambda_i): (\lambda_{\sigma(i)},\tilde\lambda_{\sigma(i)})\in \mathcal{W}_{n,k}\} \,.
\end{equation}
We denote by $\Omega^{(\sigma)}_{n,k}$ the canonical differential form of $\mathcal{M}^{(\sigma)}_{n,k}$. In particular,
\begin{equation}
\Omega_{n,k}^{(\sigma)}=\Omega_{n,k}^{(12\ldots n)}\big|_{\lambda_i\to\lambda_{\sigma(i)},\tilde\lambda_i\to\tilde\lambda_{\sigma(i)}} \,,
\end{equation}
 and the scattering amplitude with the ordering $\sigma$ can be simply calculated as
\begin{equation}
A_{n,k}[\sigma]=\delta^{(4)}(p)d^4p\wedge\Omega_{n,k}^{(\sigma)}\Big|_{d\lambda\to\eta,d\tilde\lambda\to\tilde\eta}\,.
\end{equation}
In the following, we will not use the differential forms $\Omega_{n,k}^{(\sigma)}$ to derive the KK relations. Instead, we will show how to derive them using the boundary stratifications of  momentum amplituhedra for different orderings. To do that, we will use the fact that the boundary stratifications of $\mathcal{M}_{n,k}^{(\sigma)}$ are combinatorially isomorphic to the known boundary structure of the momentum amplituhedron in the standard ordering. 
Moreover, the zero-dimensional boundaries or vertices of  $\mathcal{M}_{n,k}$, of which there are precisely $\binom{n}{k}$, are shared by all particle orderings. In fact, in order to derive the KK relations from geometry, it is sufficient to study the boundary structure of momentum amplituhedra for different orderings around a single, shared vertex.

%%%%%%%%%%%%%%%
% Geometric Realization
%%%%%%%%%%%%%%%

\section{Kleiss-Kuijf Relations from the Momentum Amplituhedron Geometry}
\label{sec:geometry}

In the previous section, we discussed how the momentum amplituhedron for a given particle ordering is defined as the intersection of two regions: a proper-dimensional subspace of the spinor helicity space and a winding space which depends on the ordering. This definition 
does not make any explicit reference to information about color structure in the gauge theory, and it is interesting to understand how the KK relations between different color-ordered amplitudes arise in this purely geometric setting. 

A first attempt in this direction was presented in  \cite{Arkani-Hamed:2014bca}, where the KK relations were obtained for MHV amplitudes from the combinatorial properties of the positive Grassmannian. In this section, we derive the KK relations from the geometry of the momentum amplituhedron instead. We begin by studying the MHV case where it is easy to visualise these  relations for four and five particles. Thereafter, we present a general procedure for deriving the KK relations in any helicity sector and for any number of particles. This algorithm is homological in nature and it is based on the structure of boundaries of the momentum amplituhedron for different particle orderings.

\subsection{Simplicial Realization for MHV Amplitudes}
\label{sec:kk-mom-alpha}

Let us begin by considering the $k=2$ momentum amplituhedron $\mathcal{M}_{n,2}$ for the standard ordering. The proper-dimensional subspace $\mathcal{V}_{n,2}$ of the spinor helicity space given in \eqref{subspace.definition} is defined in terms of $(2n-4)$ $y$ variables and $4$ $\tilde{y}$ variables. These variables are constrained by $4$ equations coming from momentum conservation, which we can use to fix all $\tilde{y}$ variables in terms of $y$'s. Since the latter parametrise $\lambda$, then $\mathcal{V}_{n,2}$ is fully determined by $\lambda$. The winding space $\mathcal{W}_{n,2}$ given in \eqref{winding.definition} forces all ordered maximal minors of $\lambda$ to be positive: $\langle ij\rangle \geq0$ for all $1\le i <j\le n$. In this case, a natural parametrization for $\lambda$ is given by the $\alpha$-parametrization of the positive Grassmannian $G_+(2,n)$. For example, in the patch for which $\langle 12 \rangle \ne0$, we can parametrise $\lambda$ as 
\begin{equation}
\label{sub12n}
\mathcal{V}_{n,2}:\quad\lambda=\left(
\begin{array}{ccccccc}
1 & \sum_{i=1}^{n-2} \alpha_{2i} & \left(\sum_{i=1}^{n-3} \alpha_{2i} \right) \alpha_{2(n-2)-1}&   \left(\sum_{i=1}^{n-4} \alpha_{2i} \right) \alpha_{2(n-3)-1} & \ldots & \alpha_2 \alpha_3 & 0\\
0 & 1 & \alpha_{2(n-2)-1} &  \alpha_{2(n-3)-1}  & \ldots & \alpha_3 & \alpha_1 \\
\end{array}
\right),
\end{equation}
which we obtained using the Mathematica$\textsuperscript{TM}$ package \texttt{positroids} \cite{bourjaily2012positroids}. Notice that the origin of the space of $\alpha$'s corresponds to the zero-dimensional cell of the Grassmannian for which $\langle 12 \rangle \ne0$. For the standard ordering all $\alpha_i$ are non-negative.  In order to find an appropriate region for some different particle ordering $\sigma$, we take the subset of $\mathcal{V}_{n,2}$ for which $\langle \sigma(i)\sigma(j)\rangle\geq0$ for all $1\leq i<j\leq n$.

In Sec. \ref{sec:relations} we introduced $\mathcal{O}_n$ as the set of all $(n-1)!$ different $n$-particle orderings, i.e.~the set of $n$-tuples up to cyclic permutations. Without loss of generality we choose the position of particle $1$ to be fixed to the first position in each tuple. Then for each ordering $\sigma\in\mathcal{O}_n$ we have that $\langle 1 i\rangle>0$ for $1<i\le n$, according to the definition of $\mathcal{W}_{n,2}^{(\sigma)}$. This implies that all the odd $\alpha$'s are always positive and we do not need to consider them when comparing different orderings. 
This simplification halves the dimensionality of the space of $\alpha$ parameters leaving us with an $(n-2)$-dimensional real space $\mathbb{R}^{n-2}$ of only even $\alpha$'s.
Consequently, in the neighbourhood of the vertex for which $\langle12\rangle\neq0$ (and indeed any vertex), it is sufficient to describe the $k=2$ momentum amplituhedron for each ordering in terms of $(n-2)$ parameters. The equations $\langle ij\rangle=0$ for $1<i<j\le n$ define $\binom{n-1}{2}$ co-dimension-one hyperplanes in $\mathbb{R}^{n-2}$ which pass through the origin. These hyperplanes are defined in terms of even $\alpha$'s as
\begin{align}
\langle ij\rangle=0 \iff \sum_{l=n-j+1}^{n-i}\alpha_{2l} = 0,\qquad\text{(for $1<i<j\le n$)}\,.
\end{align}
Let us denote the set of these co-dimension-one hyperplanes by $\mathcal{H}_n$. The hyperplanes divide $\mathbb{R}^{n-2}$ into $(n-1)!$ regions, which we will call \emph{positive sectors}. These positive sectors are precisely the regions in $\mathbb{R}^{n-2}$ cut out by the remaining positivity conditions in  $\mathcal{W}_{n,2}^{(\sigma)}$ for each ordering $\sigma\in\mathcal{O}_n$. Moreover, each positive sector is an {oriented simplicial cone}, spanned by $(n-2)$ rays and having $(n-2)$ hyperplane facets,  and its orientation is inherited from the orientation of the coordinate system of even $\alpha$'s. We will denote each positive sector by $c[\sigma]$ where $\sigma\in\mathcal{O}_n$. These positive sectors form a {complete fan} in $\mathbb{R}^{n-2}$.

For four and five particles, positive sectors correspond to cones in two- and three-dimensions, respectively, and we shall study them in the examples below. In these examples, we will also see how the KK relations arise geometrically. In particular, we will see that the KK relations correspond to collections of positive sectors whose oriented sum (see Sec.\ \ref{sec:posgeometries}) no longer contains a zero-dimensional boundary. In such cases, the oriented sum of positive sectors is no longer a positive geometry and the corresponding sum of canonical differential forms must vanish.

Before proceeding to these examples, we also note an interesting relationship between the complete fan of positive sectors in $\mathbb{R}^{n-2}$ and the permutohedron. The permutohedron of order $(n-1)$ is an $(n-2)$-dimensional polytope whose vertices correspond to the permutations of $(n-1)$ symbols and whose edges correspond to transpositions that relate two permutations. We find that the dual to the complete fan of positive sectors for $n$ particles is (isomorphic to) the permutohedron of order $(n-1)$. In particular, each positive sector is dual to a vertex of the permutohedron while rays correspond to facets. Consequently, this construction of positive sectors from the $\alpha$-parametrization of the positive Grassmannian $G_+(2,n)$ gives a new and explicit realization of the permutohedron.

\paragraph{Four-particle MHV Amplitudes.} Let us consider the parametrization for $\lambda$ in the subspace $\mathcal{V}_{4,2}$ given by the $\alpha$-parametrization for the positive Grassmannian $G_+(2,4)$ in the patch for which $\langle12\rangle\ne 0$:
\begin{equation}
\label{sub1234}
\mathcal{V}_{4,2}:\quad\lambda=\left(
\begin{array}{cccc}
1 & \alpha_2+\alpha_4 & \alpha_2 \alpha_3 & 0 \\
0 & 1 & \alpha_3 & \alpha_1 \\
\end{array}
\right).
\end{equation}
The positivity conditions coming from $\mathcal{W}^{(\sigma)}_{4,2}$ for each ordering $\sigma\in\mathcal{O}_4$ are summarised in the table below.

\begin{center}
	\begin{tabular}{lcccccc}
		\toprule
		& (1234) & (1243) & (1324) & (1342) & (1423) & (1432) \\ \hline\hline
		$\langle 12\rangle = 1 $ & + & + & + & + & + & + \\ \hline
		$\langle 13\rangle = \alpha_3$ & + & + & + & + & + & + \\ \hline
		$\langle 14\rangle = \alpha_1 $ & + & + & + & + & + & + \\ \hline
		$\langle 23\rangle = \alpha_3 \alpha_4$ & + & + & - & - & + & - \\ \hline
		$\langle 24\rangle  = \alpha_1(\alpha_2+\alpha_4) $ & + & + & + & - & - & - \\ \hline
		$\langle 34\rangle = \alpha_1 \alpha_2\alpha_3$ & + & - & + & + & - & -\\
		\bottomrule
	\end{tabular}
\end{center}

\noindent
As explained before, since $\langle 1i\rangle>0$ for each ordering, the odd $\alpha$'s are always positive and therefore do not need to be considered. The remaining positivity conditions on $\langle ij\rangle$ for $1<i<j\le4$ produce the next table.

\begin{center}
	\begin{tabular}{lcccccc}
		\toprule
		& (1234) & (1243) & (1324) & (1342) & (1423) & (1432) \\ \hline\hline
		$\langle 23\rangle \sim \alpha_4$ & + & + & - & - & + & - \\ \hline
		$\langle 24\rangle  \sim \alpha_2+\alpha_4 $ & + & + & + & - & - & - \\ \hline
		$\langle 34\rangle \sim \alpha_2$ & + & - & + & + & - & -\\
		\bottomrule
	\end{tabular}
\end{center}

\begin{figure}[!h]
	\centering
	\includegraphics[scale=0.5]{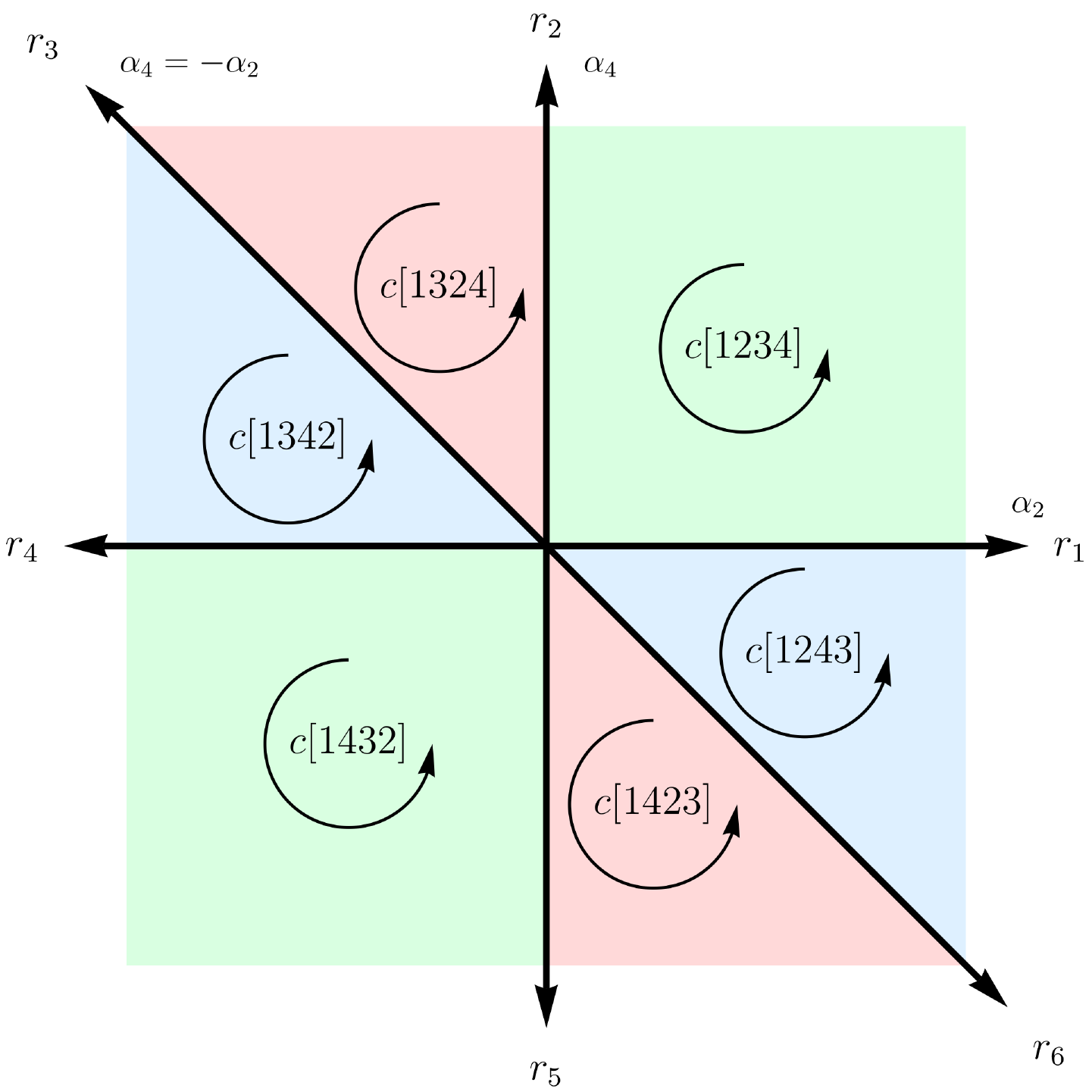}	
	\caption{Positive sectors corresponding to four-particle MHV amplitudes for each ordering.}
	\label{Allorders4pt}
\end{figure}

\noindent
From this table, we see that each positive sector is cut out by three inequalities, one of which is always redundant. In particular, each positive sector is an oriented simplicial cone. Together they form a complete fan in $\mathbb{R}^2$ as displayed in Fig.\ \ref{Allorders4pt}. % 
We orient each positive sector counter-clockwise.

In this simplified setting, it is easy to understand how the KK relations  arise geometrically from positive sectors. Firstly, consider the three reflection identities for $n=4$ which are given by 
\begin{subequations}
	\label{eq:4-2-alpha-reflection}
	\begin{align}
	A[1432] = A[1234]\,,\\
	A[1423] = A[1324]\,,\\
	A[1243] = A[1342]\,.
	\end{align}
\end{subequations}
It is clear from Fig.\ \ref{Allorders4pt} that these reflection relations stem from the fact that the corresponding positive sectors (identified by the same colors in the figure) define equivalent positive geometries with identical canonical differential forms:
\begin{subequations}
	\begin{align}
	\Omega(c[1432])&=\Omega(c[1234])=d\log(\alpha_4)\wedge d\log(\alpha_2)\,,\\
	\Omega(c[1423])&=\Omega(c[1324])=d\log(\alpha_2)\wedge d\log(\alpha_2+\alpha_4)\,,\\
	\Omega(c[1243])&=\Omega(c[1342])=d\log(\alpha_2+\alpha_4)\wedge d\log(\alpha_4)\,.
	\end{align}
\end{subequations}
Alternatively, let us denote by $c^-[\sigma]$ the cone $c[\sigma]$ carrying the opposite orientation. Then $\Omega(c^-[\sigma])=-\Omega(c[\sigma])$. It is easy to see that the oriented sums
$c[1432]\oplus c^-[1234]$, $c[1423]\oplus c^-[1324]$, and $c[1243]\oplus c^-[1342]$ have no zero-dimensional boundaries and hence the sum of the corresponding canonical differential forms in each case must vanish:
\begin{subequations}
	\begin{align}
	0&=\Omega(c[1432])+\Omega(c^-[1234])=\Omega(c[1432])-\Omega(c[1234])\,,\\
	0&=\Omega(c[1423])+\Omega(c^-[1324])=\Omega(c[1423])-\Omega(c[1324])\,,\\
	0&=\Omega(c[1243])+\Omega(c^-[1342])=\Omega(c[1243])-\Omega(c[1342])\,.
	\end{align}
\end{subequations}

\begin{figure}[!h]
	\centering
	\includegraphics[scale=0.5]{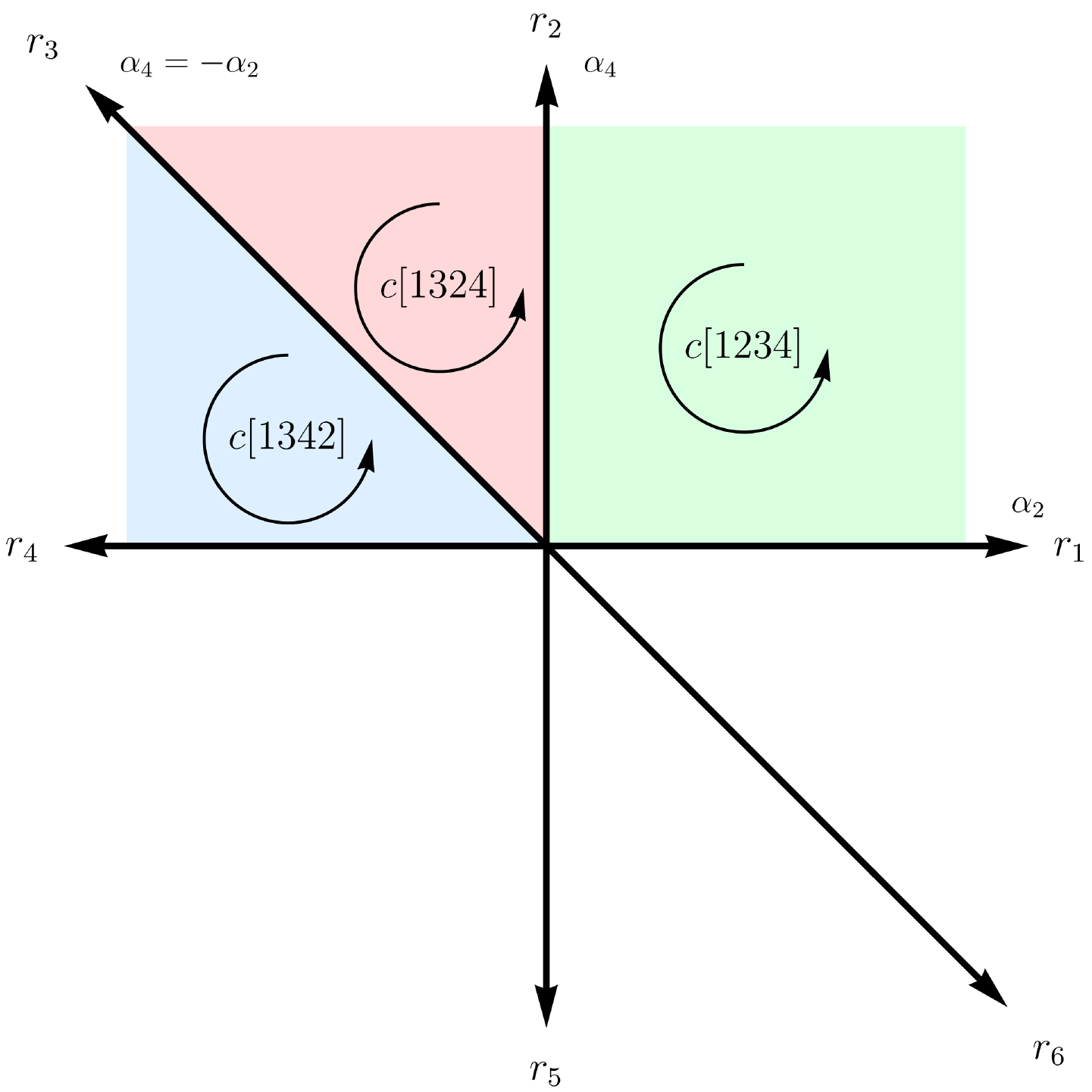}
	\caption{The three positive sectors $c[1234], c[1324], c[1342]$ appearing in the $U(1)$ decoupling relation \eqref{eq:4-2-alpha-u1a}.}
	\label{fig:4-2-alpha-u1a}
\end{figure}

\begin{figure}[!h]
	\centering
	\includegraphics[scale=0.5]{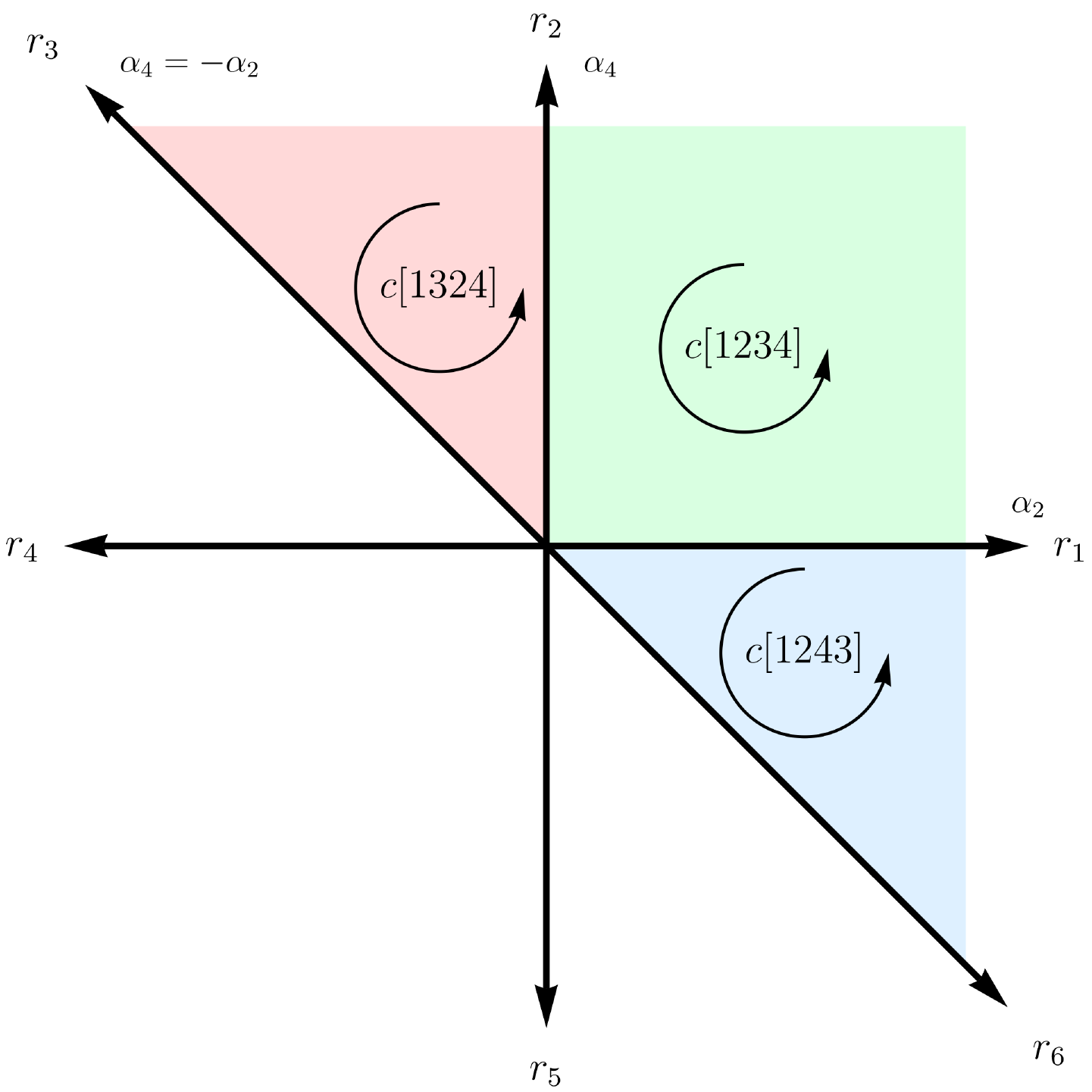}
	\caption{The three positive sectors $c[1234], c[1324], c[1243]$ appearing in the $U(1)$ decoupling relation \eqref{eq:4-2-alpha-u1b}.}
	\label{fig:4-2-alpha-u1b}
\end{figure}

Secondly, consider the two $U(1)$ decoupling relations coming from \eqref{KK} when $n_\beta=1$ which are given by 
\begin{subequations}
	\label{eq:4-2-alpha-u1}
	\begin{align}
	A[1342] + A[1324] + A[1234]=0\,,\label{eq:4-2-alpha-u1a}\\
	A[1243] + A[1234] + A[1324]=0\,.\label{eq:4-2-alpha-u1b}
	\end{align}
\end{subequations}
These relations correspond to the configurations of positive sectors depicted in Fig.\ \ref{fig:4-2-alpha-u1a} and Fig.\ \ref{fig:4-2-alpha-u1b}. %
In each case, the oriented sum of positive sectors produces a geometry without zero-dimensional boundaries and hence the sum of the corresponding canonical differential forms in each case must vanish.

\begin{figure}[!h]
	\centering
	\includegraphics[scale=0.5]{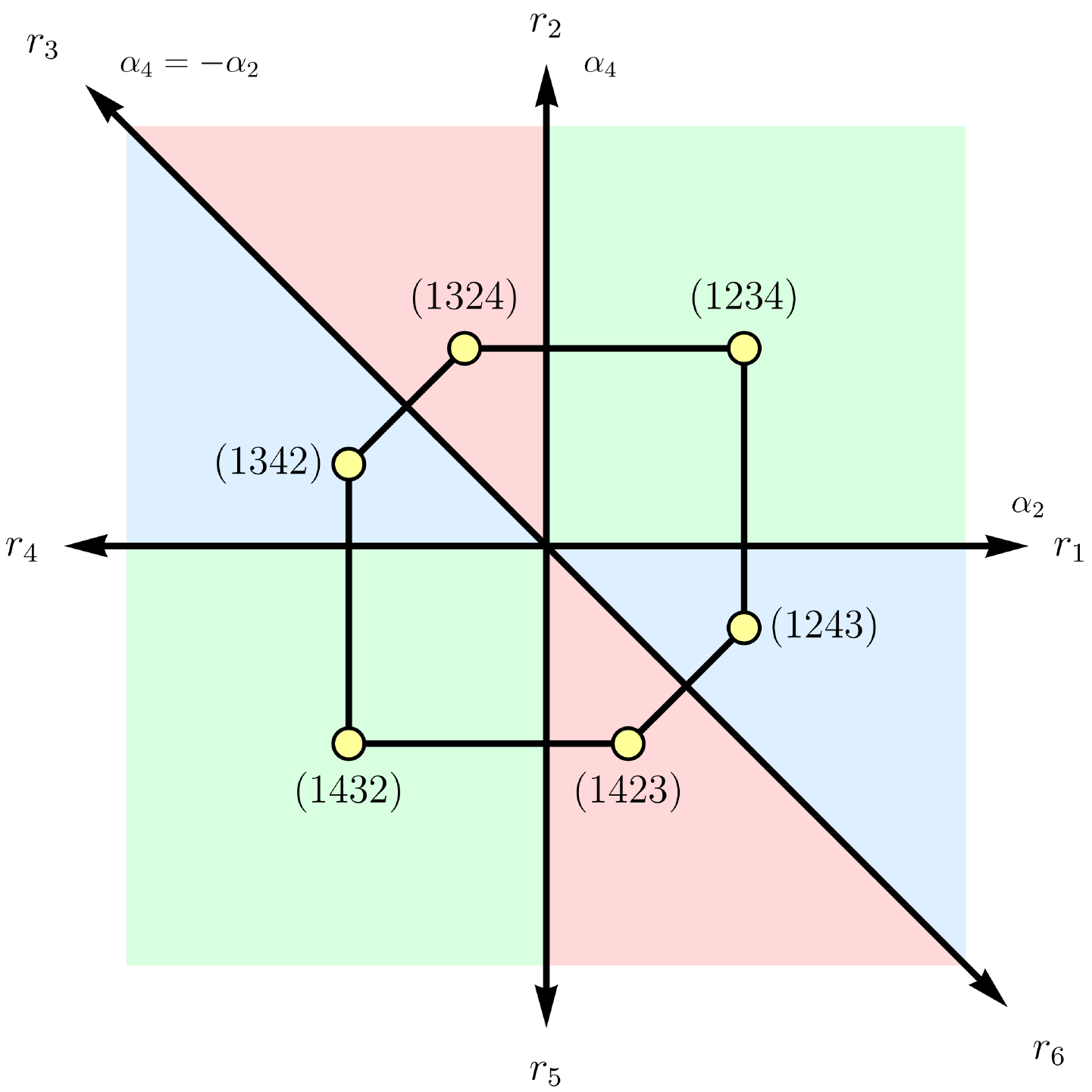}	
	\caption{The permutohedron of order $3$ is dual to the complete fan of positive sectors for $G_+(2,4)$.}
	\label{fig:4-2-alpha-permutohedron}
\end{figure}

Finally, notice that the polytope dual to the complete fan of positive sectors depicted in Fig.\ \ref{Allorders4pt} is isomorphic to the permutohedron of order $3$ which we draw in Fig.\ \ref{fig:4-2-alpha-permutohedron}. %

\begin{figure}[!h]
	\centering
	\includegraphics[scale=0.5]{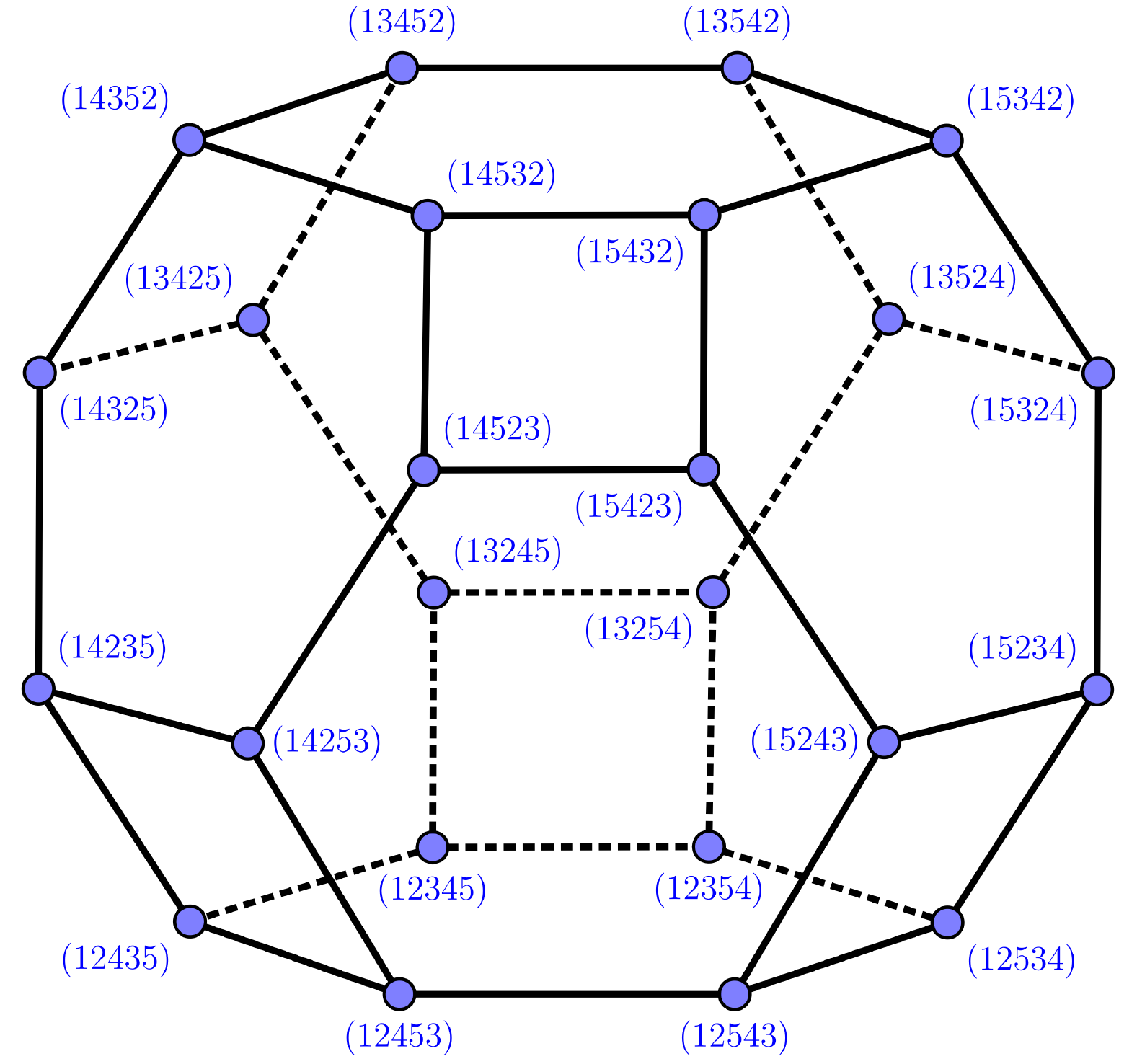}	
	\caption{The permutohedron of order $4$ is dual to the complete fan of positive sectors for $G_+(2,5)$.}
	\label{fig:5-2-alpha-permutohedron}
\end{figure}

\paragraph{Five-particle MHV Amplitudes.} 
The same analysis from the previous example can be applied to the five-particle case. 
Here the space of non-trivial (even) $\alpha$'s is the three-dimensional space $\mathbb{R}^3$ and it is divided by $6$ hyperplanes into precisely $24$ regions, each of which is a simplicial cone. These positive sectors form a complete fan whose dual is the permutohedron of order $4$ depicted in Fig.\ \ref{fig:5-2-alpha-permutohedron}. %

\begin{figure}[!h]
	\centering
	\begin{minipage}[b]{0.325\textwidth}
		\includegraphics[width=\textwidth]{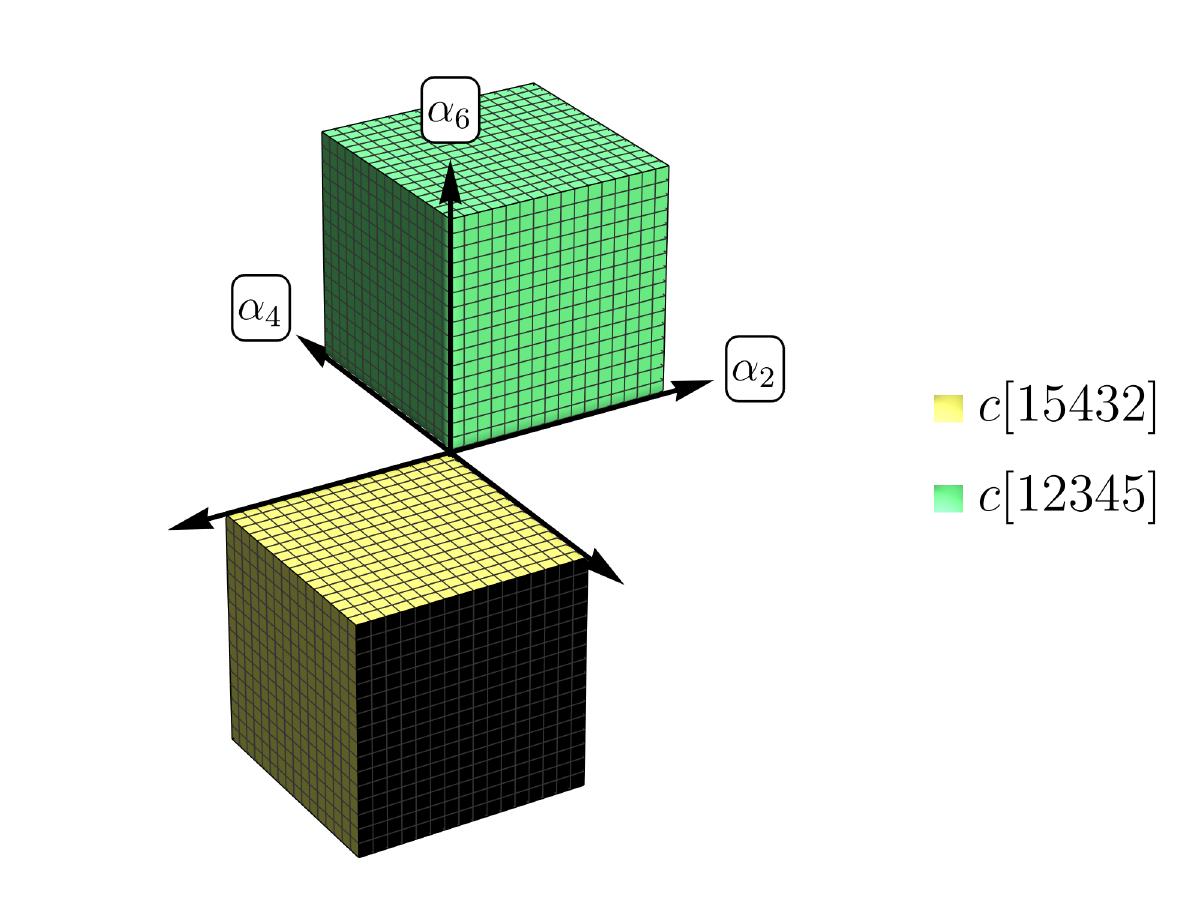}
	\end{minipage}
	\begin{minipage}[b]{0.325\textwidth}
		\includegraphics[width=\textwidth]{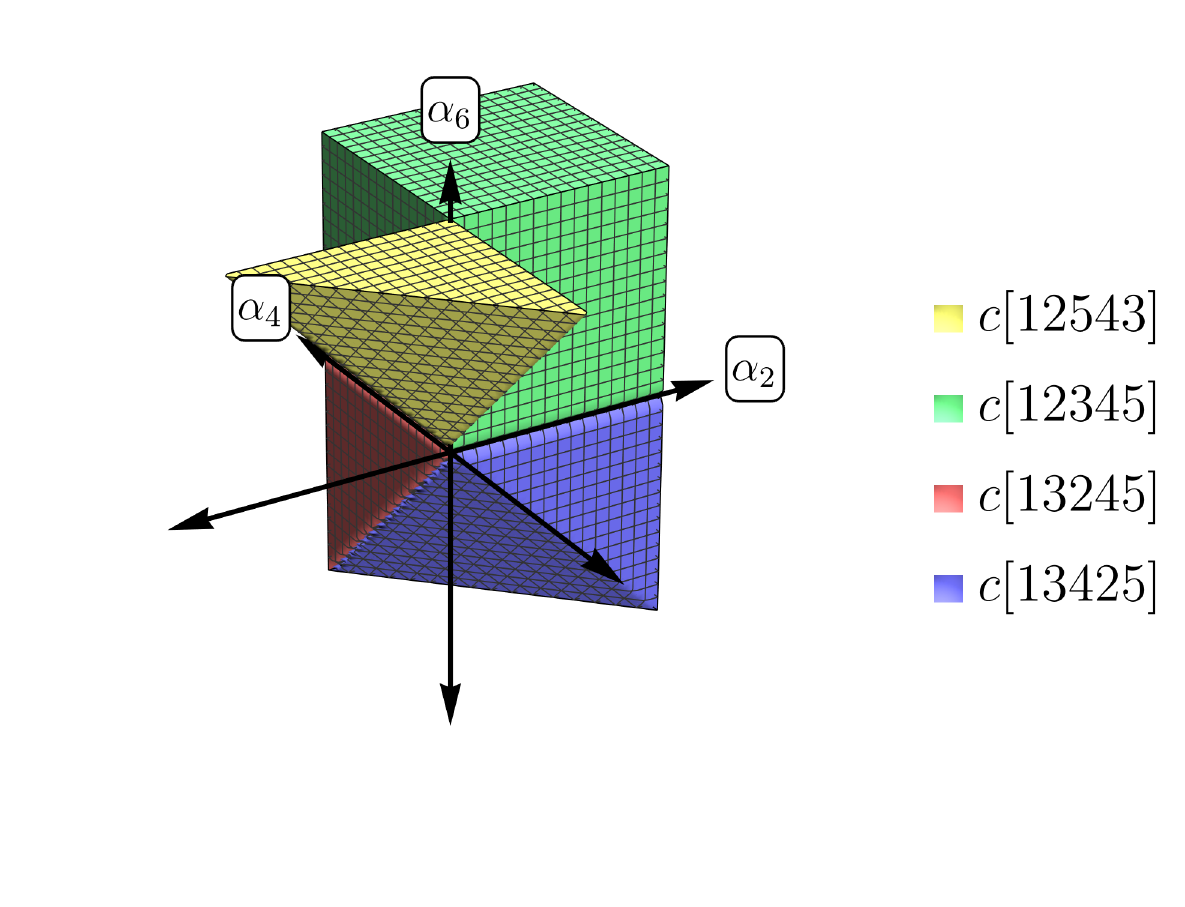}
	\end{minipage}
	\begin{minipage}[b]{0.325\textwidth}
		\includegraphics[width=\textwidth]{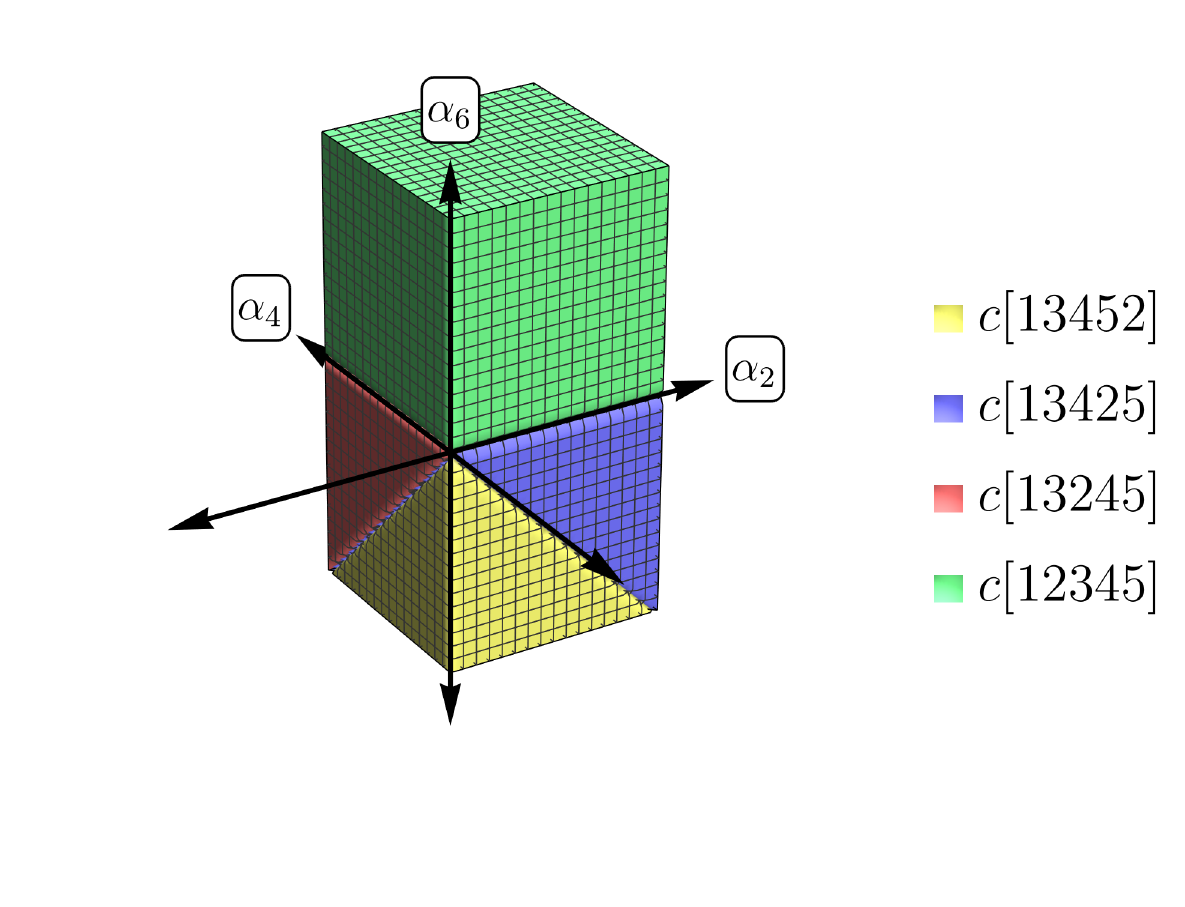}
	\end{minipage}
	\caption{Geometric realizations of three different types of KK relations for $n=5$: (left) the reflection relation in \eqref{eq:5-2-alpha-relations-1}; (middle) the KK relation in \eqref{eq:5-2-alpha-relations-2}; (right) the $U(1)$ decoupling identity in \eqref{eq:5-2-alpha-relations-3}. Each region is a cone emanating from the origin.}
	\label{fig:5-2-alpha-relations}
\end{figure}

There are three types of KK relations for $n=5$ coming from \eqref{KK} corresponding to the three different lengths of $\beta$ and examples of each are given below
\begin{subequations}\label{eq:5-2-alpha-relations}
	\begin{alignat}{3}
	&A[15432]=-A[12345]\,,& \qquad&(n_\beta=3)\,,\label{eq:5-2-alpha-relations-1}\\
	&A[12543]=A[12345]+A[13245]+A[13425]\,,& \qquad&(n_\beta=2)\,,\label{eq:5-2-alpha-relations-2}\\
	&A[13452]=-A[13425]-A[13245]-A[12345]\,,& \qquad&(n_\beta=1)\,.\label{eq:5-2-alpha-relations-3}
	\end{alignat}
\end{subequations}
The geometric realizations of these three relations as configurations of positive sectors is given in Fig.\ \ref{fig:5-2-alpha-relations}. The first relation is an example of a reflection relation, the third relation is an example of a $U(1)$ decoupling relation, and the second relation can be thought of as a combination of a reflection relation and a $U(1)$ decoupling relation. In each case, the KK relation manifests geometrically as a collection of positive sectors (some possibly carrying a reverse orientation to that inherited from the coordinate system, which explains the minus signs) whose oriented sum has no zero-dimensional boundaries and hence the corresponding sum of the canonical differential forms for these positive sectors must vanish. %

\subsection{Ray-based Homological Description for MHV Amplitudes}
\label{sec:kk-mom-ray}

The analysis from the previous subsection can be extended beyond $n=5$ and in the following we construct a homological algorithm for deriving all KK relations between MHV amplitudes from the geometry of positive sectors in $\mathbb{R}^{n-2}$. It exploits the simplicial structure of each positive sector as the positive span of rays. We will see that the algorithm relies on the ability to identify which pairs of rays point in opposite directions. Furthermore, it is possible to abstract this notion of ``pairs of rays pointing in opposite directions''. This will be done in the next subsection and it will allow us to derive the KK relations in any helicity sector from the geometry of the momentum amplituhedron.

Recall that $\mathcal{H}_n$ is the set of $\binom{n-1}{2}$ co-dimension-one hyperplanes in $\mathbb{R}^{n-2}$ which divide it into precisely $(n-1)!$ positive sectors. Let us enumerate these hyperplanes as $\mathcal{H}_n=\{h_i\}_{i=1}^{|\mathcal{H}_n|}$ where $|\mathcal{H}_n|=\binom{n-1}{2}$. In certain cases, the intersections of $(n-3)$ of these hyperplanes are one-dimensional and defines a line through the origin. Each line defines two unit vectors in $\mathbb{R}^{n-2}$ which point in opposite directions along the line and we call these vectors \emph{rays}. There are exactly $2^{n-1}-2$ rays in $\mathbb{R}^{n-2}$ defined by the one-dimensional intersection of $(n-3)$ hyperplanes, which is  precisely the number of facets of the permutohedron of order $(n-1)$. We will denote the set of all rays in $\mathbb{R}^{n-2}$ by $\mathcal{R}_n$ and we will enumerate them by $\mathcal{R}_n=\{r_j\}_{j=1}^{|\mathcal{R}_n|}$ where $|\mathcal{R}_n|=2^{n-2}-2$.

As we have already pointed out, each positive sector is a simplicial cone, which implies that for each ordering $\sigma\in\mathcal{O}_n$, the positive sector $c[\sigma]$ is given by the positive span of $(n-2)$ rays $\{r_{j^{\sigma}_{1}},r_{j^{\sigma}_{2}},\ldots,r_{j^{\sigma}_{n-2}}\}\subseteq\mathcal{R}_n$:  
\begin{align}
\label{csigma}
c[\sigma]=\text{span}_{\mathbb{R}_{\ge 0}}\{r_{j^{\sigma}_{1}},r_{j^{\sigma}_{2}},\ldots,r_{j^{\sigma}_{n-2}}\}\,.
\end{align}
Moreover, we can associate a formal form to each ray, and by extension a formal form to each positive sector, which will allow us to discuss boundary operations in the language of linear algebra. In particular, to each $c[\sigma]$ we can assign a formal $(n-2)$-form
\begin{align}
\omega(c[\sigma])=\frac{\det(r_{j^{\sigma}_{1}}, r_{j^{\sigma}_{2}},\ldots,r_{j^{\sigma}_{n-2}})}{|\det(r_{j^{\sigma}_{1}}, r_{j^{\sigma}_{2}},\ldots,r_{j^{\sigma}_{n-2}})|}\, \tilde{r}_{j^{\sigma}_{1}}\wedge \tilde{r}_{j^{\sigma}_{2}}\wedge\cdots\wedge \tilde{r}_{j^{\sigma}_{n-2}}\,, 
\label{eq:-n-2-ray-omega}
\end{align}
where $\det(r_{j^{\sigma}_{1}}, r_{j^{\sigma}_{2}},\ldots,r_{j^{\sigma}_{n-2}})$ is the determinant of the matrix whose columns are the rays $r_{j^{\sigma}_{1}}, r_{j^{\sigma}_{2}},\ldots,r_{j^{\sigma}_{n-2}}$ and, given any ray $r$, we denote by $\tilde{r}$ a formal one-form labelled by it. We will call any such formal one-form a \emph{ray one-form} and we will call the $p$-fold wedge product of ray one-forms a \emph{ray $p$-form}.
The definition of $\omega(c[\sigma])$ is manifestly invariant under a relabelling and rescaling of the rays in $\mathcal{R}_n$ and hence it is well-defined. We can define a boundary operator with respect to any hyperplane $h\in\mathcal{H}_n$, denoted by $\partial_{h}$, which acts on ray $p$-forms as follows: given a single ray $r$ define
\begin{align}
\label{boundary-op1}
\partial_{h} \tilde{r}=\bar{\Theta}_{h}(r) \equiv \left\{\begin{array}{ll}
0,&\quad \text{if $r\in h$}\\
1,&\quad \text{otherwise}
\end{array}\right.,
\end{align}
where $r\in h$ means $r$ is contained in the hyperplane $h$, and for $p>1$ rays $\{r_{j_1},r_{j_2},\ldots,r_{j_p}\}$ define
\begin{align}
\label{boundary-op2}
\begin{split}
\partial_{h}(\tilde{r}_{j_1}\wedge \tilde{r}_{j_2}\wedge\cdots\wedge \tilde{r}_{j_p})=\bar{\Theta}_h(r_{j_1})\Theta_h(r_{j_2},\ldots,r_{j_p})\,\tilde{r}_{j_2}\wedge\ldots\wedge \tilde{r}_{j_p}\\
- \Theta_h(r_{j_1})\,\tilde{r}_{j_1}\wedge\partial_h(\tilde{r}_{j_2}\wedge\cdots\wedge \tilde{r}_{j_p})\,,
\end{split}
\end{align}
where
\begin{align}
\Theta_h(r_{j_2},\ldots,r_{j_p})\equiv\Theta_h(r_{j_2})\cdots\Theta_h(r_{j_p})\,,\qquad\Theta_h(r_{j_1})\equiv 1-\bar{\Theta}_h(r_{j_1})\,.
\label{boundaryTheta}
\end{align}
By definition, the boundary operator with respect to any hyperplane $h\in\mathcal{H}_n$ is nilpotent (i.e.~$\partial_h^2=0$) because $\bar{\Theta}_h(r)\Theta_h(r) = 0$ for all rays $r\in\mathcal{R}_n$. Moreover, the result of applying the boundary operator $\partial_h$ to the ray form of a given positive sector is the ray form of the simplicial cone obtained as an intersection of $h$ with the positive sector.

Finally, let us construct a graded vector space which combines all ray forms for positive sectors and their boundaries. We define 
\begin{align}
\label{vspace}
V_n\equiv\bigoplus_{i=0}^{n-2}V_n^{(i)}\,,
\end{align}
where
\begin{align}
V_n^{(0)}\equiv\text{span}_{\mathbb{Z}}\left\{\omega(c[\sigma]):\sigma\in\mathcal{O}_n\right\} \,
\end{align}
is the vector space of integer linear combinations of the ray forms of degree $(n-2)$ given in \eqref{eq:-n-2-ray-omega} corresponding to positive sectors, and for $1\le i \le n-2$ 
\begin{align}
V_n^{(i)}\equiv\text{span}_{\mathbb{Z}}\left\{\bigwedge_{j\in J}\tilde{r}_j:J\in\binom{[|\mathcal{R}_n|]}{n-2-i}\right\} \,
\end{align}
is the vector space of integer linear combinations of all ray forms of degree $(n-2-i)$. The notation $\binom{[|\mathcal{R}_n|]}{n-2-i}$ denotes the collection of $(n-2-i)$ element subsets of $[|\mathcal{R}_n|]\equiv \{1,2,\ldots,|\mathcal{R}_n|\}$. Clearly $V_n^{(n-2)}=\mathbb{Z}$. For any hyperplane $h\in\mathcal{H}_n$ we have the exact sequence
\begin{align}
V_n^{(0)}\xrightarrow{\partial_h}V_n^{(1)}\xrightarrow{\partial_h} \cdots\xrightarrow{\partial_h}V_n^{(n-3)}\xrightarrow{\partial_h}V_n^{(n-2)}=\mathbb{Z}\xrightarrow{\partial_h}0\,.
\end{align}

With these definitions in place, we can now determine all KK relations between MHV amplitudes from the geometry of the positive sectors in $\mathbb{R}^{n-2}$. In the previous subsection, we saw that each KK relation was realized geometrically as a collection of positive sectors whose oriented sum was void of any zero-dimensional boundaries. We can equivalently express this observation as follows: a KK relation corresponds to a vector $\nu\in V_n^{(0)}$ such that for every $(n-2)$ element subset $I$ of $[|\mathcal{H}_n|]=\{1,2,\ldots,|\mathcal{H}_n|\}$ (i.e.~$I\in\binom{[|\mathcal{H}_n|]}{n-2}$) we have that 
\begin{align}
\big(\prod_{i\in I}\partial_{h_i}\big)\nu=0\,.
\end{align}
The above condition simply expresses the fact that the geometry corresponding to $\nu$ does not have the origin as a zero-dimensional boundary. Let $M_n$ denote the $\binom{|\mathcal{H}_n|}{n-2}\times(n-1)!$ matrix whose rows are labelled by $(n-2)$ element subsets $I$ of $[|\mathcal{H}_n|]$, whose columns are labelled by orderings $\sigma\in\mathcal{O}_n$, and whose entries are given by 
\begin{align}
\label{eq:-n-2-ray-M}
\tensor{(M_n)}{^{I}_{\sigma}}\equiv
\big(\prod_{i\in I}\partial_{h_i}\big)\omega(c[\sigma])\,.
\end{align} 
We will call this matrix $M_n:V_n^{(0)}\to V_{n}^{(n-2)}$ the \emph{boundary matrix}. Then the kernel of the boundary matrix is the space of all KK relations amongst the different particle orderings. In order to clarify this discussion, we will apply it to the four-particle case.

\paragraph{Four-particle MHV Amplitudes.} For four particles we have $3$ hyperplanes (which are lines) and $6$ rays. Let us label the hyperplanes by
\begin{align}
h_1: \alpha_4=0\,,\qquad h_2: \alpha_2+\alpha_4=0\,,\qquad h_3: \alpha_2=0\,,
\end{align}
and the rays by
\begin{align}
r_1=(1,0)=-r_4\,,\qquad r_2=(0,1)=-r_5\,,\qquad r_3=\frac{1}{\sqrt{2}}(-1,1)=-r_6\,.
\end{align}
 These rays correspond to those drawn in Fig.~\ref{Allorders4pt}. Using the definition given in \eqref{eq:-n-2-ray-omega}, the ray forms for each positive sector are given by
\begin{align}
\begin{split}
&\omega(c[1234])=\tilde{r}_1\wedge\tilde{r}_2\,,\qquad \omega(c[1324])=\tilde{r}_2\wedge\tilde{r}_3\,,\\
&\omega(c[1342])=\tilde{r}_3\wedge\tilde{r}_4\,,\qquad
\omega(c[1432])=\tilde{r}_4\wedge\tilde{r}_5\,,\\
&\omega(c[1423])=\tilde{r}_5\wedge\tilde{r}_6\,,\qquad
\omega(c[1243])=\tilde{r}_6\wedge\tilde{r}_1\,.
\end{split}
\end{align}
For the above enumeration of hyperplanes, the boundary matrix computed according to \eqref{eq:-n-2-ray-M} reads
\begin{align}
M_4=
\begin{blockarray}{l *{6}{c} l}
&(1234) & (1243) & (1324) & (1342) & (1423) & (1432) &\\
\begin{block}{l(*{6}{c})l}
\{1,2\}&0 & -1 & 1 & -1 & 1 & 0& \partial_{h_1}\partial_{h_2} \bigstrut[t] \\
\{1,3\}&1 & 0 & -1 & 0 & -1 & 1 & \partial_{h_1}\partial_{h_3} \\
\{2,3\}&1 & 0 & -1 & 0 & -1 & 1 & \partial_{h_2}\partial_{h_3} \bigstrut[b]\\
\end{block}
&\tilde{r}_1\wedge\tilde{r}_2 & \tilde{r}_6\wedge\tilde{r}_1 & \tilde{r}_2\wedge\tilde{r}_3 & \tilde{r}_3\wedge\tilde{r}_4 & \tilde{r}_5\wedge\tilde{r}_6 & \tilde{r}_4\wedge\tilde{r}_5 
\end{blockarray}%
.
\end{align}
We find the following basis for the kernel of the boundary matrix
\begin{align}
\begin{split}
-\omega(c[1234])+\omega(c[1432])\,,\\
-\omega(c[1324])+\omega(c[1423])\,,\\
\omega(c[1234])+\omega(c[1324])+\omega(c[1342])\,,\\
\omega(c[1234])+\omega(c[1243])+\omega(c[1324])\,.
\end{split}
\end{align}
By replacing $\omega(c[\sigma])\mapsto A[\sigma]$ in each of the above vectors and then equating each vector to zero we reproduce the KK relations for four particles as listed in \eqref{eq:4-2-alpha-reflection} and \eqref{eq:4-2-alpha-u1}.

We have used this ray-based homological algorithm for up to  $n=7$ and found all KK relations for MHV amplitudes. Since we know that the KK relations for a given $n$ hold across helicity sectors, we have therefore found all KK relations for $n\leq 7$ for any helicity. However, we can explicitly check that this is the case by deriving the KK relations for non-MHV sectors. To do this requires more sophisticated machinery which is the topic of the next subsection.

\subsection{Poset-based Homological Algorithm} 
\label{sec:kk-mom-poset}
In Sec.\ \ref{sec:kk-mom-alpha}, we explained how the geometry of the momentum amplituhedron for $k=2$ naturally leads to a polytopal realization for MHV amplitudes in terms of positive sectors/oriented simplicial cones. These positive sectors capture which boundaries of the momentum amplituhedron are shared between different particle orderings. Exploiting this description, we then presented a ray-based homological algorithm for determining the KK relations between MHV amplitudes in Sec.\ \ref{sec:kk-mom-ray}. Unfortunately, the simplifications which  
produced cones for MHV amplitudes do not extend to other helicities. For $n$ particles and $k>2$ we cannot reduce the space of $\alpha$'s parametrizing $\lambda$ by fixing some subset of them while still satisfying all positivity constraints for every particle ordering, and the inequalities coming from these constraints are no longer linear. In fact, for $2<k<n-2$, momentum conservation between $\lambda$'s and $\tilde{\lambda}$'s produces rational inequalities for $\alpha$'s. These inequalities define complicated, curvy hypersurfaces and in the neighbourhood of any vertex the positive geometries for different particle orderings can no longer be described as cones. This being said, the description of MHV amplitudes for different particle orderings in terms of cones is not essential for deriving the KK relations. Indeed, the only information we used was that (1) we knew which rays were shared by different particle orderings and (2) we knew which pairs of rays lived in the same one-dimensional intersection of hyperplanes --- we knew which pairs of rays pointed in opposite directions. The latter point, namely being able to identify one-dimensional boundaries of momentum amplituhedra which inhabit the same one-dimensional intersection of hypersurfaces will prove to be the crucial point for deriving the KK relations beyond $k>2$, albeit in a more abstract guise.

In this subsection, we abstract the derivation of the KK relations given previously for $k=2$ and present a poset-based homological algorithm 
which can be applied to any helicity sector. Our algorithm does not depend on any detailed analysis of parametrizations for each positive geometry, but rather it takes as inputs the combinatorial structure of boundaries of the momentum amplituhedron for different particle orderings which we generate using the Mathematica$^\text{TM}$ package \texttt{amplituhedronBoundaries} \cite{Lukowski:2020bya}. We will introduce this algorithm first by example and re-derive all KK relations for four-particle MHV amplitudes. In order to simplify our discussion and to make direct contact with the previous subsection, we will initially continue to work as we did before and parametrise $\lambda$ in terms of $\alpha$'s according to \eqref{sub1234}. Thereafter, we will describe how to move away from this simplified setting. Throughout our presentation, we will introduce  new concepts and terminology which will ultimately allow us to abstract the notion of ``pairs of rays pointing in opposite directions''.

\subsubsection{Revisiting MHV Amplitudes}
Recall that $\mathcal{O}_4$ is the set of four-tuples describing the $(4-1)!=6$ different four-particle orderings. As detailed previously, for each $\sigma\in\mathcal{O}_4$ we can describe the geometry of the $k=2$ momentum amplituhedron as an oriented simplicial cone $c[\sigma]$ in the two-dimensional $(\alpha_2,\alpha_4)$-space as depicted in Fig.\ \ref{Allorders4pt}. Each cone has two co-dimension one boundaries which are semi-infinite lines spanned by rays and a single co-dimension two boundary which is the vertex $v$ at the origin. We have labelled the six rays in Fig.\ \ref{Allorders4pt} by $r_i$ where $i=1,\ldots,6$.

Let us denote by $\mathcal{P}^{(\sigma)}$ the set containing $c[\sigma]$ together with all of its boundaries (of all co-dimensions). For example,
\begin{align}
\mathcal{P}^{(1234)}=\{c[1234],r_1,r_2,v\}\,.
\end{align}
We will generically refer to elements of $\mathcal{P}^{(\sigma)}$ as boundaries. $\mathcal{P}^{(\sigma)}$ defines a \emph{partially ordered set} or \emph{poset} where the \emph{partial order $\preceq$} is defined for any two boundaries $\mathcal{B}_1,\mathcal{B}_2\in\mathcal{P}^{(\sigma)}$ by 
\begin{align}
\text{$\mathcal{B}_1\preceq \mathcal{B}_2$ if $\mathcal{B}_1=\mathcal{B}_2$ or $\mathcal{B}_1$ is a boundary (of any co-dimension) of $\mathcal{B}_2$}\,.
\end{align}
If $\mathcal{B}_1\preceq \mathcal{B}_2$ and $\mathcal{B}_1\ne \mathcal{B}_2$, then we write $\mathcal{B}_1\prec\mathcal{B}_2$. Each boundary $\mathcal{B}\in\mathcal{P}^{(\sigma)}$ has a well-defined dimension given by $\dim(\mathcal{B})$, which turns $\mathcal{P}^{(\sigma)}$ into a graded poset.

More generally, given a positive geometry $(X,X_{\ge0})$, let $\mathcal{P}[X_{\ge 0}]$ be the set consisting of $X_{\ge 0}$ and all of its boundaries (of all co-dimensions) in $X$. Then $\mathcal{P}[X_{\ge 0}]$ forms a graded poset which we will call the \emph{boundary stratification} of $X_{\ge 0}$. In what follows, we will write $\mathcal{P}=\mathcal{P}[X_{\ge 0}]$ for brevity. The combinatorial relationships between boundaries in $\mathcal{P}$ can be depicted graphically as a Hasse diagram. A \emph{Hasse diagram} is a graph where each node corresponds to a boundary and two nodes are connected by an edge if one of the nodes is a co-dimension one boundary of the other. Specifically, if $\mathcal{B}_1$ is a co-dimension one boundary of $\mathcal{B}_2$, we draw a \emph{directed edge} $e=(\mathcal{B}_2,\mathcal{B}_1)$ from $\mathcal{B}_2$ to $\mathcal{B}_1$. Here $\mathcal{B}_2$ is the source node of $e$, denoted by $\partial^-(e)$, and $\mathcal{B}_1$ is the target node of $e$, denoted by $\partial^+(e)$. We will always use a lowercase $e$ to denote a directed edge of a Hasse diagram. Let $H[\mathcal{P}]$ label the Hasse diagram corresponding to $\mathcal{P}$ and let $E[\mathcal{P}]$ be the set of its directed edges. 

\begin{figure}[h!]
	\centering
	\includegraphics[scale=0.6]{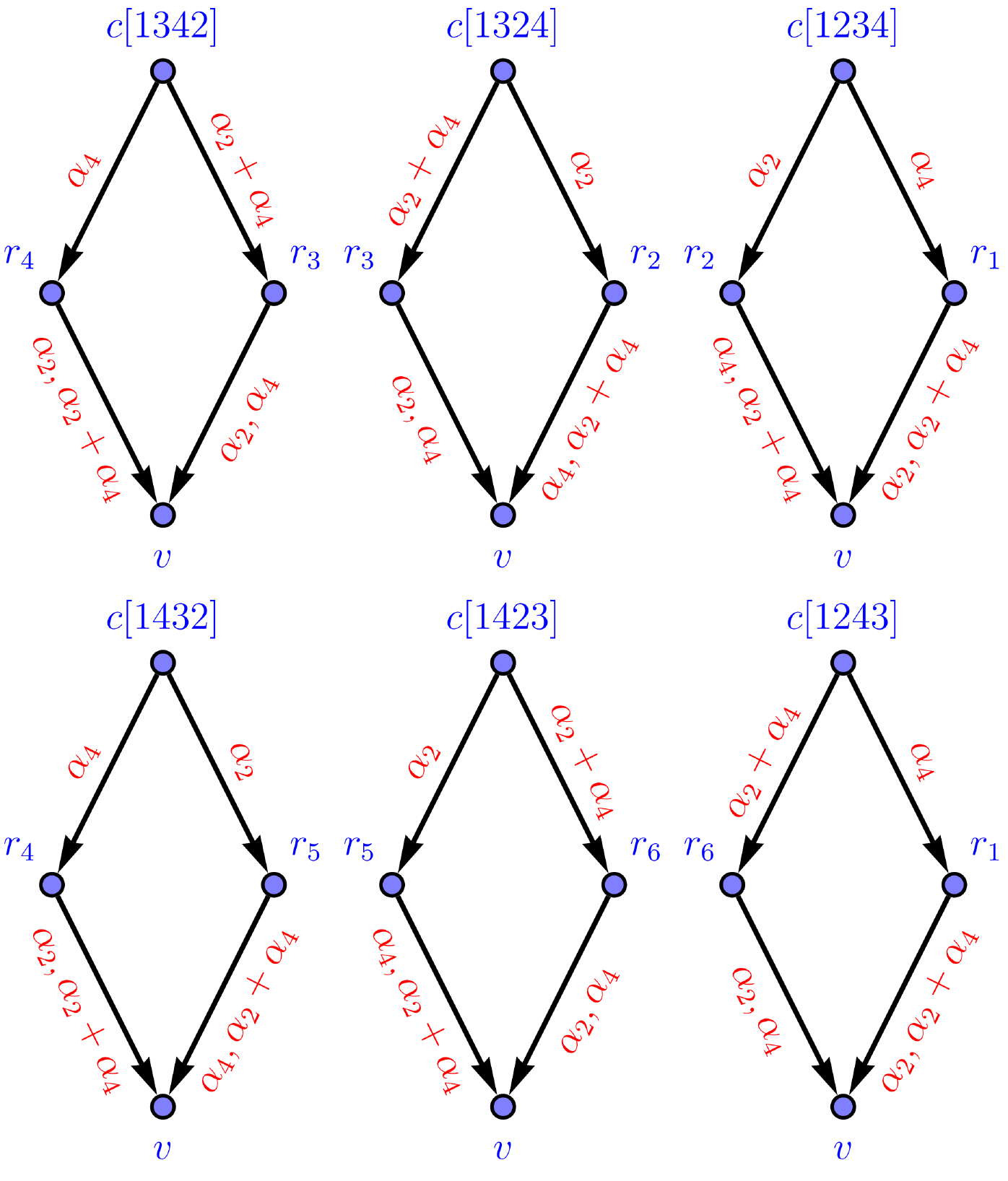}
	\caption{The Hasse diagrams $H[\mathcal{P}^{(\sigma)}]$ for all four-particle orderings.}
	\label{hom-4-2-posets}
\end{figure}

The Hasse diagrams $H[\mathcal{P}^{(\sigma)}]$ for each $\sigma\in\mathcal{O}_4$ are drawn in Fig.\ \ref{hom-4-2-posets}. Note that we have labelled each directed edge $e$ by a subset of the expressions in $\{\alpha_2,\alpha_4,\alpha_2+\alpha_4\}$. We will refer to these as edge labels. Given a directed edge $e=(\mathcal{B}_2,\mathcal{B}_1)$, an expression $l=l(\vec{\alpha})$ in $\alpha$ parameters is an \emph{edge label} for $e$ if $\mathcal{B}_1$ is a boundary of $\mathcal{B}_2$ in the limit $l\to0$. We will always use a lowercase $l$ for edge labels and we will use $L(e)$ to denote the set of edge labels for  $e$.  For example, $H[\mathcal{P}^{(1234)}]$ has edges
\begin{align}
e_1=(c[1234],r_1)\,,\quad 
e_2=(c[1234],r_2)\,,\quad
e_3=(r_1,v)\,,\quad 
e_4=(r_2,v)\,,
\end{align}
which are labelled by
\begin{align}
L(e_1)=\{\alpha_4\}\,,\quad
L(e_2)=\{\alpha_2\}\,,\quad L(e_3)=\{\alpha_2,\alpha_2+\alpha_4\}\,,\quad L(e_4)=\{\alpha_4,\alpha_2+\alpha_4\}\,.
\end{align}

Since we are ultimately interested in deriving relations between momentum amplituhedra for different particle orderings, it will be useful to introduce some terminology for families of positive geometries. From now on, we will assume that $n$ and $k$ are fixed (and $k=2$ in this subsection) which will allow us to label relevant positive geometries using permutations from $\mathcal{O}_n$. Then $\{\mathcal{P}^{(\sigma)}\}_{\sigma\in \mathcal{O}_n}$ is an indexed family of boundary stratifications of positive geometries $X_{\geq 0}^{(\sigma)}$ and let us suppose we have assigned edge labels to every edge in every Hasse diagram. We will denote by $\mathcal{E}$ the set of all edges in all Hasse diagrams:
\begin{align}
\mathcal{E}\equiv\bigcup_{\sigma\in \mathcal{O}_n}E[\mathcal{P}^{(\sigma)}]\,.
\end{align}
Given an edge label $l$ we define
\begin{align}
\check{E}(l) \equiv \left\{e\in \mathcal{E}:l\in L(e)\right\}
\end{align}
to be the set of all edges for which $l$ is an edge label. 
For example for $n=4$, the posets $\{\mathcal{P}^{(\sigma)}\}_{\sigma\in\mathcal{O}_4}$ are such that the sets $\check{E}(\alpha_2)$, $\check{E}(\alpha_4)$, and $\check{E}(\alpha_2+\alpha_4)$ each contain 12 edges as can be verified from Fig. \ref{hom-4-2-posets}.

Having introduced the definitions for Hasse diagrams and edge labels, we now turn our attention to defining boundary operators in analogy with what we did in Sec.\ \ref{sec:kk-mom-ray}. Here the notion of ``boundary operators with respect to hyperplanes'' will be replaced by ``boundary operators with respect to edge labels''. We will then use these to define a ``boundary matrix'' as we did before, and the kernel of this matrix will be spanned precisely by all KK relations. In order to define our boundary operators, it will prove useful to  assign signs to edges in each Hasse diagram subject to certain compatibility criteria. In order to clarify these compatibility criteria, we need to define poset intervals and diamonds. 

Let $\mathcal{B}_1$ and $\mathcal{B}_2$ be two boundaries in some boundary stratification $\mathcal{P}$ and suppose that $\mathcal{B}_1\preceq \mathcal{B}_2$. Then their \emph{interval}, denoted $[\mathcal{B}_1,\mathcal{B}_2]$, is defined as the set of all boundaries $\mathcal{B}$ such that $\mathcal{B}_1\preceq \mathcal{B}\preceq \mathcal{B}_2$:
\begin{align}
[\mathcal{B}_1,\mathcal{B}_2]=\left\{\mathcal{B}\in\mathcal{P}\big|\mathcal{B}_1\preceq \mathcal{B}\preceq \mathcal{B}_2\right\}\,.
\end{align}
Let $\mathcal{I}=[\mathcal{B}_1,\mathcal{B}_2]$. 
If $\dim(\mathcal{B}_2)=\dim(\mathcal{B}_1)+2$ and $\mathcal{I}=\{\mathcal{B}_1,\mathcal{B},\mathcal{B}',\mathcal{B}_2\}$ where $\mathcal{B}$ and $\mathcal{B}'$ are both co-dimension-one boundaries of $\mathcal{B}_2$, and $\mathcal{B}\ne\mathcal{B}'$, then we will call this interval a \emph{diamond}. The terminology reflects the fact that the Hasse diagram for this interval is diamond-shaped. We will also use the term diamond to refer to any subgraph of a Hasse diagram which represents a diamond.
The Hasse diagrams in Fig.\ \ref{hom-4-2-posets} are all examples of diamonds. 

\begin{figure}
	\centering
	\includegraphics[scale=0.6]{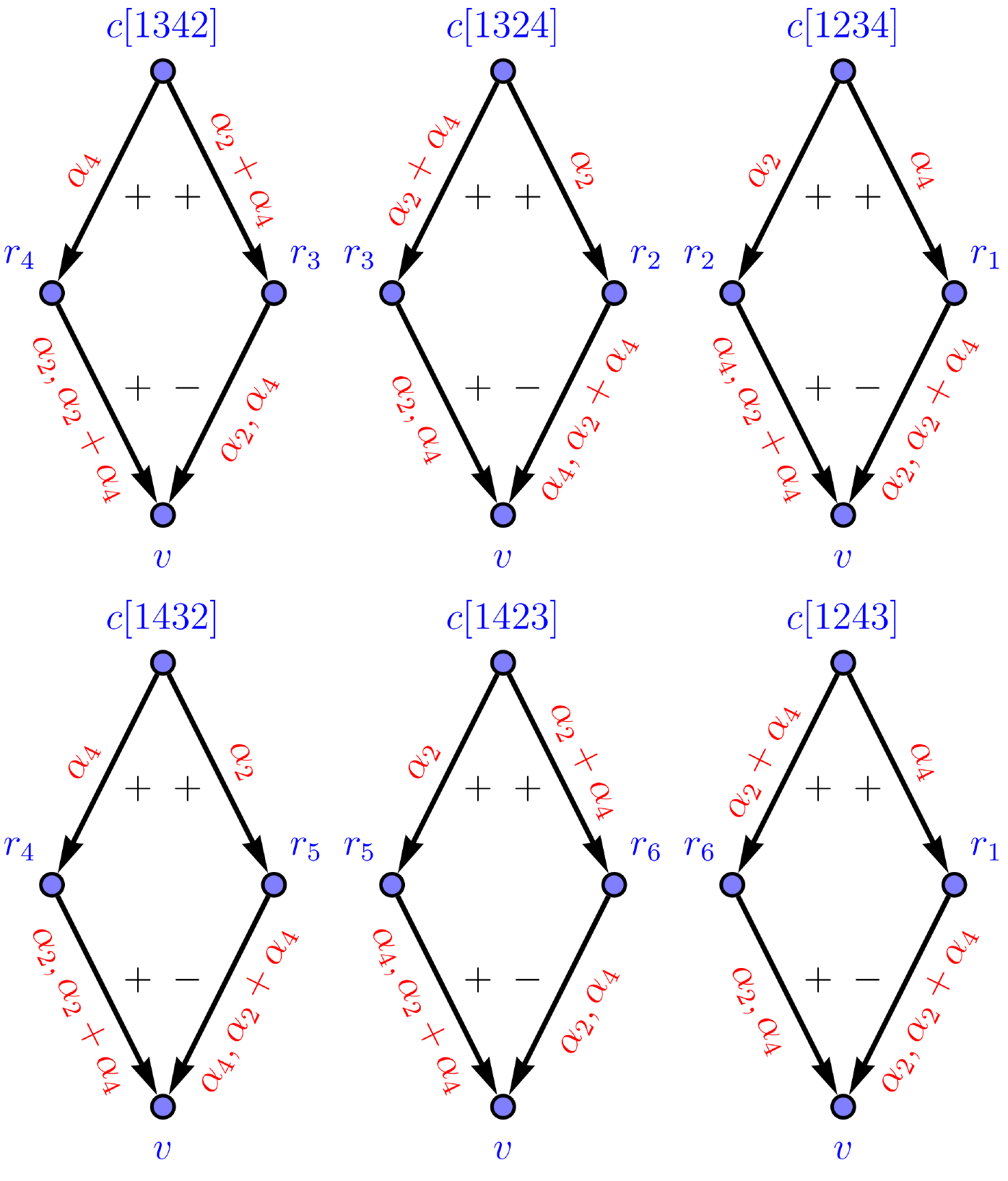}	
	\caption{Example of diamond-compatible sign assignment for the Hasse diagrams $H[\mathcal{P}^{(\sigma)}]$, $\sigma \in \mathcal{O}_4$.}
	\label{hom-4-2-posets-signs}
\end{figure}

Returning to our generic interval $\mathcal{I}$, it is possible to assign a sign to each edge in the corresponding Hasse diagram $H[\mathcal{I}]$ such that for every diamond $\mathcal{D}$:
\begin{align}
\prod_{\mathclap{e\in E[\mathcal{D}]}} \text{sgn}(e)=-1\,,
\end{align}
where the product is over all edges $e$ in the diamond $\mathcal{D}$ and $\text{sgn}(e)$ is the sign assigned to $e$. We will refer to an assignment of signs satisfying this condition as being \emph{diamond compatible}. We demand this condition to make the boundary operator nilpotent, and therefore our construction homological. An example of a diamond compatible sign assignment for the Hasse diagrams in Fig.\ \ref{hom-4-2-posets} is given in Fig.\ \ref{hom-4-2-posets-signs}. With these sign assignments, we can now define boundary operators with respect to edge labels. 

For an edge label $l$, we define the boundary operator with respect to $l$, denoted by $\partial_{l}$, as follows: given a boundary $\mathcal{B}\in\mathcal{P}^{(\sigma)}$ for some $\sigma\in \mathcal{O}_n$ 
\begin{align}
\partial_{l}\mathcal{B}\equiv\sum_{\mathclap{{e:\mathcal{B}\to \mathcal{B}'\in \check{E}(l)}}}\text{sgn}(e)\mathcal{B}'\,,
\end{align}
where the sum is over all directed edges $e$ with source node $\partial^-(e)=\mathcal{B}$ which can be labelled by $l$. For example, from Fig.\ \ref{hom-4-2-posets-signs} we see that
\begin{align}
\partial_{\alpha_4}\partial_{\alpha_2}c[1234]=\partial_{\alpha_4}r_2=v \quad\text{and}\quad \partial_{\alpha_2}\partial_{\alpha_4}c[1234]=\partial_{\alpha_2}r_1=-v\,.
\end{align}
By construction, $\partial_{l}^2=0$ for any edge label $l$. This follows from that fact that for every pair of edges $(e,e')$ where $\partial^+(e)=\partial^-(e')$, their sets of edge labels are always disjoint: $L(e)\cap L(e')=\emptyset$. Additionally, if we define the total boundary operator 
\begin{align}
\partial\equiv\sum_{l}\partial_{l}\,,
\end{align}
with the sum over all distinct edge labels $l$, then using the diamond compatible sign assignment given in Fig.\ \ref{hom-4-2-posets-signs}, it is easy to check that for all particle orderings $\sigma\in\mathcal{O}_4$
\begin{align}
\partial^2c[\sigma] = \left(\partial_{\alpha_2}+\partial_{\alpha_4}+\partial_{(\alpha_2+\alpha_4)}\right)^2c[\sigma] = 0 \,,
\end{align}
i.e.~the total boundary operator $\partial$ is nilpotent. In fact, the diamond compatibility condition was chosen precisely such that the total boundary operator would be nilpotent.

The boundary operator with respect to an edge label gives us a way to move from a \emph{level} (all boundaries of the same dimension) in a Hasse diagram to a level of one lower dimension. We next consider chains of these boundary operators which take us from the top of a Hasse diagram to the bottom. To this end it is useful to introduce the notion of complete paths, complete path labels, and boundary operators with respect to these labels.
 
Let $\mathcal{P}$ be the boundary stratification of a $d$-dimensional positive geometry $X_{\ge0}$ and let $v$ be one of the zero-dimensional boundaries in $\mathcal{P}$ -- a vertex. We will denote by $\mathcal{I}_v=[v,X_{\ge 0}]$ an interval with the lowest element $\min(\mathcal{I}_v)=v$ and the top-dimensional element $\max(\mathcal{I}_v)=X_{\ge 0}$. We define a \emph{complete path $\gamma$} in $\mathcal{I}_v$ to be a path in $H[\mathcal{I}_v]$ from $X_{\ge 0}$ to $v$. Each complete path $\gamma$ can be expressed as a $d$-tuple of edges $\gamma=(e_1,\ldots,e_d)$ where the edges form a connected chain: $\partial^-(e_1)=X_{\ge 0}$, $\partial^+(e_d)=v$, and every pair of adjacent edges $(e_i,e_{i+1})$ in $\gamma$ satisfies $\partial^+(e_i)=\partial^-(e_{i+1})$. We will use $\Gamma[\mathcal{I}_v]$ to denote the set of complete paths in $\mathcal{I}_v$.

Now let us consider $\{\mathcal{I}_v^{(\sigma)}\}_{\sigma\in \mathcal{O}_n}$ = $\{[v,X_{\geq 0}^{(\sigma)}]\}_{\sigma\in \mathcal{O}_n}$ --  an indexed family of intervals where \mbox{$\min(\mathcal{I}_v^{(\sigma)})=v$} is the same for all $\sigma\in \mathcal{O}_n$. Additionally, suppose we have assigned edge labels to all edges as well as signs compatible with diamonds. We will denote by $\Gamma_v$ the set of complete paths in all Hasse diagrams:
\begin{align}
\Gamma_v\equiv\bigcup_{\sigma\in \mathcal{O}_n}\Gamma[\mathcal{I}_v^{(\sigma)}]\,.
\end{align}
Let $\vec{l}=(l^{(1)},\ldots,l^{(d)})$ be a $d$-tuple of edge labels. We define
\begin{align}
\check{\Gamma}(\vec{l}\,)=\bigcup_{\sigma\in \mathcal{O}_n}\left\{\gamma=(e_1,\ldots,e_d)\in \Gamma[\mathcal{I}_v^{(\sigma)}]\,:\,  e_i\in\check{E}(l^{(i)})\right\}\subseteq \Gamma_v\,
\end{align}
to be the set of all complete paths in each interval which can be identified by $\vec{l}$. If $\check{\Gamma}(\vec{l}\,)\ne\emptyset$, then we will refer to $\vec{l}$ as a \emph{complete path label}. 
For example, for $n=4$ there are $4$ complete paths which can be labelled by $\vec{l}=(\alpha_4,\alpha_2)$, therefore $(\alpha_4,\alpha_2)$ is a complete path label and $\check{\Gamma}(\alpha_4,\alpha_2)$ has $4$ elements. 

Returning to $\{\mathcal{I}_v^{(\sigma)}\}_{\sigma\in \mathcal{O}_n}$, the significance of these complete path labels is that they formalise the notion of ``pairs of rays pointing in opposite directions''. More precisely, given a complete path label $\vec{l}=(l^{(1)},\ldots,l^{(d-1)},l^{(d)})$, the complete paths in $\check{\Gamma}(\vec{l}\,)$ allow us to identify all one-dimensional boundaries inhabiting the same one-dimensional variety defined by $l^{(1)}=\cdots=l^{(d-1)}=0$. The one dimensional boundaries for which $l^{(1)}=\cdots=l^{(d-1)}=0$ are given by the source nodes $\partial^-(e_{d})$ of the final edges $e_{d}$ in each path $\gamma=(e_1,\ldots,e_d)\in\check{\Gamma}(\vec{l}\,)$. Consequently, using complete path labels allows us to identify one-dimensional boundaries from different intervals which (1) are either the same one-dimensional boundary or (2) join together to form the one-dimensional variety given by $l^{(1)}=\cdots=l^{(d-1)}=0$ without needing to solve any equations.   

We define the boundary with respect to the complete path label $\vec{l}=(l^{(1)},\ldots,l^{(d)})$, denoted by $\partial_{\vec{l}}$, as
\begin{align}
\partial_{\vec{l}} \equiv \partial_{l^{(d)}}\cdots\partial_{l^{(1)}}\,,
\end{align}
the product (written in reverse order) of the boundary operators with respect to each edge label appearing in $\vec{l}$. For example, in our example for $n=4$
\begin{align}
\partial_{(\alpha_2,\alpha_4)}c[1234]=\partial_{\alpha_4}\partial_{\alpha_2}c[1234]=v \quad\text{and}\quad \partial_{(\alpha_4,\alpha_2)}c[1234]=\partial_{\alpha_2}\partial_{\alpha_4}c[1234]=-v\,.
\end{align}

\begin{figure}
	\centering
	\includegraphics[scale=0.6]{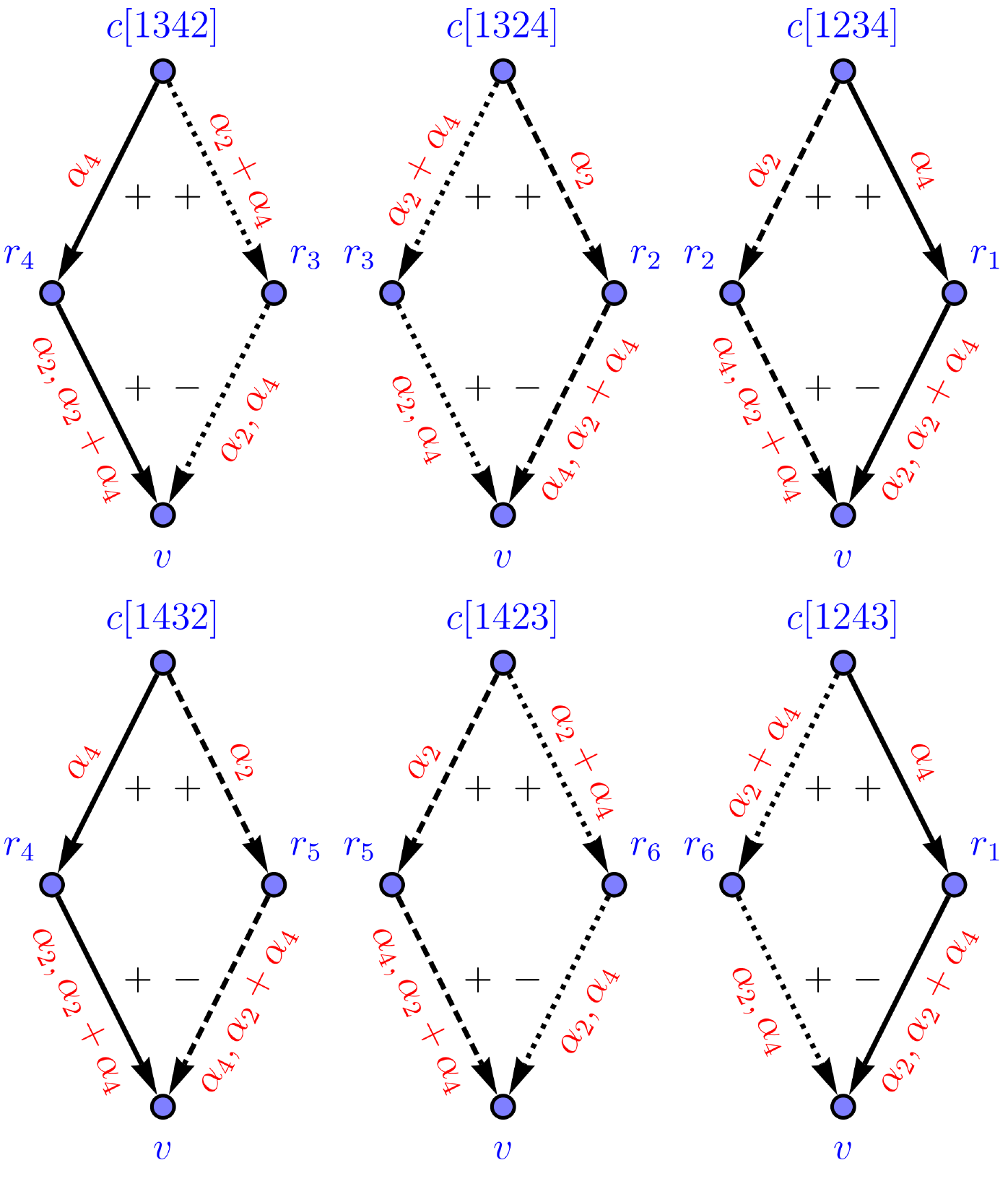}	
	\caption{Complete paths for $n=4$ and the minimal collection of labels which identify them.}
	\label{hom-4-2-posets-paths}
\end{figure}

Having defined these boundary operators, we now want to define the analogue of the ``boundary matrix'' introduced in Sec.\ \ref{sec:kk-mom-ray}. We do this by first identifying a minimal collection of complete paths needed in order for the kernel of the resulting boundary matrix to be congruent with the space of all KK relations. 
A set of complete path labels $\Gamma_{\text{min}}\subset \Gamma_v$ is called a \emph{minimal collection} if for every $\sigma\in \mathcal{O}_n$ and for every one-dimensional boundary $\mathcal{B}\in\mathcal{I}_v^{(\sigma)}$, there exists a complete path $\gamma=(e_1,\ldots,e_d)$ labelled by one of the labels in $\Gamma_{\text{min}}$, that is $\gamma\in\bigcup_{\vec{l}\in\Gamma_{\text{min}}}\check{\Gamma}(\vec{l}\,)$, such that $\gamma$ passes through $\mathcal{B}$ (i.e.~$\partial^-(e_d)=\mathcal{B}$). Generically, the sets of complete paths in $\{\check{\Gamma}(\vec{l}\,)\}_{\vec{l}\in\Gamma_{\text{min}}}$ are not mutually disjoint and their union is a strict subset of $\Gamma_v$.
An example of a minimal collection for $n=4$ is given by $\{\vec{l}_1,\vec{l}_2,\vec{l}_3\}$ where 
\begin{subequations}
	\label{eq:hom-4-2-minimal-collection}
	\begin{alignat}{3}
	\vec{l}_1~&\text{(solid)}&&=(\alpha_4,\alpha_2)\,,\\
	\vec{l}_2~&\text{(dashed)}&&=(\alpha_2,\alpha_4)\,,\\
	\vec{l}_3~&\text{(dotted)}&&=(\alpha_2+\alpha_4,\alpha_2)\,.
	\end{alignat}
\end{subequations}
The complete paths identified by these labels are drawn in Fig.\ \ref{hom-4-2-posets-paths} as solid, dashed and dotted paths, respectively. 

Finally, we define the \emph{boundary matrix with respect to a minimal collection of complete paths $\Gamma_{\text{min}}$}, written $M(\Gamma_{\text{min}})$, to have components
\begin{align}
M_{j\sigma}(\Gamma_{\text{min}})=\partial_{\vec{l}_j}X^{(\sigma)}_{\geq 0} = \left[\sum_{{\gamma=(e_1,\ldots,e_d)\in\check{\Gamma}(\vec{l}_j)\cap \Gamma[\mathcal{I}_v^{(\sigma)}]}}\Big(\prod_{t=1}^{d}\text{sgn}(e_t)\Big)\right]v\,,
\end{align}
where $\vec{l}_j \in \Gamma_{\text{min}}$, $\sigma \in \mathcal{O}_n$ and $v$ is the common vertex for each interval.
In the second equality we have written a sum over all complete paths in $H[\mathcal{I}_v^{(\sigma)}]$ labelled by $\vec{l}_j$ and for each complete path we have taken the product of the signs along the edges of the path. 

Crucially, the null space of $M$ determines all KK relations between the positive geometries $\{X_{\geq 0}^{(\sigma)}\}_{\sigma\in \mathcal{O}_n}$. To see this, remember that any minimal collection of complete path labels has the property that for each one-dimensional boundary $\mathcal{B}$ in an interval in $\{\mathcal{I}_v^{(\sigma)}\}_{\sigma\in \mathcal{O}_n}$, there is at least one complete path identified by one of the labels in $\Gamma_{\text{min}}$ which passes through $\mathcal{B}$. Let $\nu_\text{null}\in\ker(M)$ be a non-trivial element of the kernel of $M$. Then $\nu_\text{null}$ is a linear combination of the positive geometries $\{X_{\geq 0}^{(\sigma)}\}_{\sigma\in \mathcal{O}_n}$ such that for every complete path label $\vec{l}=(l^{(1)},\ldots,l^{(d-1)},l^{(d)})\in\Gamma_{\text{min}}$, $\partial_{\vec{l}}\,\nu_\text{null}=0$; for every one-dimensional variety defined by $l^{(1)}=\cdots=l^{(d-1)}=0$, which passes through the zero-dimensional boundary $v$ and which contains a non-empty subset of one-dimensional boundaries from the intervals in $\{\mathcal{I}_v^{(\sigma)}\}$, the one-dimensional boundaries which inhabit this variety conspire in $\nu_\text{null}$ to completely remove the zero-dimensional boundary. Consequently, $\nu_\text{null}$ represents a geometry without a zero-dimensional boundary which means that it cannot be a positive geometry and, hence, the corresponding linear combination of canonical differential forms must vanish.

For example, using the minimal collection of complete path labels given in \eqref{eq:hom-4-2-minimal-collection} for $n=4$, the corresponding boundary matrix is given by 
\begin{align}
M= \left(\begin{array}{rrrr}
\partial_{\vec{l}_1}c[1234]&\cdots&\partial_{\vec{l}_1}c[1432]\\
\partial_{\vec{l}_2}c[1234]&\cdots&\partial_{\vec{l}_2}c[1432]\\
\partial_{\vec{l}_3}c[1234]&\cdots&\partial_{\vec{l}_3}c[1432]
\end{array}\right)
=\left(\begin{array}{rrrrrr}
-1&-1&0&1&0&1 \\
1&0&-1&0&1&-1 \\
0&1&1&-1&-1&0 \\
\end{array}\right).
\end{align} 
We find the following basis for the null space of $M$:
\begin{align}
\label{eq:hom-4-2-null-vectors}
\begin{split}
c[1234]+c[1432]\,,\\
c[1324]+c[1423]\,,\\
c[1234]+c[1324]+c[1342]\,,\\
-c[1234]+c[1243]-c[1324]\,.
\end{split}
\end{align}

We now want to replace the cones in each of the four null vectors listed above by their corresponding canonical differential forms and set each linear combination to zero, but we will need to multiply each canonical differential form by an appropriate sign in order for its leading singularities (its residues on zero-dimensional boundaries) to be compatible with the signs we assigned to the edges of the corresponding Hasse diagram. To find these multiplicative weights, we begin by listing the canonical differential forms for each cone,
\begin{align}
\begin{split}
&\Omega(c[1234])=\Omega(c[1432])=d\log\alpha_4\wedge d\log\alpha_2 \,,\\
&\Omega(c[1243])=\Omega(c[1342])=d\log(\alpha_2+\alpha_4)\wedge d\log\alpha_4\,,\\
&\Omega(c[1324])=\Omega(c[1423])=d\log\alpha_2\wedge d\log(\alpha_2+\alpha_4)\,,
\end{split}
\end{align}
which can be read off from Fig.\ \ref{Allorders4pt}. 
For each of the complete path labels $\vec{l}=(l^{(1)},l^{(2)})\in\{\vec{l}_1,\vec{l}_2,\vec{l}_3\}$ given in \eqref{eq:hom-4-2-minimal-collection} we can define the \emph{residue operation along $\vec{l}$}, denoted by $\text{res}_{\vec{l}}\,$, as 
\begin{align}
\text{res}_{\vec{l}}\, = \text{res}_{l^{(2)}=0}\,\text{res}_{l^{(1)}=0}\,.
\end{align}
Then for each $\sigma\in\mathcal{O}_4$, the weight required to multiply $\Omega(c[\sigma])$, which we will denote by $w[\sigma]$, is obtained by taking a single complete path label $\vec{l}$ with respect to $\sigma$ (i.e.~$\check{\Gamma}(\vec{l}) \cap \Gamma[\mathcal{I}_v^{(\sigma)}] \ne\emptyset$) and computing
\begin{align}
\text{res}_{\vec{l}}\,\Omega(c[\sigma])\times v=w[\sigma]\times\partial_{\vec{l}}\,c[\sigma]\,.
\end{align} 
It is easy to check that
\begin{align}
w[1234]=-1=-w[1432]\,,\ \ w[1243]=1=-w[1342]\,,\ \ w[1324]=-1=-w[1423]\,.
\end{align}
If we now replace $c[\sigma]\to w[\sigma]\Omega(c[\sigma])$ in \eqref{eq:hom-4-2-null-vectors} and set each null vector to zero we obtain the four-particle KK relations for the canonical differential forms
\begin{align}
\label{eq:hom-4-2-kk-relations}
\begin{split}
-\Omega(c[1234])+\Omega(c[1432])=0\,,\\
-\Omega(c[1324])+\Omega(c[1423])=0\,,\\
\Omega(c[1234])+\Omega(c[1324])+\Omega(c[1342])=0\,,\\
\Omega(c[1234])+\Omega(c[1243])+\Omega(c[1324])=0\,,
\end{split}
\end{align}
in agreement with the KK relations for four-particle MHV amplitudes listed in \eqref{eq:4-2-alpha-reflection} and \eqref{eq:4-2-alpha-u1}.

A similar strategy can be employed in the general case when working with $\{\mathcal{I}_v^{(\sigma)}\}_{\sigma\in \mathcal{O}_n}$. For each $\sigma\in \mathcal{O}_n$, we define the weight $w[\sigma]$ by taking any complete path label $\vec{l}$ with respect to $\mathcal{I}_v^{(\sigma)}$ and computing 
	\begin{align}
	\text{res}_{\vec{l}}\,\Omega(X^{(\sigma)}_{\ge 0})\times v =w[\sigma]\times \partial_{\vec{l}}\,X^{(\sigma)}_{\ge 0}\,,
	\end{align}
	where $\Omega(X^{(\sigma)}_{\ge 0})$ is the canonical differential form for $X^{(\sigma)}_{\ge 0}$, $v$ is again the zero-dimensional boundary common to each interval, and $\text{res}_{\vec{l}}\,$ is the natural generalization of residue operation along $\vec{l}$ defined previously. Then each vector in the null space of the boundary matrix (with respect to some minimal collection) can be mapped to a KK relation by replacing each $X^{(\sigma)}_{\ge 0}$ with $w[\sigma]\Omega(X^{(\sigma)}_{\ge 0})$ and setting the null vector to zero.

\subsubsection{All Helicity Sectors}
The poset-based homological algorithm presented in the previous subsection was used to derive the KK relations for MHV amplitudes in the simplified setting of Sec.\ \ref{sec:kk-mom-alpha}, where each positive sector was an oriented simplicial cone in $(n-2)$ dimensions as opposed to the full $(2n-4)$-dimensional space. However, we can also derive the same KK relations using the full boundary stratification of the momentum amplituhedron for different orderings. In fact, this algorithm can be readily applied to any helicity sector. The only steps which need to be clarified in order to do this are (1) how to find the boundary stratification of the momentum amplituhedron for different orderings, and (2) how to generate edge labels.

For a general $n$ and $k$, with $2\le k \le n-2$, the zero-dimensional boundaries (or vertices) of the standard-ordering momentum amplituhedron $\mathcal{M}_{n,k}=\mathcal{M}_{n,k}^{(12\dots n)}$ are shared by all particle orderings and there are $\binom{n}{k}$ of them. These vertices are in one-to-one correspondence with vertices of the non-negative Grassmannian $G_{\ge 0}(k,n)$ via a linear map \cite{Damgaard:2019ztj}. Each vertex of $\mathcal{M}_{n,k}$ can be labelled by a $k$-element subset $I$ of $[n]=\{1,2,\ldots,n\}$ which identifies the only non-zero maximal minor of the matrix representing the corresponding vertex in $G_{\ge 0}(k,n)$. We will denote the vertex of $\mathcal{M}_{n,k}$ identified by $I\in\binom{[n]}{k}$ as $v_I$. Fix $I=\{i_1,i_2,\ldots,i_k\}\in\binom{[n]}{k}$. The poset interval between $\mathcal{M}_{n,k}$ and $v_I$, denoted by $[v_{I},\mathcal{M}_{n,k}]$, can be easily obtained using the function \texttt{momInterval} from the Mathematica$^\text{TM}$ package \texttt{amplituhedronBoundaries} \cite{Lukowski:2020bya}. Given any ordering $\sigma\in\mathcal{O}_n$, the interval between $\mathcal{M}_{n,k}^{(\sigma)}$ and $v_I$ is isomorphic to the interval between the standard ordering momentum amplituhedron and the vertex $v_{\sigma^{-1}(I)}$ labelled by $\sigma^{-1}(I)\equiv\{\sigma^{-1}(i_1),\sigma^{-1}(i_2),\ldots,\sigma^{-1}(i_k)\}$;
\begin{align}
[v_{I},\mathcal{M}_{n,k}^{(\sigma)}]\cong[v_{\sigma^{-1}(I)},\mathcal{M}_{n,k}^{(12\cdots n)}]\,.
\end{align}
Consequently, \texttt{momInterval} can again be used to obtain intervals for different orderings around each vertex $v_I$. Examples of such intervals for $\mathcal{M}^{(\sigma)}_{4,2}$ can be found in App.~\ref{app:posets42}. 

Turning our attention to edge labels, for any boundary $\mathcal{B}$ of the momentum amplituhedron $\mathcal{M}_{n,k}^{(\sigma)}$, we can determine which spinor brackets and which multi-particle Mandelstam variables vanish for $\mathcal{B}$. Additionally, starting from $n=6$ and $k=3$, some elements in the boundary poset of $\mathcal{M}_{n,k}$ might have boundaries corresponding to a sum of more than two external momenta going soft. We will denote the set of vanishing spinor brackets, multi-particle Mandelstam variables and multi-particle momenta $p_{i_1}+\ldots+p_{i_r}$ for $r>2$, by $Z(\mathcal{B})$. Now, given a directed edge $e=(\mathcal{B}_2,\mathcal{B}_1)$ 
in the Hasse diagram for the interval $[v_{I},\mathcal{M}_{n,k}^{(\sigma)}]$ we fix the set of edge labels for $e$ to be $S(e)=Z(\mathcal{B}_1)\setminus Z(\mathcal{B}_2)$. It contains all spinor brackets, Mandelstam variables and sums of momenta which vanish for $\mathcal{B}_1$ but are not zero for $\mathcal{B}_2$. 

Once all intervals have been generated and edge labels have been determined, our poset-based homological algorithm can be used to derive the KK relations in all helicity sectors. Importantly, it is sufficient to consider just one vertex, say $v_{\{1,2,\ldots,k\}}$, to derive all KK relations. This can be attributed to the fact that all momentum amplituhedra share all vertices and moreover that the geometries around a given vertex $v_I$ are identical with the ones around $v_{\{1,2,\ldots,k\}}$, after a relabelling. We have explicitly checked that this reproduces the correct KK relations for all $2\le k\le n-2$ and for $n\le 7$, and we expect it to work for all $n$.

%%%%%%%%%%%%%%%
% KK associahedron
%%%%%%%%%%%%%%%

\section{Kleiss-Kuijf Relations from the Kinematic Associahedron Geometry}
\label{sec:associahedron}
Our construction from the previous section can also be adapted to bi-adjoint scalar $\phi^3$ theory to derive KK relations using the kinematic associahedron. The kinematic associahedron $\mathcal{A}_n$ is the positive geometry associated with tree-level amplitudes of scalars in the adjoint representation of the product of two color groups $SU(N) \times SU(\tilde{N})$ with cubic interactions. A comprehensive discussion of this theory can be found in \cite{Cachazo:2013iea} and the kinematic associahedron was first introduced in \cite{Arkani-Hamed:2017mur}. The $n$ particle tree-level amplitude $\mathbb{M}_n$ in this theory has a color decomposition with respect to both color groups: 
\begin{align}\label{phi3amp}
	\mathbb{M}_n = \sum_{\alpha \in S_n /Z_n} \sum_{\beta \in S_n /Z_n}  \text{ Tr}\left(T^{a_{\alpha (1)}} T^{a_{\alpha (2)}}\cdots T^{a_{\alpha(n)}}\right) \text{Tr}\left(\tilde{T}^{b_{\beta (1)}} \tilde{T}^{b_{\beta (2)}} \cdots \tilde{T}^{b_{\beta(n)}}\right) m_n(\alpha|\beta) \,,
\end{align}
where  $\alpha$ and $\beta$ encode the orderings, and $m_n(\alpha |\beta) $ are referred to as \emph{double-partial} amplitudes.
Since the KK relations refer to a single color structure,  in the following we will fix $\alpha$ to be the standard ordering and define $m_n(1,2,\ldots,n |\beta) \equiv m_n(\beta )$. In this case, the double-partial amplitudes can be written as 
\begin{align}\label{eq:scalar-amp-beta}
	m_n(\beta ) = (-1)^{n-3+n_{\text{flip}}(\beta)} \sum_{\beta} \prod_{a = 1 }^{n-3} \frac{1}{X_{i_a, j_a}} \,,
\end{align}
where the sum is over planar Feynman diagrams which can also be ordered with respect to $\beta$, $X_{i,j} \equiv s_{i, i+1,\ldots, j-1}$ are the planar Mandelstam variables formed of momenta of consecutive particles and are the propagators in each Feynman diagram, and $n_\text{flip}(\beta)\equiv n_\text{flip}(1,2,\ldots,n|\beta)$ is defined in \cite{Cachazo:2013iea}.

The fact that the amplitudes \eqref{phi3amp} have a similar color structure to that of $SU(N)$ gauge theories described earlier allows for a derivation of similar KK relations. 
In particular, in analogy to \eqref{KK} the double-partial ordered amplitudes satisfy 
\begin{align}
	m_n(1, \{\alpha\}, n, \{\beta\}) = (-1)^{n_\beta} \sum_{\sigma\in\{\alpha\}\shuffle\{\beta^T\}} m_n(1 ,\{\sigma\},n)  \,.
	 \label{eq:kkScalar}
\end{align}
In this section we show how these relations can be derived from the kinematic associahedron.

Let us start by recalling the usual definition of the kinematic associahedron for the standard ordering \cite{Arkani-Hamed:2017mur}.
Similarly to the momentum amplituhedron, it is defined as an intersection:
\begin{equation} \label{defassstandard}
\mathcal{A}_n \equiv \Delta_n \cap H_n \,,
\end{equation}
where $\Delta_n$ is the positive region defined by the requirement that all planar Mandelstam variables are positive, i.e.~$X_{i,j} \geq 0$, and  $H_n$ is the affine subspace defined by demanding the following constants to be positive:
\begin{equation} \label{asscs}
	c_{i,j} \equiv  X_{i,j} + X_{i+1, j+1} - X_{i, j+1} -X_{i+1, j}  \geq 0 \,,
\end{equation}
for all non-adjacent $1\leq i <j < n$. Since the subspace $H_n$ is $(n-3)$-dimensional, when solving \eqref{asscs} we can choose which planar variables parametrize $H_n$. In the following we will parametrize $H_n$ by $X_{1,i}$ for $3\leq i\leq n-1$.

\begin{figure}
\centering
\begin{minipage}{.4\textwidth}
  \centering
  \includegraphics[scale=0.25]{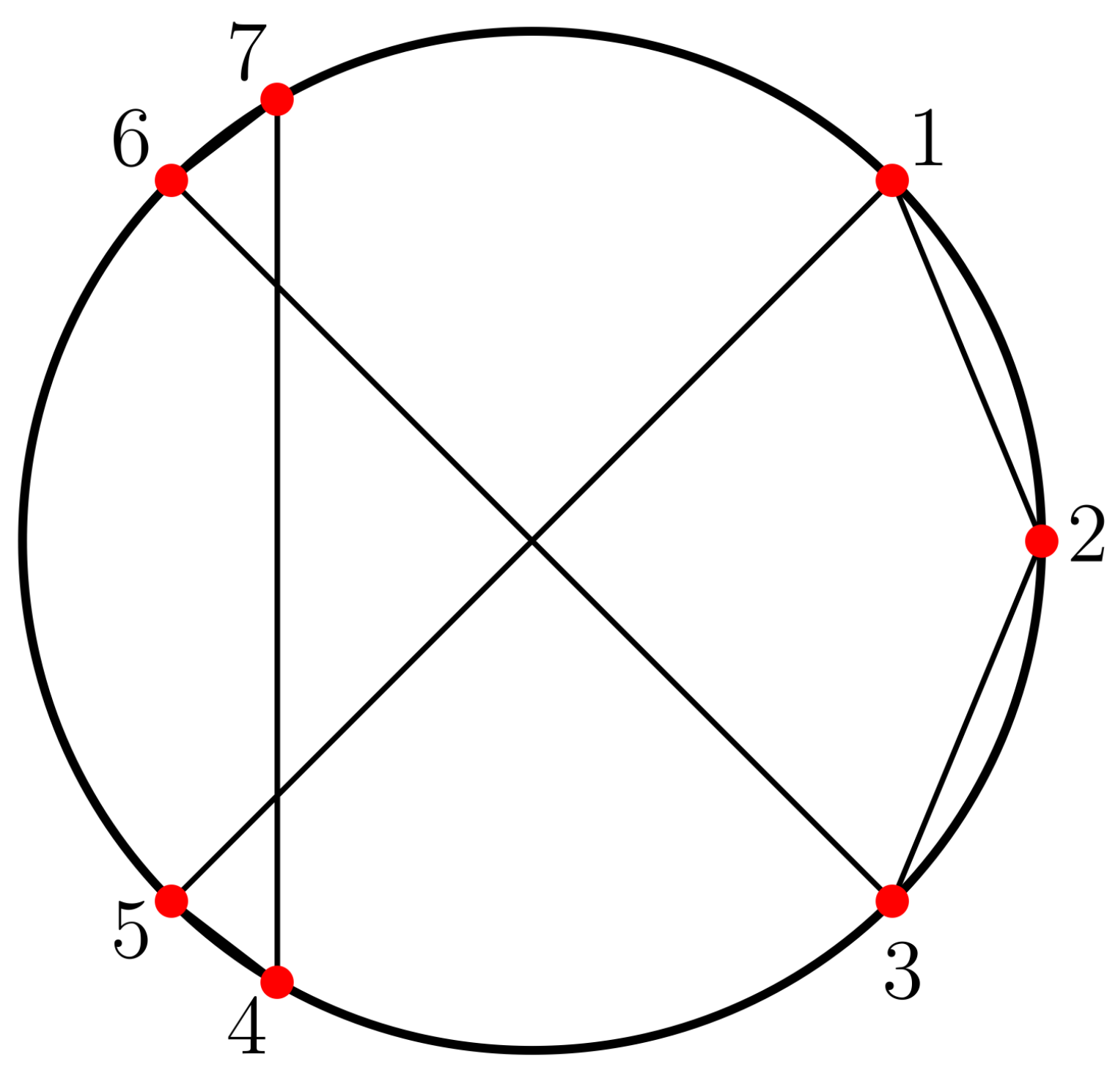}
\end{minipage}%
$\rightarrow$
\begin{minipage}{.4\textwidth}
  \centering
  \includegraphics[scale=0.25]{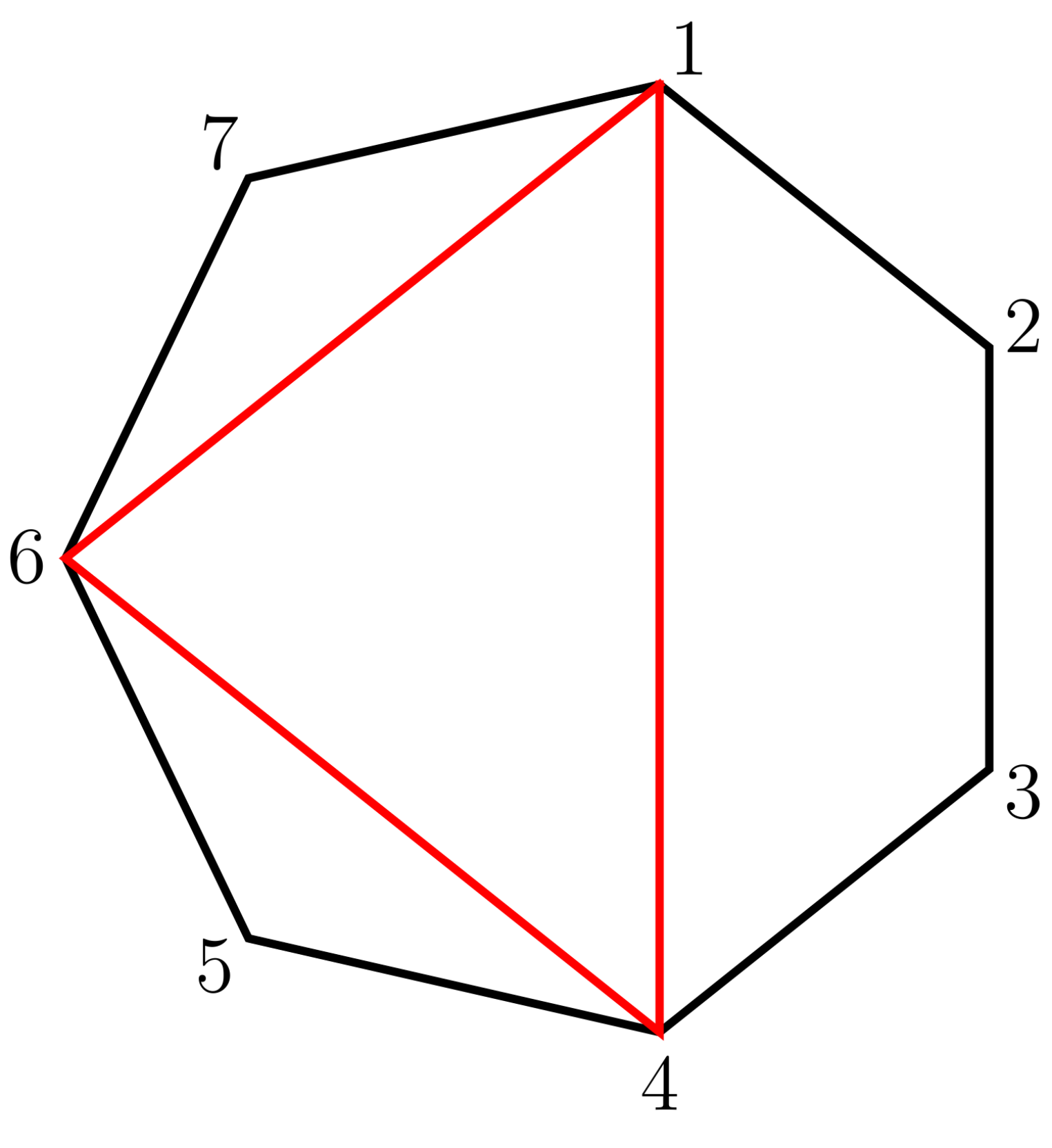}
\end{minipage}
\caption{Definition of the positive region $\Delta_7(1547632)$. We get a partial triangulation of a regular $7$-gon with diagonals $\mathcal{D}(1547632)=\{(1,4),(1,6),(4,6)\}$. This leads to $\Delta_7(1547632)=\{X_{1,3}\geq 0,X_{1,4}\geq 0,X_{1,6}\geq 0,X_{2,4}\geq 0,X_{4,6}\geq 0\}$.}
\label{fig:chord}
\end{figure}

For a non-standard ordering $\beta$, certain variables $X_{i,j}$ do not appear in the amplitude $m(\beta)$ since they are not planar with respect to this ordering. At the level of the underlying geometry, this corresponds to taking the corresponding boundaries of the (standard-ordering) associahedron to infinity. This can be accomplished in the definition of the associahedron for the ordering $\beta$ in two different ways. One option is to change the definition of the affine subspace $H_n$ by modifying the constraints in \eqref{asscs}, as was done in \cite{Arkani-Hamed:2017mur}. This however leads to associahedra for different orderings living on different subspaces. Instead, and in order to apply the results of previous sections, here we will define the associahedron $\mathcal{A}_n(\beta)$ for the ordering $\beta$ by modifying the definitions of the positive regions. 
Then
\begin{align}\label{defass}
	\mathcal{A}_n(\beta) = \Delta_n(\beta) \cap H_n \,,
\end{align} 
where the positive region $\Delta_n(\beta)$ can be obtained using a method which closely follows the construction of double-partial amplitudes in \cite{Cachazo:2013iea}. First, draw a circle with $n$ nodes on its boundary, labelled by the standard ordering, and link the nodes with a loop of line segments according to the ordering $\beta$. Thereafter, one proceeds iteratively as follows: start by locating a set $\{ i,i+1,\ldots ,i+r\}$ of at least two consecutive external labels, $r>1$, which are also consecutive in the $\beta$-ordering. We assume that this set is maximal and cannot be extended by adding other consecutive labels. If there is no such set then $\Delta_n(\beta)=\emptyset$ and if $r=n$ then $\Delta_n(\beta)=\Delta_n$. Next, redraw the graph by moving all points
in the set along the boundary of the disk, until they are close to each other. If the lines emanating from the nodes labelled by $i$ and $i+r$ intersect, call this intersection point by $R$. The points $\{i,i+1,\ldots,i+r,R\}$ form a convex polygon which should now be removed by bringing the point $R$ to the boundary of the disk. This leads to a new graph, with $R$ as an external point, and one can repeat the same procedure. If at any given point one fails to find a consecutive set of $r$ external labels with $r>1$ then $\Delta_n(\beta)=\emptyset$. Finally, every time we remove a polygon we define a pair of labels $(i_j,i_j+r_j+1)$. These labels provide a partial triangulation of a regular $n$-gon with diagonals  given by
	\begin{equation}
	\mathcal{D}(\beta)=\{(i_1,i_1+r_1+1),(i_2,i_2+r_2+1),\ldots ,(i_q,i_q+r_q+1)\}\,,
	\end{equation}
	where $q$ indicates the number of iterations of the above procedure before it halts.
	 To define the positive region $\Delta_n(\beta)$, for each diagonal $(i,j)\in \mathcal{D}(\beta)$ we demand $X_{i,j} \geq 0$, and for each diagonal $(a,b)$ which does not intersect any diagonal $(i,j)\in \mathcal{D}(\beta)$, we demand $X_{a,b} \geq 0$ also. In this way, given an ordering $\beta$ we can construct the positive region $\Delta_n(\beta)$ from the partial triangulation of a regular $n$-gon $\mathcal{D}(\beta)$ corresponding to $\beta$.
We illustrate our definition of the positive regions with the example in Fig.~\ref{fig:chord}. It is worth emphasizing that in our construction $\Delta_n(\beta)=\Delta_n(\beta^{-1})$ and therefore $\mathcal{A}_n(\beta)=\mathcal{A}_n(\beta^{-1})$. Moreover, in contradistinction to the momentum amplituhedron case, all associahedra do overlap.  

Definition \eqref{defass} allows us to define an oriented sum of associahedra for different orderings since they all live in the same affine subspace $H_n$. In order to determine the KK relations for bi-adjoint scalar $\phi^3$ amplitudes, we will look for oriented sums which do not have vertices in their boundary stratifications. Importantly, given an ordering $\beta$, the associahedron $\mathcal{A}_n(\beta)$ is an $(n-3)$-dimensional polyhedron in $H_n$, whose vertices are a subset of the vertices of the associahedron for the standard ordering. Let us call this set of vertices $\mathcal{V}_n$. In particular $|\mathcal{V}_n|=C_{n-2}$, where $C_n$ is the $n$-th Catalan number.

Having defined the associahedron $\mathcal{A}_n(\beta)$ for any ordering $\beta$, one can find a canonical differential form $\omega^{(\beta)}_n=\Omega(\mathcal{A}_n(\beta))$ with logarithmic singularities on all its boundaries. 
For the standard ordering this can be written as \cite{Arkani-Hamed:2017mur}
\begin{align}\label{assform}
	\omega_n^{(12\ldots n)} =\sum_{\text{planar } g } \text{sign}(g)  \bigwedge_{a=1}^{n-3} d\text{log} X_{i_a, j_a}\,,
\end{align}
where the sum is over all planar cubic graphs 
and the signs are determined by requiring that the form must be projective. The canonical forms $\omega_n^{(\beta)}$ can be found from \eqref{assform} by setting to zero all terms involving $d\text{log} X_{i,j}$ for which the boundaries $X_{i,j}=0$ are pushed to infinity in $\mathcal{A}^{(\beta)}_n$. 
In particular, we define
\begin{align}
\omega_n^{(\beta)}=(-1)^{n_\text{flip}(\beta)}\omega_n^{(12\ldots n)}\Big|_{d\text{log}X_{i,j}\to 0 \text{ if } X_{i,j}=0 \text{ is not a boundary of } \Delta_n(\beta)}\,,
\end{align}
where $n_\text{flip}(\beta)$ was introduced earlier in \eqref{eq:scalar-amp-beta} and the prefactor $(-1)^{n_\text{flip}(\beta)}$ determines the orientation of the associahedron $\mathcal{A}^{(\beta)}_n$ relative to the standard ordering. This orientation is chosen such that the double-partial amplitudes $m_n(\beta)$ can be consistently extracted from the canonical forms $\omega_n(\beta)$ via the pull-back onto the subspace $H_n$:
\begin{align}\label{ampfromass}
	\omega_n^{(\beta)}\Big{|}_{H_n}= m_{n}(\beta )\bigwedge_{j=3}^{n-1} d X_{1, j}  \,.
\end{align}

We conclude this subsection with an interesting remark on the number of non-empty positive regions $\Delta_n(\beta)$ for a given $n$. In the definition of $\Delta_n(\beta)$, we specified that if at any stage we cannot find a set of $r$ consecutive external labels with $r>1$ then $\Delta_n(\beta)=\emptyset$. This specification was made because in such a case the corresponding amplitude $m_n(\beta)=0$, and we say that the ordering $\beta$ is not compatible with the standard ordering. This implies that
the number of non-empty positive regions is not simply $(n-1)!=|\mathcal{O}_n|$, but can be smaller. In particular, we found by direct enumeration that the number of non-empty positive regions $p_n$ for $n$ particles is given (up to $n=8$) by
\begin{center}
	\begin{tabular}{c|c|c|c|c|c}
	\toprule
		$	n$ & $4$  & $5$ &  $6$ & $7$ & $8$ \\
		\hline \hline
		$p_n$ & $6 $ & $22$  &$90 $  & $394$ & $ 1806$ \\
		\bottomrule
	\end{tabular}
\end{center}
This sequence of numbers is called the Large Schr\"oder Numbers and has already been found in the context of positive geometries in the study of generalized triangles for the amplituhedron $\mathcal{A}_{n,k}^{(2)}$, see \cite{Lukowski:2019sxw} for details. In particular, the partial triangulations which we construct in the definition of the positive region $\Delta_n(\beta)$ correspond to the graphical labels for generalized triangles described in \cite{Lukowski:2019sxw}.

\subsection{Ray-based Homological Description}
In order to determine the KK relations \eqref{eq:kkScalar} for $m_n(\beta)$  we will apply the ray-based homological construction from Sec.~\ref{sec:kk-mom-ray} to each vertex of the associahedron $\mathcal{A}_n$, namely all vertices $v\in \mathcal{V}_n$. Once again, we will find that the KK relations correspond to oriented sums of associahedra which do not have any vertices in their boundary stratification. For such oriented sums, the corresponding sum of canonical forms must vanish. We will exploit the fact that when one zooms in on any vertex $v\in \mathcal{V}_n$, then around this point every associahedron $\mathcal{A}_n(\beta)$ which contains $v$ as a vertex looks like an $(n-3)$-dimensional simplicial cone spanned by exactly $(n-3)$ rays. We denote this cone by 
\begin{align}
c^{(v)}[\beta]=
v+\text{span}_{\mathbb{R}\geq 0}\left\{r^v_{j_1^{\beta}},r^v_{j_2^{\beta}},\ldots,r^v_{j_{n-3}^{\beta}}\right\},
\end{align} if $v\in\mathcal{A}_n(\beta)$ and otherwise it is empty. Here $r^v_{j_i^{\beta}}$ are rays. To each such cone we can associate a ray $(n-3)$-form similarly to what was done in \eqref{eq:-n-2-ray-omega} with
\begin{align}
\omega(c^{(v)}[\beta])=
(-1)^{n_\text{flip}(\beta)}\frac{\det\left(r^v_{j^{\beta}_{1}}, r^v_{j^{\beta}_{2}},\ldots,r^v_{j^{\beta}_{n-2}}\right)}{\left|\det\left(r^v_{j^{\beta}_{1}}, r^v_{j^{\beta}_{2}},\ldots,r^v_{j^{\beta}_{n-2}}\right)\right|}\, \tilde{r}^v_{j^{\beta}_{1}}\wedge \tilde{r}^v_{j^{\beta}_{2}}\wedge\cdots\wedge \tilde{r}^v_{j^{\beta}_{n-2}}\,,
\end{align}
if $v\in\mathcal{A}_n(\beta)$ and otherwise zero.
 Again, we can construct boundary operators which act on these ray forms as explained in Sec.\ \ref{sec:kk-mom-ray} and proceed to construct boundary matrices $M_n^{(v)}$ around each vertex $v\in \mathcal{V}_n$. Then the KK relations are given by the common null space of all matrices $M_n^{(v)}$. Importantly, there are exactly $p_n-|\mathcal{V}_n|=p_n-C_{n-2}$ independent KK relations, where $p_n$ is given by the number of non-empty positive regions, and $C_n$ is the $n$-th Catalan number.

We illustrate this discussion by providing kinematic associahedra for $n=4,5$ and we show how the KK relations emerge in these examples.

\begin{figure}
\centering
\begin{tabular}{ccc}
\includegraphics[scale=0.2]{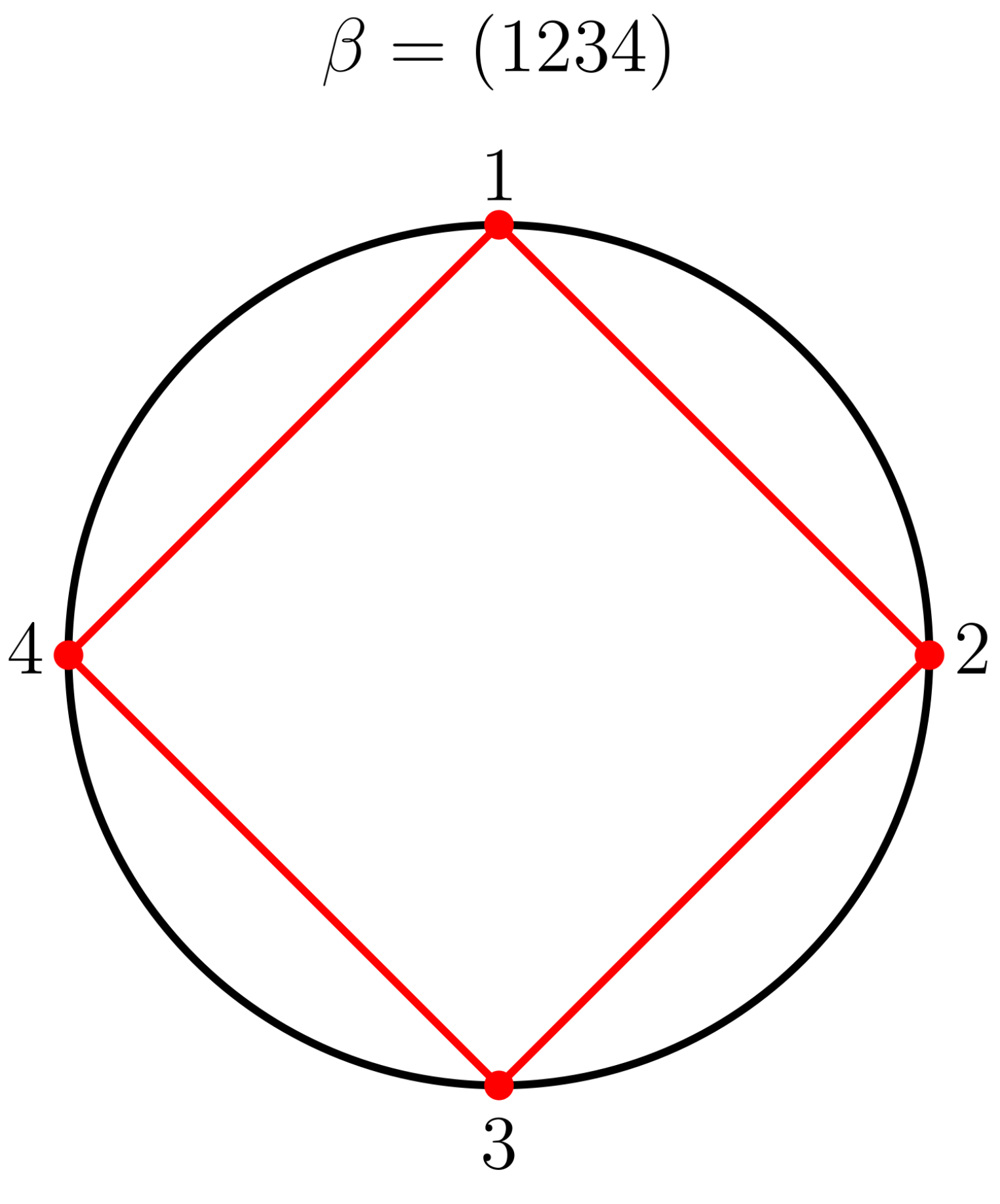}&
\includegraphics[scale=0.2]{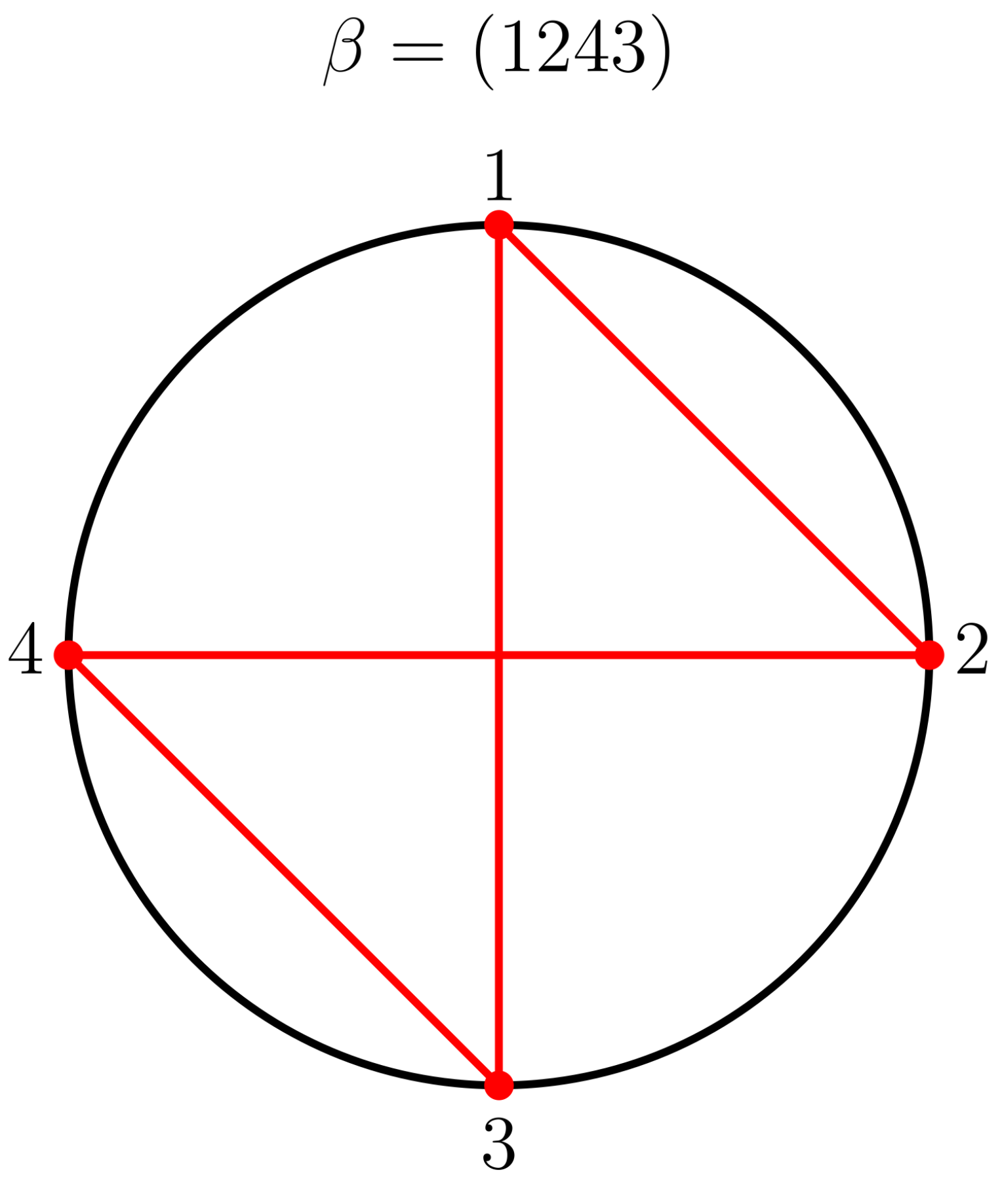}& 
\includegraphics[scale=0.2]{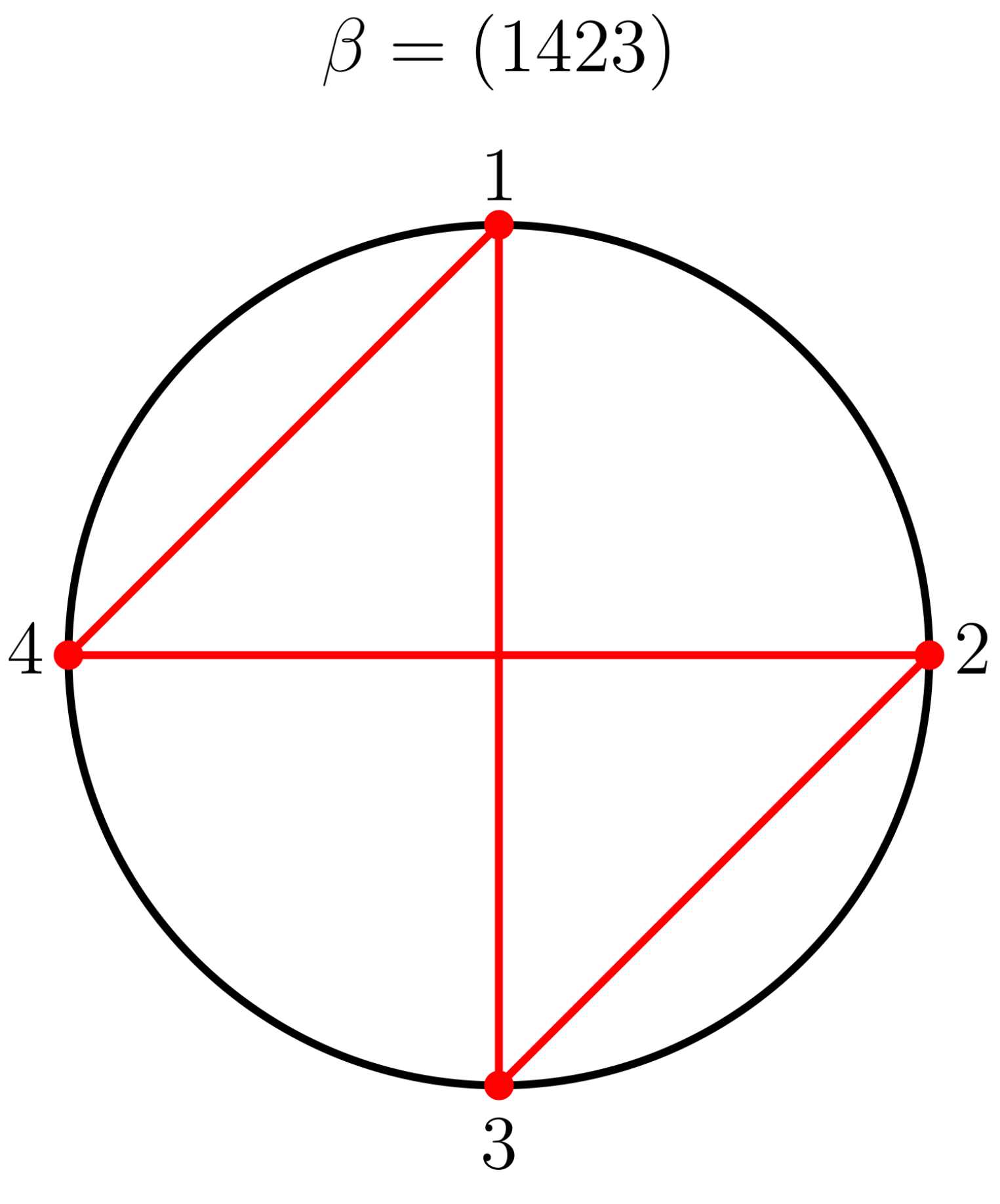}\\
$\downarrow$ & $\downarrow$ & $\downarrow$ \\
\includegraphics[scale=0.18]{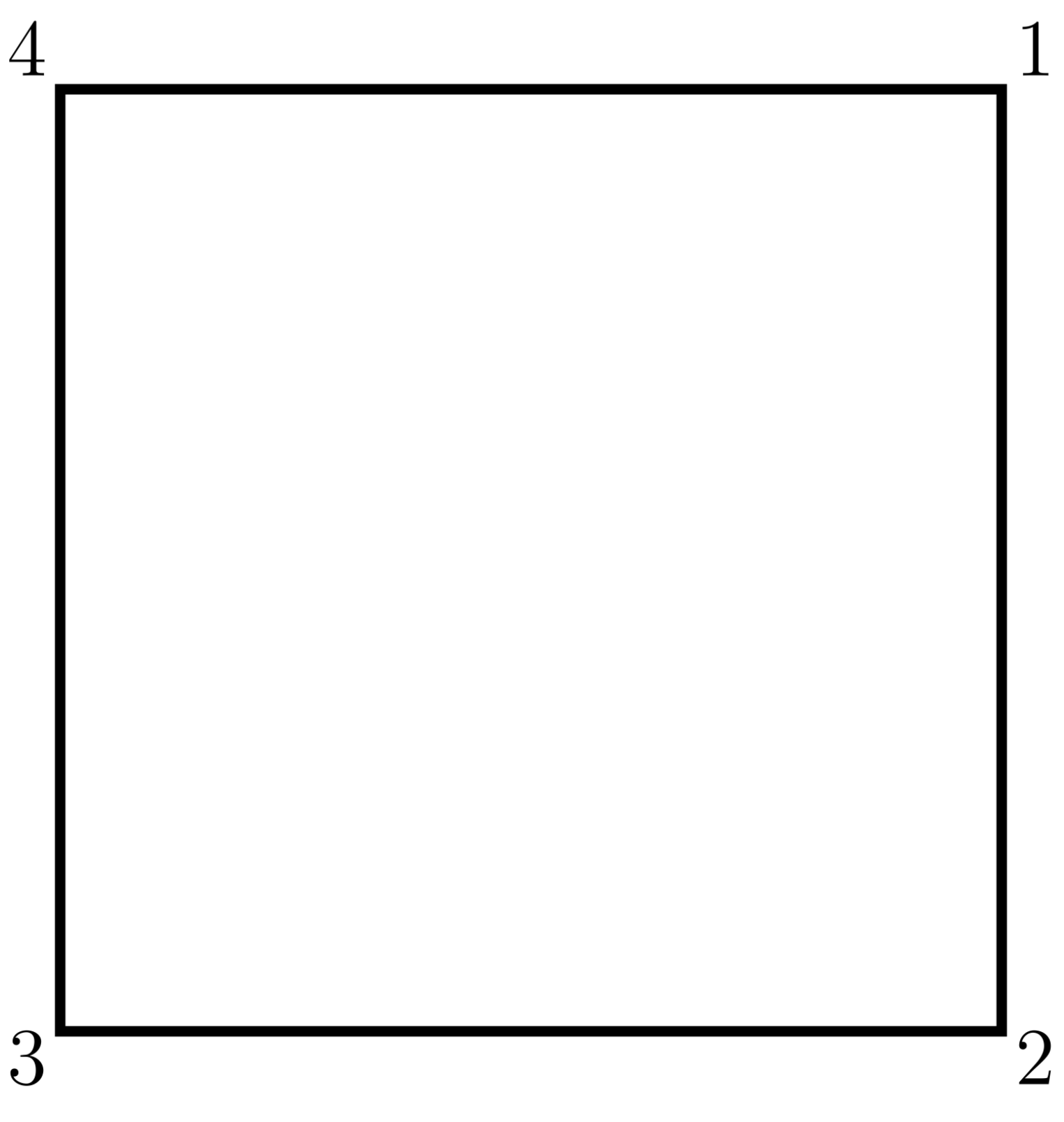}&
\includegraphics[scale=0.18]{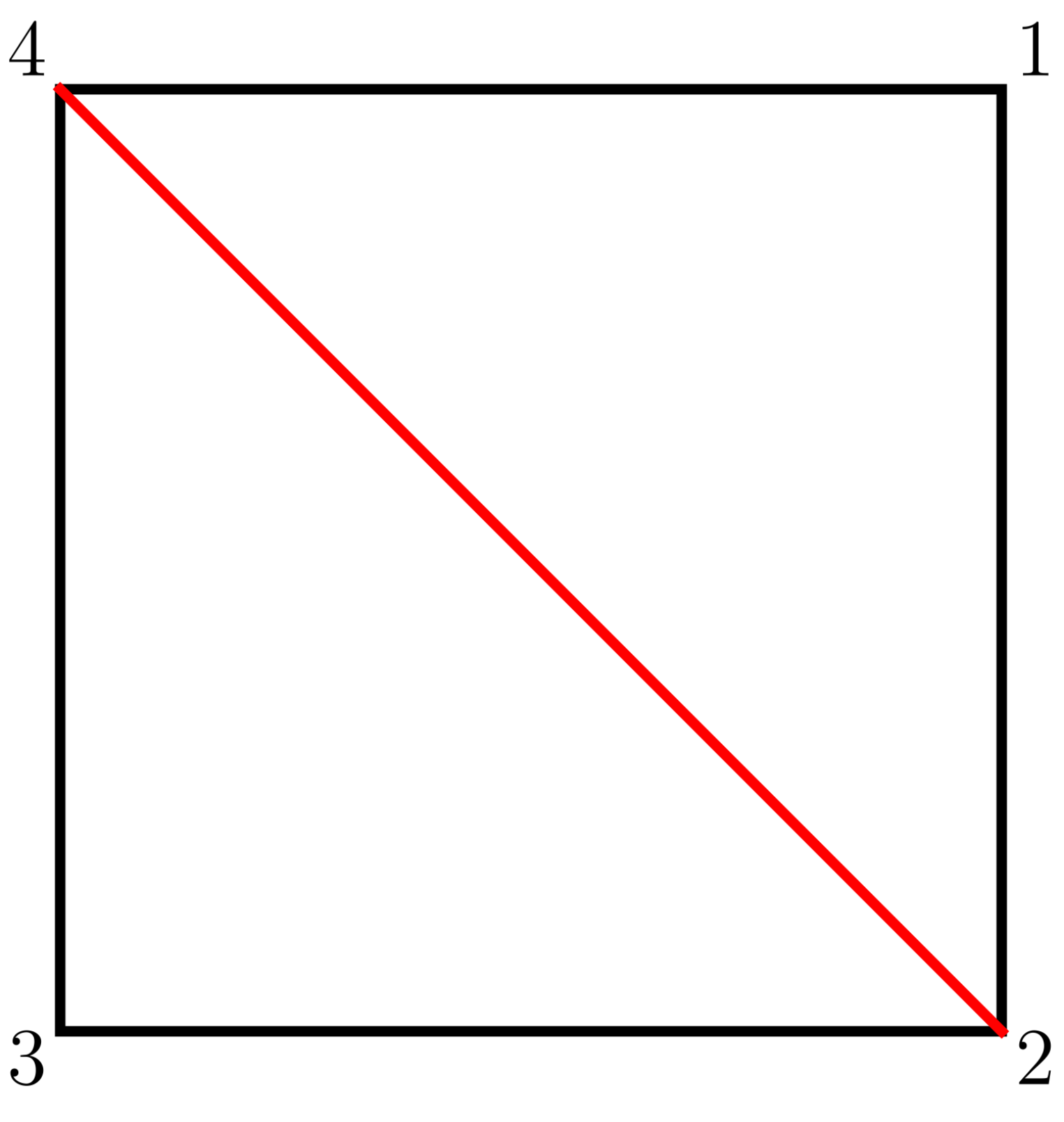}&
\includegraphics[scale=0.18]{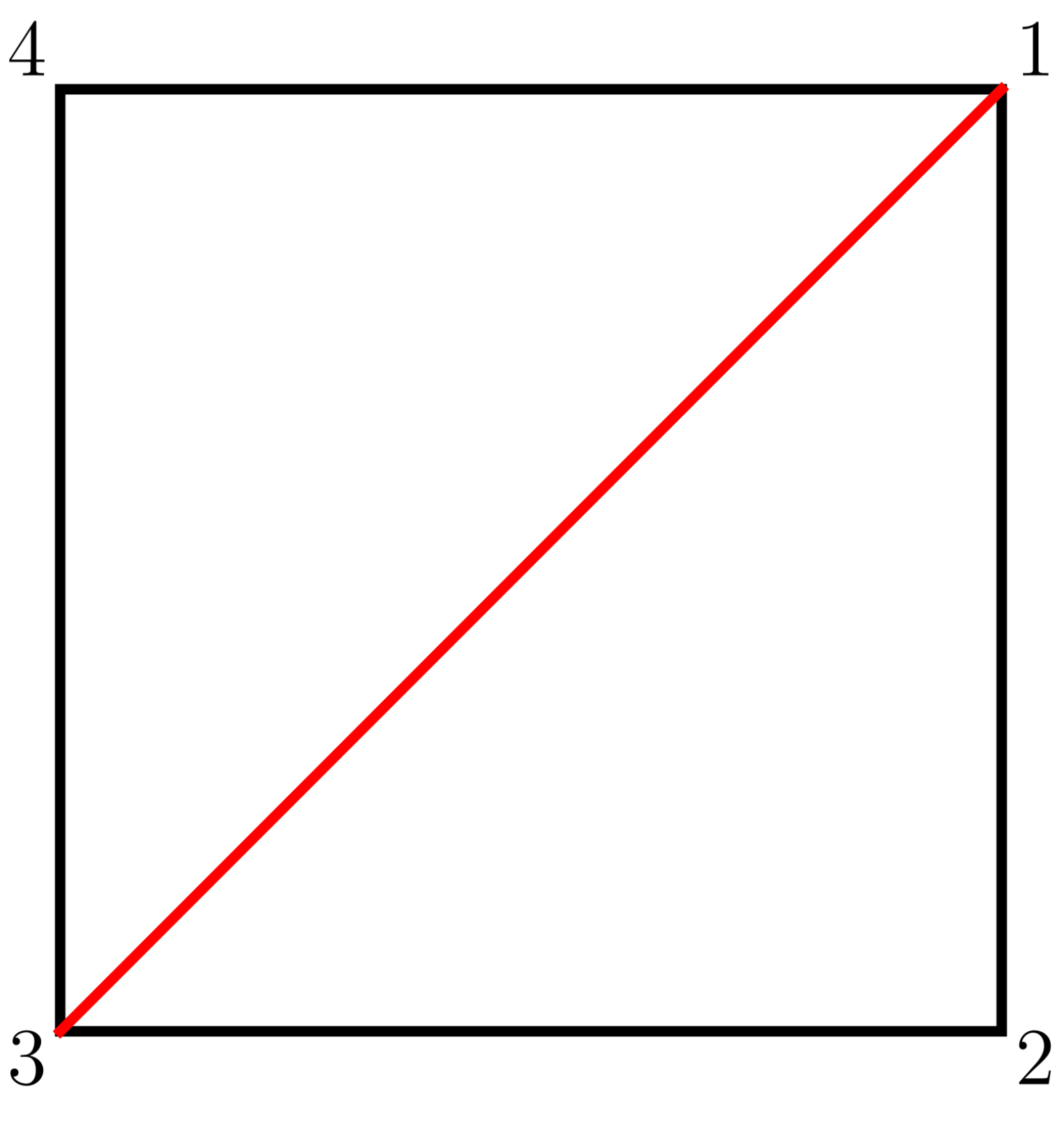}
\end{tabular}
\caption{Partial triangulations in the definition of positive regions $\Delta_4(\beta)$.}
\label{fig:square}
\end{figure}

\begin{figure}[!h]
\begin{align*}
\vcenter{\hbox{\includegraphics[scale=0.22]{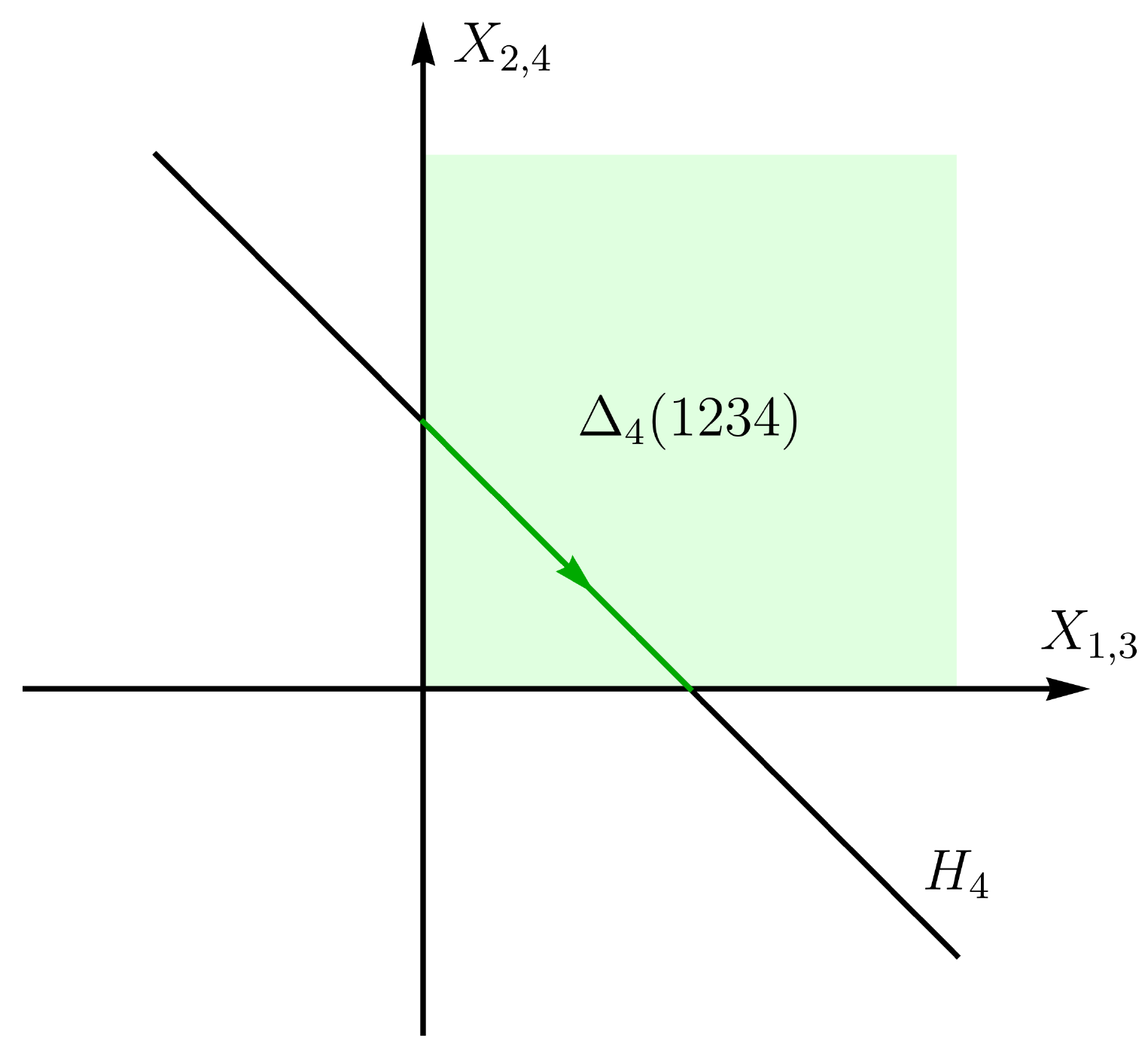}}}\oplus 
\vcenter{\hbox{\includegraphics[scale=0.22]{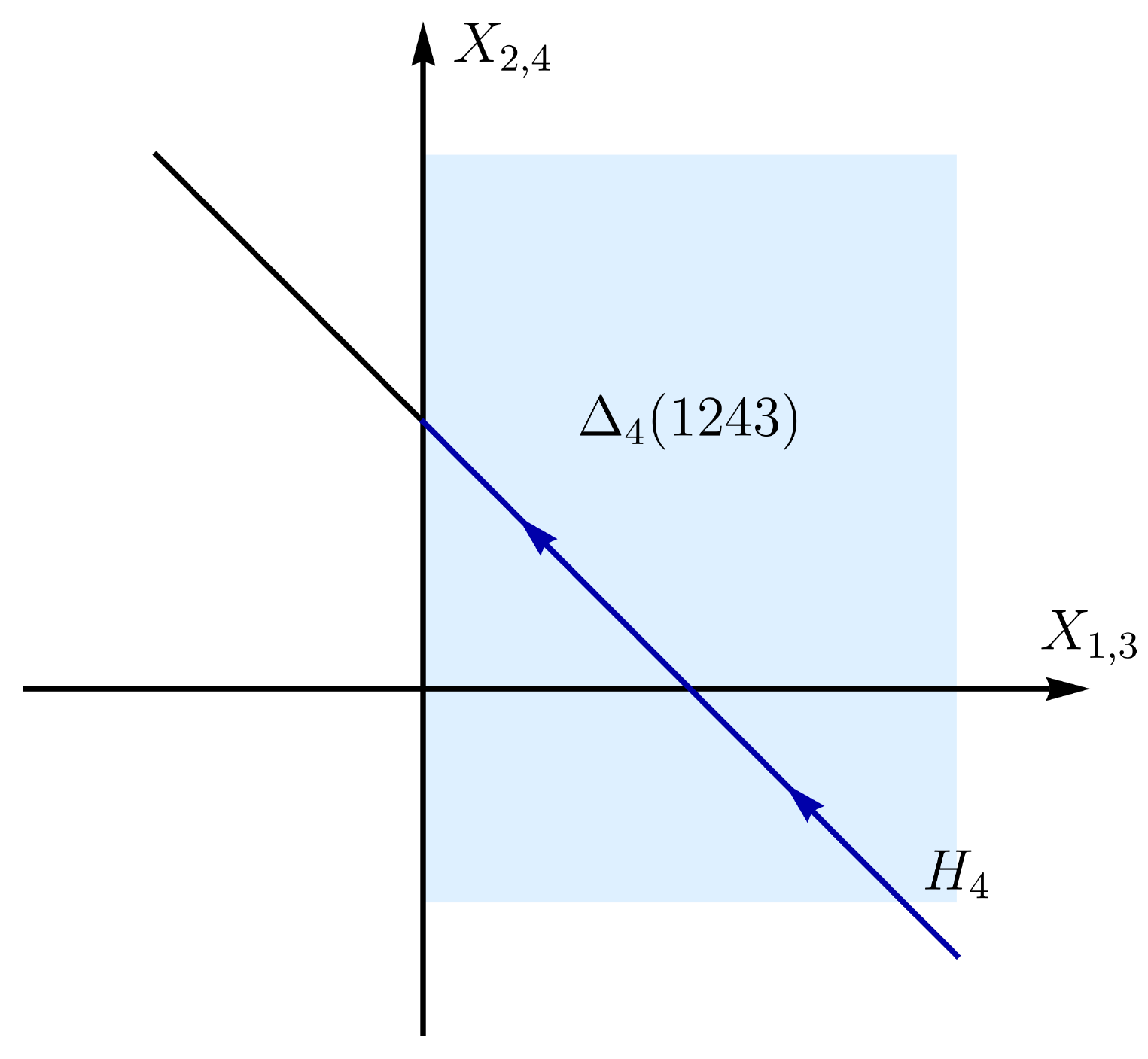}}}\oplus
\vcenter{\hbox{\includegraphics[scale=0.22]{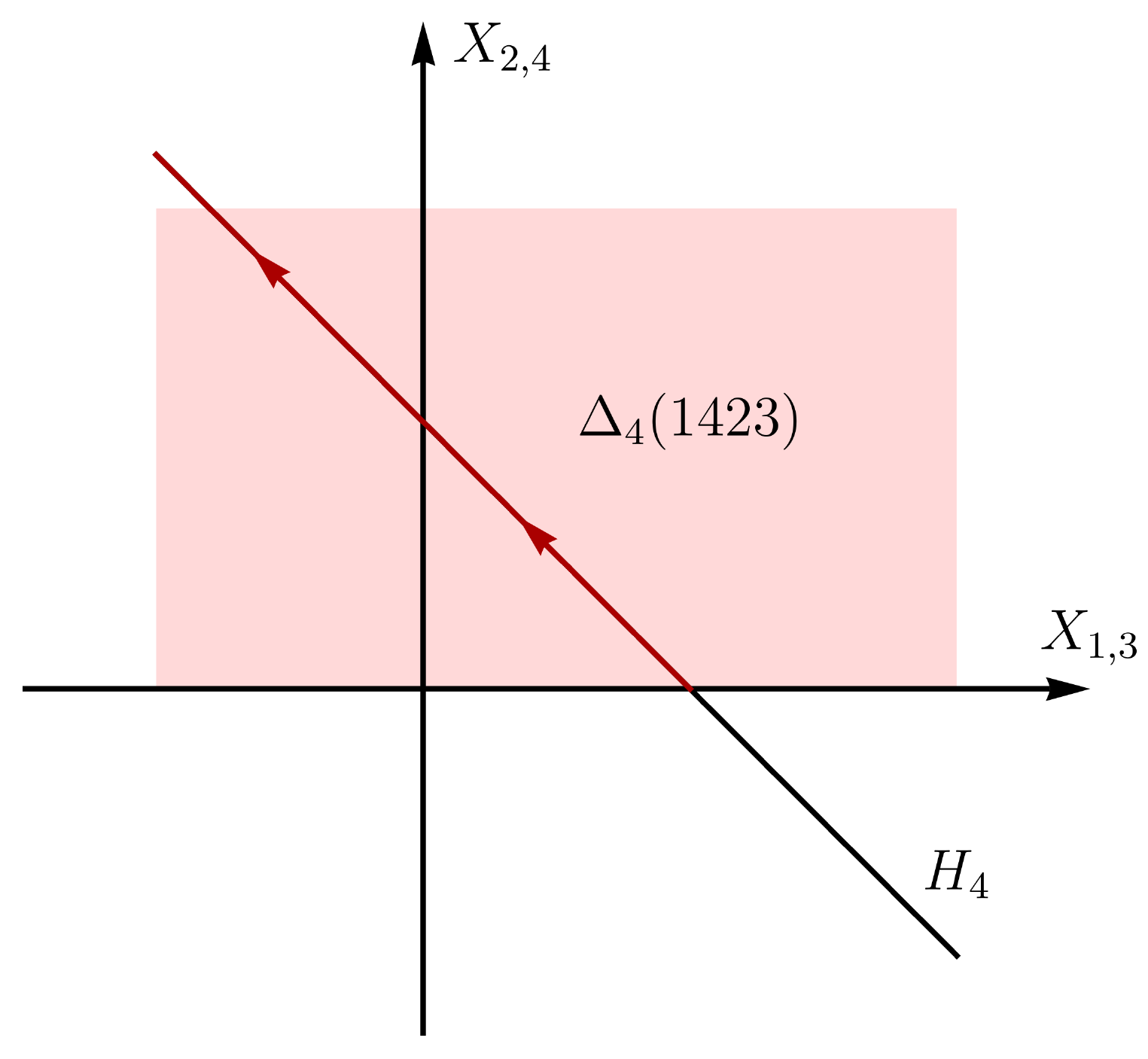}}} = 
\vcenter{\hbox{\includegraphics[scale=0.22]{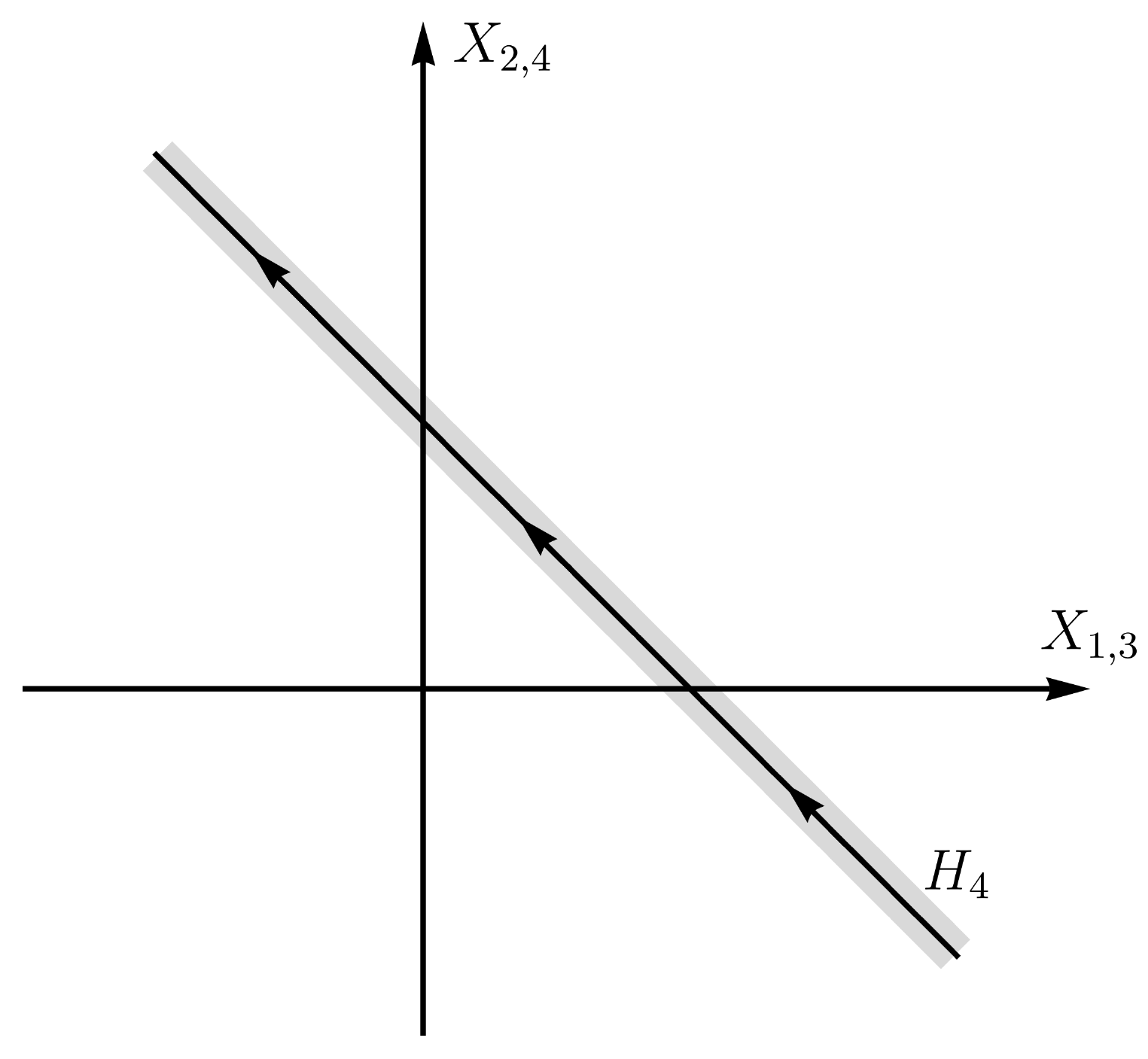}}}
\end{align*}
\caption{Oriented sum of three associahedra for $n=4$ producing an infinite line.}
\label{fig:ass-4-kk}
\end{figure}

\paragraph{Four-particle Amplitudes.}
For $n=4$ there are six non-empty positive regions labelled by the $\beta$-orderings in $\mathcal{O}_4$. Since $\Delta_n(\beta)=\Delta_n(\beta^{-1})$ we get three distinct positive regions: $\Delta_4(1234)=\Delta_4(1432)$, $\Delta_4(1243)=\Delta_4(1342)$ and $\Delta_4(1423)=\Delta_4(1324)$. Using the definition of $\Delta_n(\beta)$ we find 
\begin{align}
\Delta_4(1234)&=\{X_{1,3}\geq 0,X_{2,4}\geq 0\}\,,\\
\Delta_4(1243)&=\{X_{1,3}\geq 0\}\,,\\
\Delta_4(1423)&=\{X_{2,4}\geq 0\}\,,
\end{align}
which can be found from the partial triangulations of a square, depicted in Fig.~\ref{fig:square}.
The affine subspace $H_4$ is given by 
\begin{align}
	H_4=  \{(X_{1,3}, X_{2,4}) : X_{2,4}= c_{13} -X_{1,3}\}\subset \mathbb{R}^2\,.
\end{align}
The three associahedra for $n=4$, in the space parametrized by $X_{1,3}$, are given by a segment for the standard ordering and two semi-infinite lines for the remaining two orderings:
\begin{align}
\mathcal{A}_4[1234]=[0,c_{13}]\,,\quad\quad
\mathcal{A}_4[1243]=[0,+\infty)\,,\quad\quad
\mathcal{A}_4[1423]=(-\infty,c_{13}]\,.
\end{align} 
The set of vertices $\mathcal{V}_4=\{v_1,v_2\}$ contains two points: $v_1=(0)$ and $v_{2}=(c_{13})$.
Around each vertex in $\mathcal{V}_2$, the associahedra can be described as one-dimensional cones and we collect this information in the following table
\begin{center}
	\begin{tabular}{c|c|c|c}
	\toprule
			$\beta$ & $\mathcal{A}_4(\beta)$ & around $v_1=(0)$&around $v_2=(c_{13})$\\
		\hline \hline
		$(1234)$	& $[0,c_{13}]$ & $v_1+\text{span}_{\mathbb{R}_{\geq0}}\{r^{v_1}\}$& $v_2+\text{span}_{\mathbb{R}_{\geq0}}\{r^{v_2}\}$ \\
		\hline
		$(1243) $ 	&  $[0,+\infty)$& $v_1+\text{span}_{\mathbb{R}_{\geq0}}\{r^{v_1}\}$&$\emptyset$ \\ \hline
		$(1423) $ & $(-\infty,c_{13}]$ &$\emptyset$  & $v_2+\text{span}_{\mathbb{R}_{\geq0}}\{r^{v_2}\}$\\
		\bottomrule
	\end{tabular}
\end{center} 
where $r^{v_1}=(1)$ and $r^{v_2}=(-1)$.  The rays one-forms for each associahedron around each vertex are given in the next table %
\begin{center}
	\begin{tabular}{c|c|c|c}
		\toprule
		$\beta$ & $\mathcal{A}_4(\beta)$ & around $v_1=(0)$&around $v_2=(c_{13})$\\
		\hline \hline
		$(1234)$	& $[0,c_{13}]$ & $\tilde{r}^{v_1}$& $-\tilde{r}^{v_2}$ \\
		\hline
		$(1243) $ 	&  $[0,+\infty)$& $-\tilde{r}^{v_1}$&$0$ \\ \hline
		$(1423) $ & $(-\infty,c_{13}]$ &$0$  & $\tilde{r}^{v_2}$\\
		\bottomrule
	\end{tabular}
\end{center}% 
There are two boundary matrices, one for each vertex, given by
\begin{equation}
M_4^{(v_1)}=\left(\begin{matrix}
1&-1&0
\end{matrix}\right) \,,\qquad\qquad M_4^{(v_2)}=\left(\begin{matrix}
-1&0&1
\end{matrix}\right)\,,
\end{equation}
for which the common kernel is spanned by a single vector $(1,1,1)$.
Therefore, the canonical differential forms satisfy, in addition to the three reflection identities, the following relation
\begin{equation}\label{eq:ass-4-kk-forms}
\omega_4^{(1234)}+\omega_4^{(1243)}+\omega_4^{(1423)}=0\,,
\end{equation}
from which we can immediately extract the KK relation
\begin{equation}
m_4(1234)+m_4(1243)+m_4(1423)=0\,,
\end{equation}
using \eqref{ampfromass}. This KK relation can be understood directly in terms of the oriented sum $\mathcal{A}_4(1234)\oplus\mathcal{A}_4(1243)\oplus\mathcal{A}_4(1423)$ which is depicted in Fig.\ \ref{fig:ass-4-kk}. There we see that the oriented sum of the three associahedra produces an infinite line without any vertices, and this absence of zero-dimensional boundaries necessitates \eqref{eq:ass-4-kk-forms}.

\begin{figure}
\begin{center}
\includegraphics[scale=0.85]{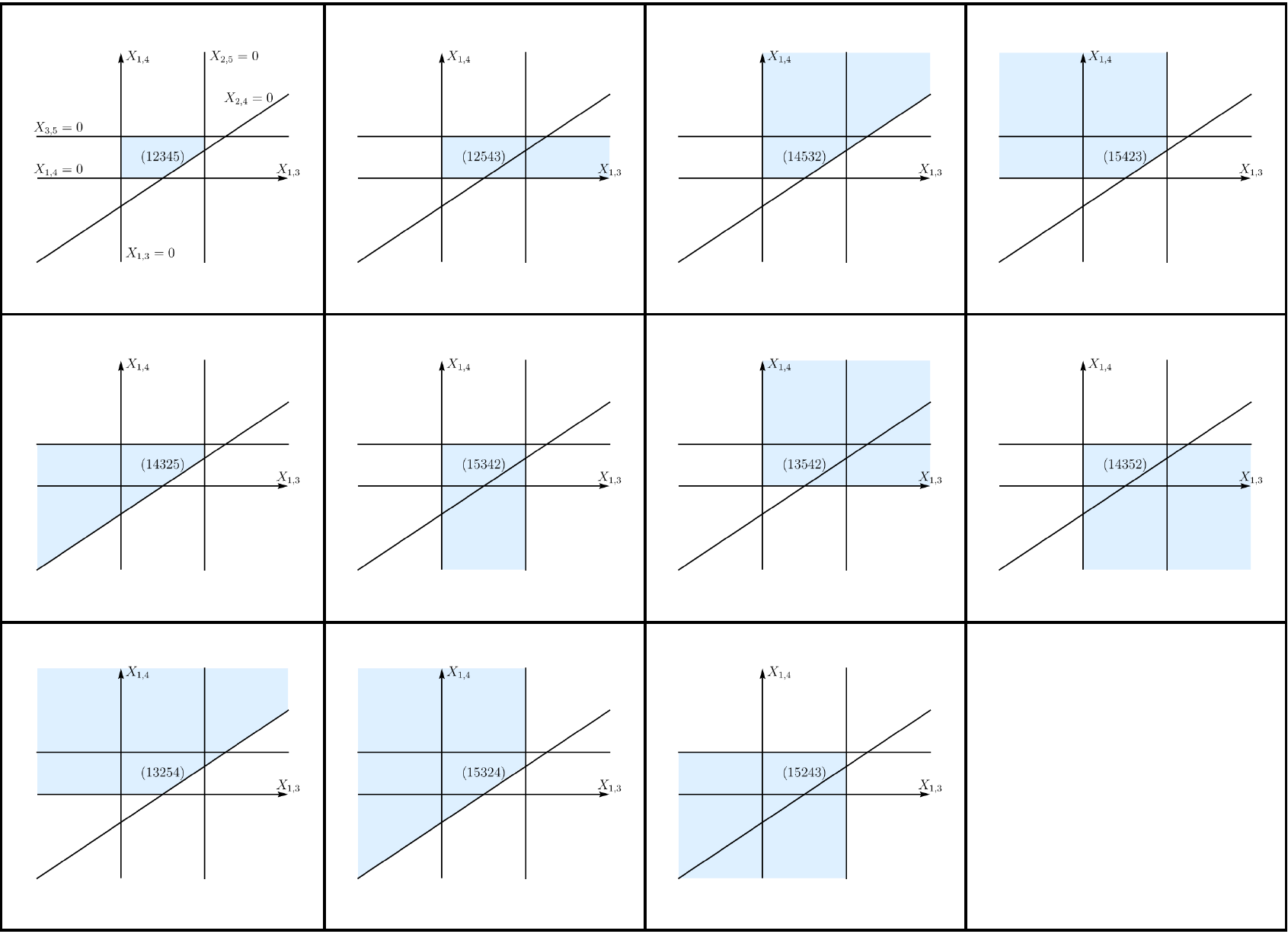}
\end{center}
\caption{Associahedra for $n=5$. The depicted geometries are all oriented counter-clockwise.}
\label{fig:ass5}
\end{figure}

\begin{figure}[!h]
	\begin{align*}
	\vcenter{\hbox{\includegraphics[scale=0.227]{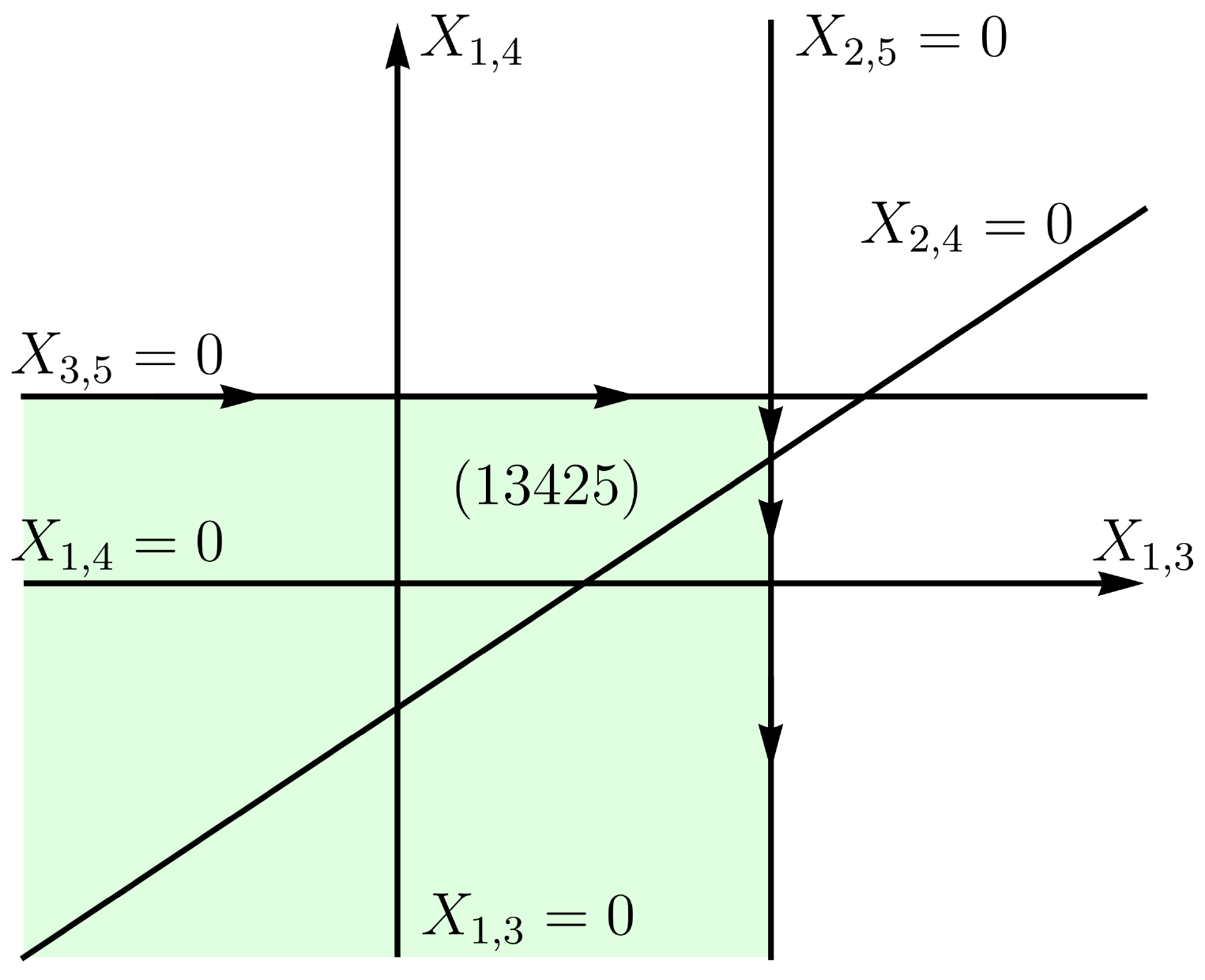}}}
	\oplus
	\vcenter{\hbox{\includegraphics[scale=0.227]{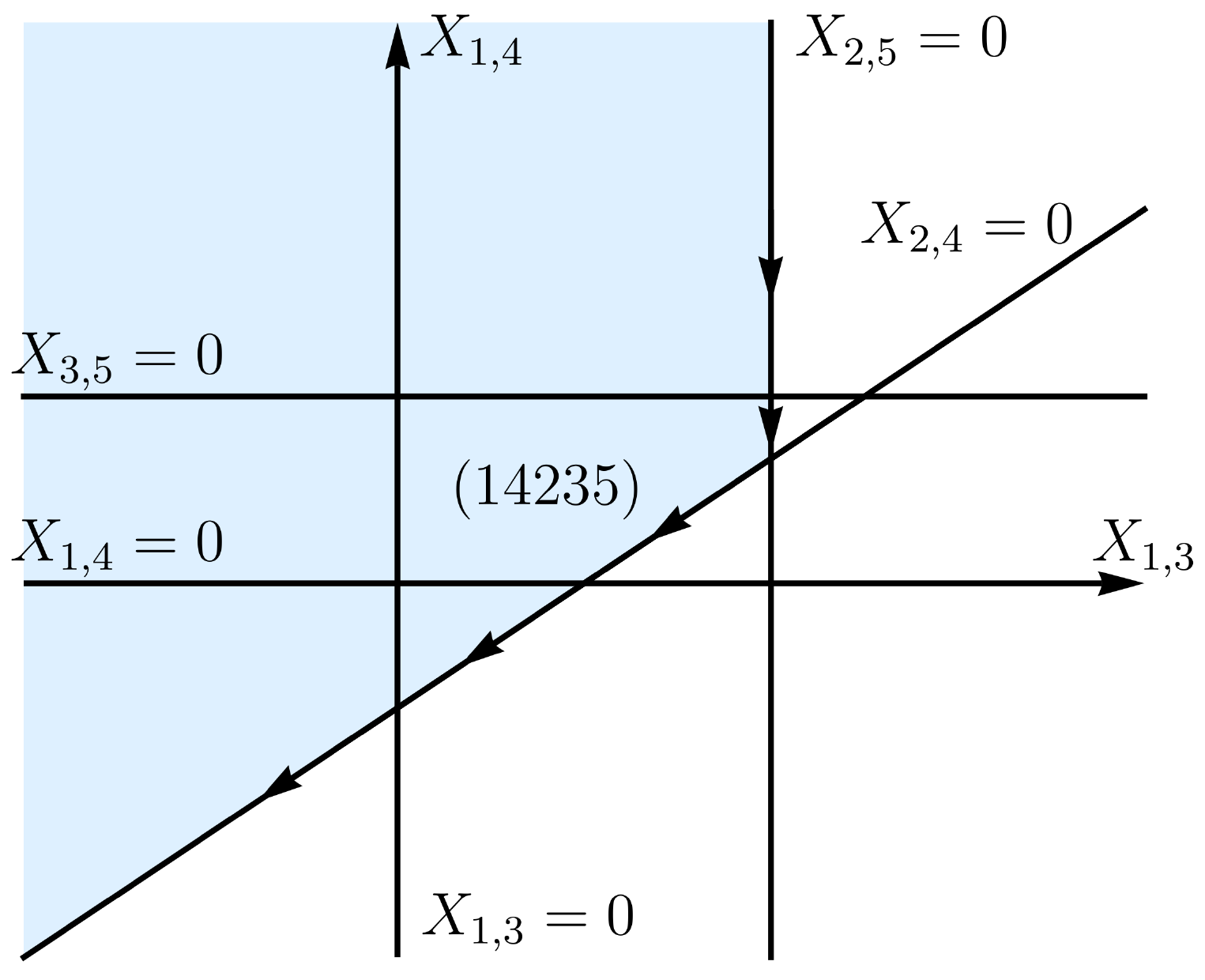}}}
	\oplus
	\vcenter{\hbox{\includegraphics[scale=0.227]{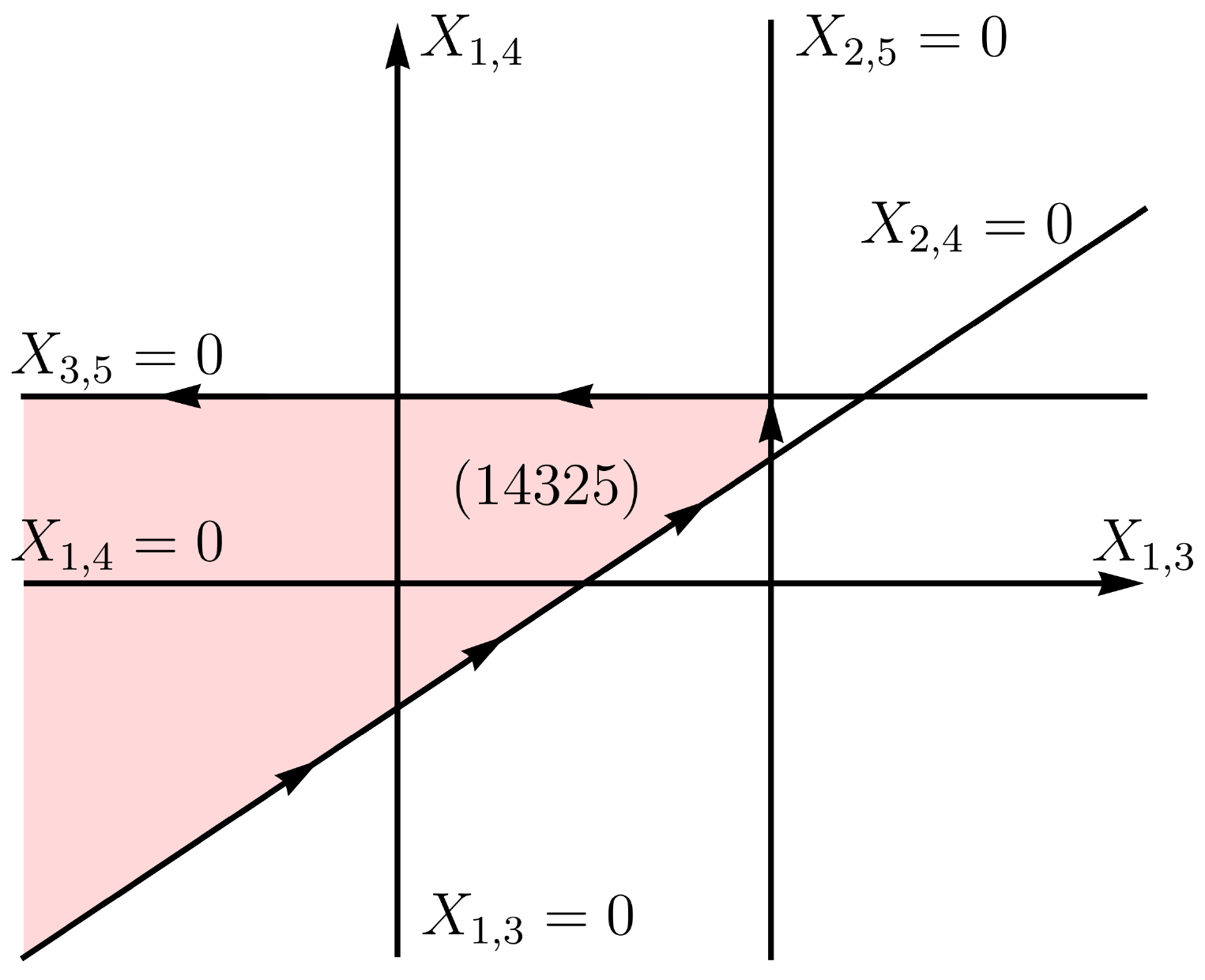}}}
	= \vcenter{\hbox{\includegraphics[scale=0.227]{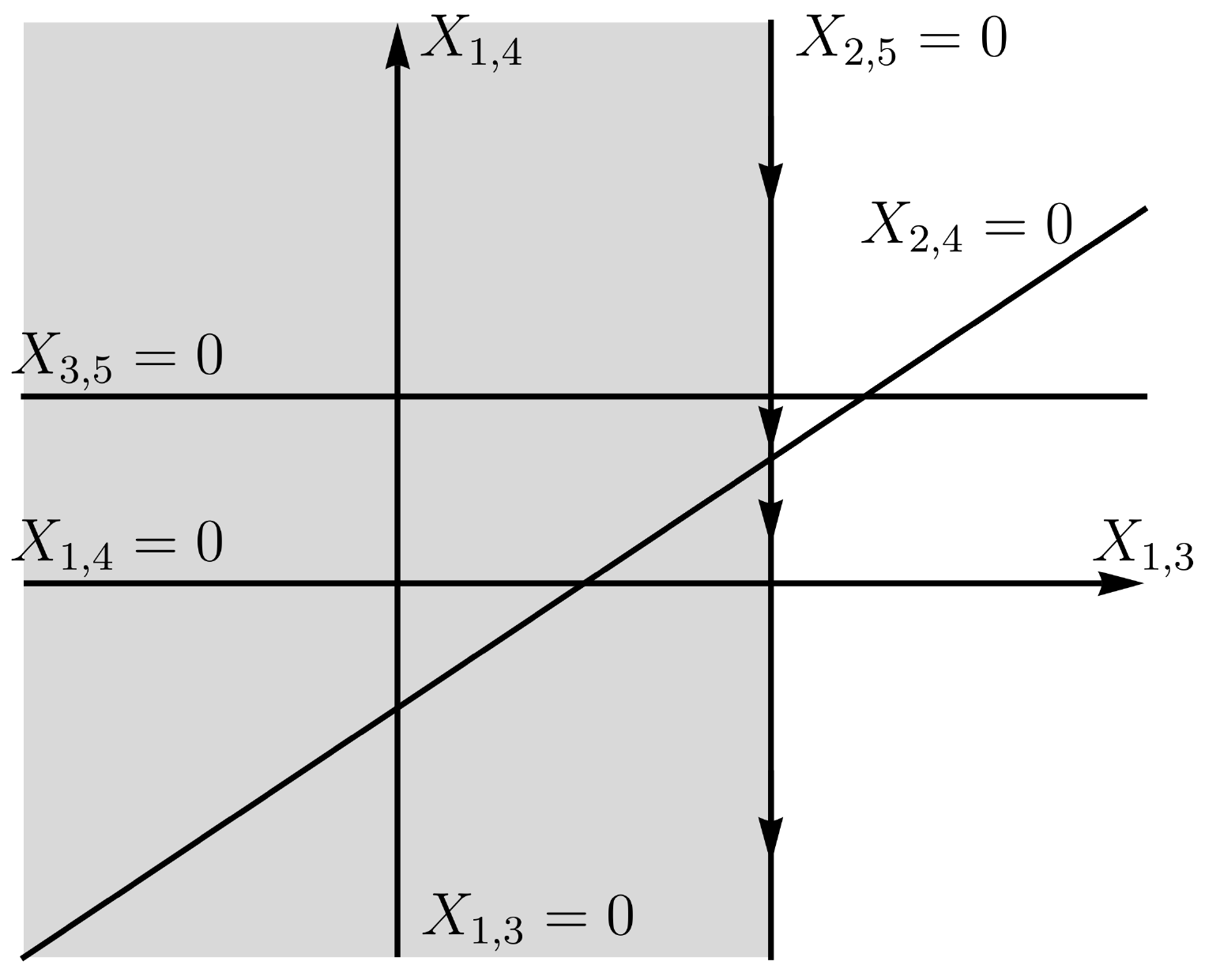}}}
	\end{align*}
	\caption{Oriented sum of three associahedra for $n=5$ giving rise to the KK relation $\omega_5^{(13425)}+\omega_5^{(14235)}+\omega_5^{(14325)}=0$.}
	\label{fig:ass-5-kk}
\end{figure}

\paragraph{Five-particle Amplitudes.} 
The five-particle case is the first time when not all permutations $\beta$ in $\mathcal{O}_5$ lead to non-empty positive regions $\Delta_5(\beta)$. In particular, $\Delta_{5}(13524)=\Delta_5(14253)=\emptyset$ and we are left with only 22 non-trivial associahedra, and $22/2=11$ distinct positive regions, which we depicted in Fig.~\ref{fig:ass5}.
The associahedron $\mathcal{A}_5^{(12345)}$ in the standard ordering is a pentagon and therefore the set of all vertices which we need to study is $\mathcal{V}_5=\{v_1,v_2,v_3,v_4,v_5\}$. These vertices can be found in the two-dimensional space parametrized by $X_{1,3}$ and $X_{1,4}$, having coordinates
\begin{align}
\underbrace{(0,0)}_{v_1}\,,\quad \underbrace{(c_{13},0)}_{v_2}\,,\quad \underbrace{(c_{13}+c_{14},c_{14})}_{v_3}\,,\quad \underbrace{(c_{13}+c_{14},c_{14}+c_{24})}_{v_4}\,,\quad\underbrace{(0,c_{14}+c_{24})}_{v_5}\,,
\end{align}
where $c_{ij}$ are positive constants.
To define boundary operators we need to consider the set of five hyperplanes $\mathcal{H}_5=\{h_1,h_2,h_3,h_4,h_5\}$ where
\begin{equation}
\underbrace{X_{1,3} =0}_{h_1}\,,\ \
\underbrace{X_{1,4} = 0}_{h_2}\,,\ \
\underbrace{X_{1,3} -c_{13} -c_{14}=0}_{h_3}\,,\ \
\underbrace{X_{1,4} - c_{14} - c_{24}=0}_{h_4}\,, \ \ 
\underbrace{X_{1,3} - X_{1,4} - c_{13}=0}_{h_5}\,.
\end{equation} 
Around each vertex $v\in\mathcal{V}_5$, every positive region $\Delta_5(\beta)$ is either a two-dimensional cone $c^{(v)}[\beta]$ spanned by two rays or it is empty. Therefore we can assign a ray two-form $\omega(c^{v}[\beta])$ to each positive region and define five boundary matrices $M^{(v)}_5$ as in \eqref{eq:-n-2-ray-M}, each of which is of size $\binom{|\mathcal{H}_5|}{2}\times \frac{p_5}{2}=10\times 11$. We need to find all elements which are common to the kernels of these matrices. To do this we stack these matrices together and define $M_5=(M_5^{(v_1)},M_5^{(v_2)},M_5^{(v_3)},M_5^{(v_4)},M_5^{(v_5)})$. We find that the kernel of $M_5$ is 6-dimensional and, given a particular choice of $11$ orderings from $\mathcal{O}_5$ labelling distinct positive regions, the kernel of $M_5$ gives rise to the following relations
\begin{align}
\begin{split}
\omega_5^{(12345)}+\omega_5^{(12354)}+\omega_5^{(12435)}+\omega_5^{(14235)}=0\,,\\
\omega_5^{(12345)}+\omega_5^{(12435)}+\omega_5^{(12453)}+\omega_5^{(13245)}=0\,,\\
\omega_5^{(12345)}+\omega_5^{(13245)}+\omega_5^{(13425)}+\omega_5^{(13452)}=0\,,\\
\omega_5^{(13425)}+\omega_5^{(14235)}+\omega_5^{(14325)}=0\,,\\
\omega_5^{(12435)}-\omega_5^{(13425)}+\omega_5^{(14352)}=0\,,\\
\omega_5^{(13245)}+\omega_5^{(13254)}-\omega_5^{(14235)}=0\,.
\end{split}
\end{align} 
These relations together with the reflection relations provide all $6+11=17$ KK relations between canonical forms $\omega_5^{(\beta)}$ and therefore between double-partial amplitudes $m_5(\beta)$.
To illustrate further how these relations arise geometrically, we include in Fig.~\ref{fig:ass-5-kk} the oriented sum of the associahedra participating in the fourth KK relation listed above. The oriented sum produces the bounded region where $X_{2,5}\ge 0$. Since this bounded region does not have any vertices, the corresponding sum of canonical differential forms must vanish.

\section{Conclusions and Outlook}

In this paper, we showed how the KK relations for tree-level partial amplitudes originate in the framework of positive geometries. In particular, we derived the KK relations in $\mathcal{N}=4$ sYM and in bi-adjoint $\phi^3$ theory from the geometry of the momentum amplituhedron and the kinematic associahedron, respectively. 
To this purpose, we started by defining these positive geometries for orderings of the external particles different from the standard. The momentum amplituhedron and the kinematic associahedron are defined as the intersection of two spaces -- a proper-dimensional affine subspace of the appropriate kinematic space and a winding space/positive region.  For both geometries, we can choose the positive region to depend on the particle ordering, while the affine subspace remains the same across different orderings. In this way, we can directly compare the positive geometries for different orderings and study how they fit together. 
After formulating the notion of an oriented sum for positive geometries, we were able to determine which collections of positive geometries give rise to a vanishing sum of canonical differential forms. 
We showed that, whenever the oriented sum of positive geometries has no zero-dimensional boundaries, i.e.~vertices, the corresponding sum of canonical forms must vanish. 
This serves as the principle underlying the geometric representation of the KK relations.

Using this guideline, we were able to  derive the KK relations from the geometry of the momentum amplituhedron. We presented two procedures, both homological in nature. The first algorithm was applicable only to the MHV sector, where the definition of the $k=2$ momentum amplituhedron for $n$ particles naturally gives rise to a complete fan 
of oriented simplicial cones in $\mathbb{R}^{n-2}$ (incidentally, this fan is dual to the permutohedron of order $(n-1)$). This algorithm exploited the fact that each positive sector -- each simplicial cone -- could be written as the positive span of $(n-2)$ rays. The essence of this method was that we were able to identify which pairs of rays pointed in opposite directions; these were the rays which lived in the same one-dimensional intersection of $(n-3)$ facet-defining hyperplanes. 
By abstracting this notion, and taking as input the combinatorial structure of boundaries of the momentum amplituhedron for different orderings, we presented a poset-based homological algorithm for deriving the KK relations in any helicity sector. 
Afterwards, we moved to consider the kinematic associahedron. In order to compare kinematic associahedra for different particle orderings we modified the original definition presented in \cite{Arkani-Hamed:2017mur} such that the same affine subspace could be shared. In the neighbourhood of each vertex, kinematic associahedra can be described as oriented simplicial cones and in doing this we were able to reuse the ray-based homological algorithm developed for MHV amplitudes to derive all KK relations in this context. 

The results found in this paper are surprising since the KK relations, which are group-theoretic in nature,  arise geometrically from partial amplitudes which do not carry any information about color.  Nevertheless, we see that the notion of positivity with respect to some ordering is rich enough to fully encode all KK relations between color-ordered amplitudes. 
This fits well with the idea that ``color is kinematics"  \cite{Arkani-Hamed:2017mur}.

This paper opens various directions for future research.
The first direction to consider is how to formalise the oriented sum of positive geometries introduced here. In particular, can one formulate a well-defined set-theoretic definition for the oriented sum of semi-algebraic sets in the real slice of some complex projective algebraic variety? The idea of adding positive geometries already appeared in a previous paper \cite{Damgaard:2020eox} where we considered the sum of canonical forms for momentum amplituhedra over different helicity sectors. The oriented sum also appears in the cancellation of spurious boundaries in triangulations of a positive geometry. This mathematical point certainly deserves further investigation. 

Furthermore, having demonstrated how the KK relations are realized geometrically from the momentum amplituhedron and the kinematic associahedron, it is natural to 
ask whether the Bern-Carrasco-Johansson (BCJ) relations \cite{Bern:2008qj} can also be derived from a geometric perspective.  In particular, can we find a  meaning for the kinematic pre-factors multiplying partial amplitudes in these relations? 
It would be very interesting to understand whether and how these kinematic pre-factors could arise in a geometric fashion. We leave this to future work.

\section{Acknowledgements}

We would like to thank Nima Arkani-Hamed for suggesting the problem.
This work was partially funded by the Deutsche Forschungsgemeinschaft (DFG, German Research Foundation) -- Projektnummern 404358295 and 404362017.

\appendix
\section{Poset Intervals for MHV Four-point Amplitudes}
\label{app:posets42}

\begin{figure}[!ht]
	\centering
	\includegraphics[height=10cm]{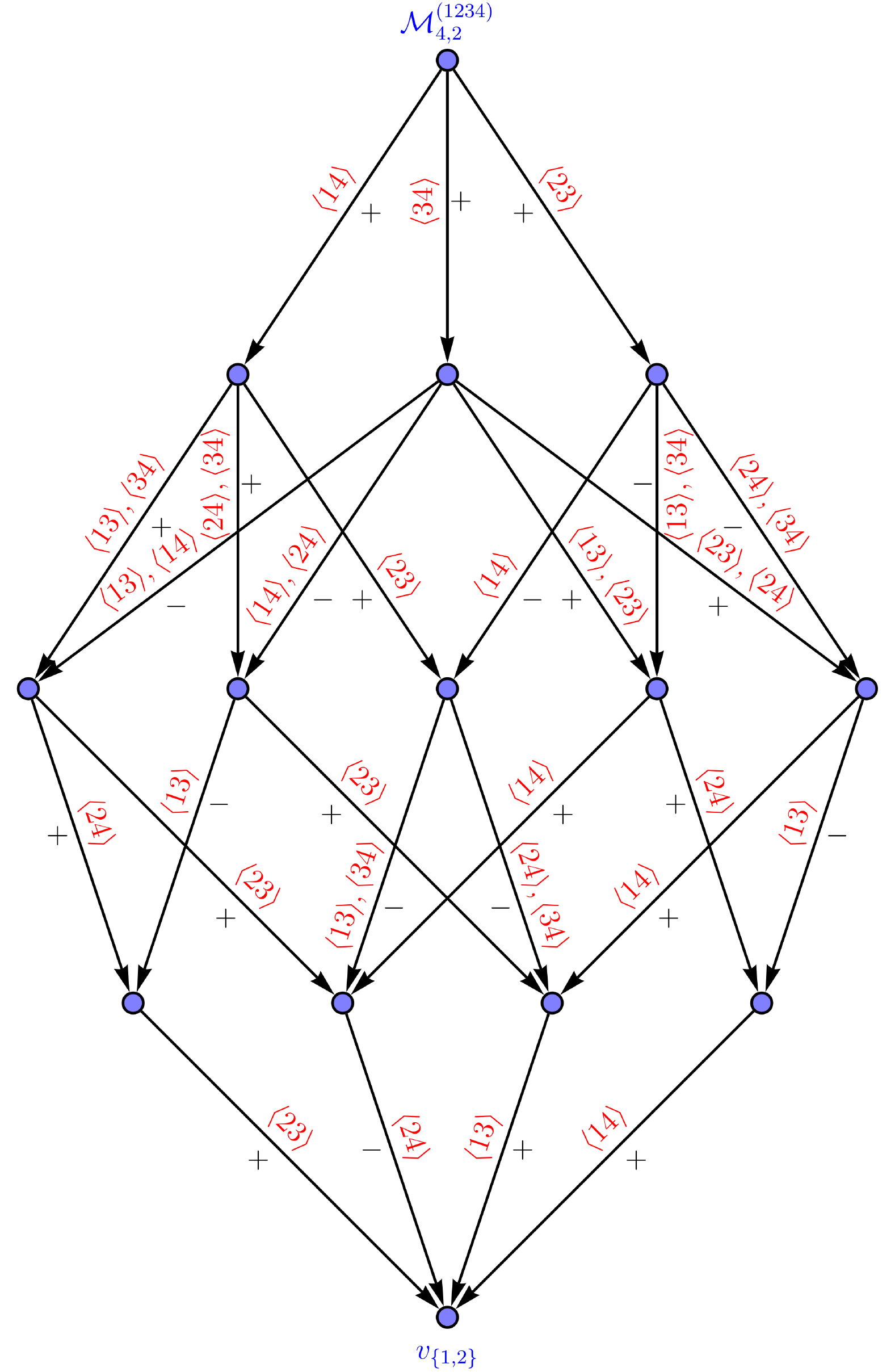}	
	\includegraphics[height=10cm]{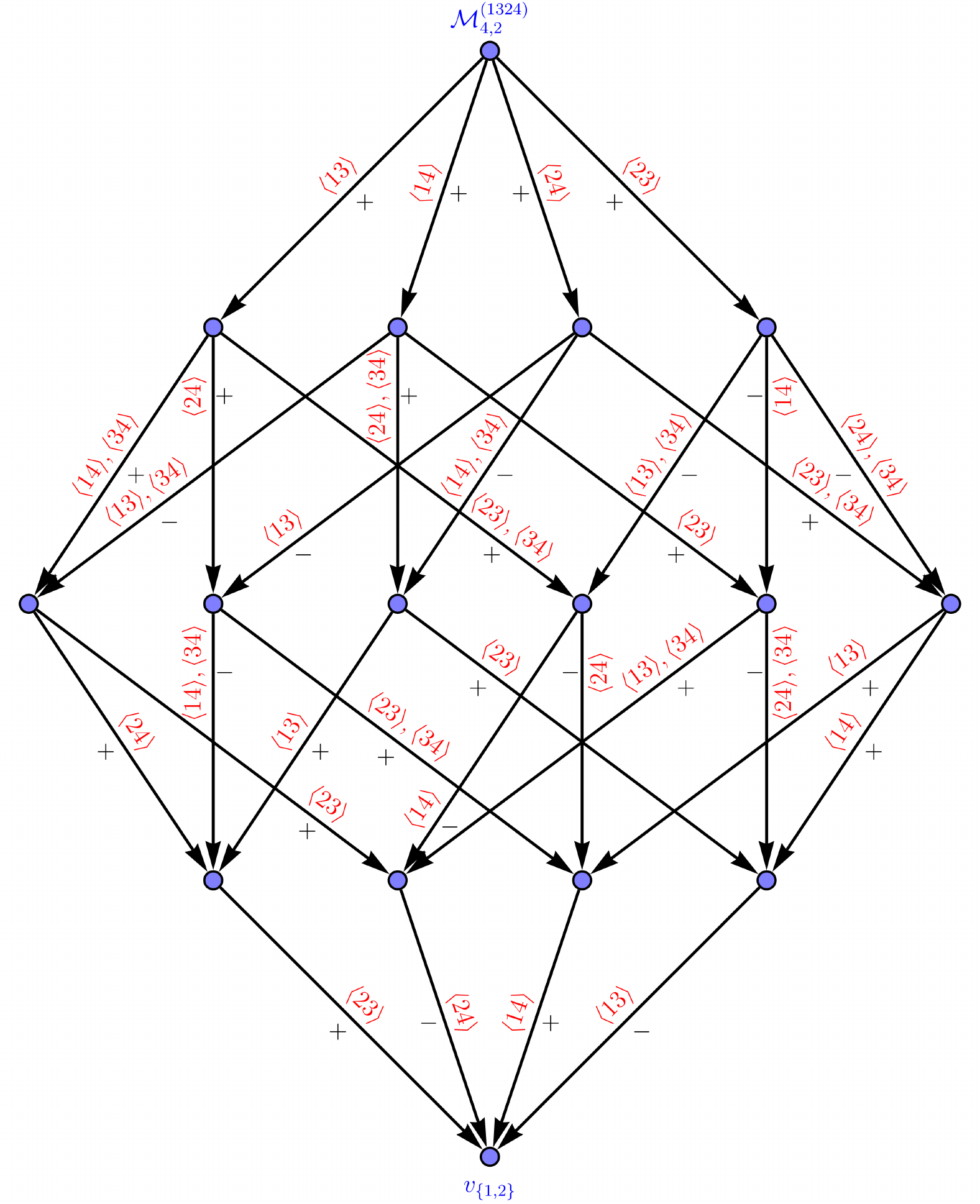}
	\caption{Hasse diagrams of the interval between $\mathcal{M}^{(1234)}_{4,2}$ (left) and $\mathcal{M}^{(1324)}_{4,2}$ (right), and the vertex $v_{\{1,2\}}$.}
	\label{full4-2-poset1}
\end{figure}

In this appendix we present in Fig.~\ref{full4-2-poset1} the Hasse diagrams of the intervals between the momentum amplituhedra $\mathcal{M}^{(1234)}_{4,2}$, $\mathcal{M}^{(1324)}_{4,2}$ and the zero-dimensional boundary, or vertex, $v_{\{1,2\}}$ using the full boundary stratifications. The intervals for the other orderings are topologically equivalent to the ones presented here. In the Hasse diagrams, we explicitly indicate the edge labels, i.e.~the spinor brackets which vanish when approaching a specific boundary, and a diamond-compatible assignment of signs to every edge. 

\bibliographystyle{nb}

\bibliography{KKmomamp}

\end{document}